\documentclass[fleqn,usenatbib]{mnras}

\usepackage[T1]{fontenc}
\usepackage{ae,aecompl}
\usepackage{lineno}

\DeclareRobustCommand{\VAN}[3]{#2}
\let\VANthebibliography\thebibliography
\def\thebibliography{\DeclareRobustCommand{\VAN}[3]{##3}\VANthebibliography}

\usepackage{graphicx}	
\usepackage{amsmath}	
\usepackage{amssymb}	
\usepackage{graphicx}
\usepackage{times}
\usepackage{mdframed}
\usepackage{xcolor}
\usepackage{url}
\newcommand{\HI}{\hbox{\rmfamily H\,{\textsc i}}}
\newcommand{\HIsub}{\hbox{{\scriptsize H}\,{\tiny I}}}
\newcommand{\MHI}{\hbox{$M_{\HIsub}$}}
\newcommand{\rhoHI}{\hbox{$\rho_{\HIsub}$}}

\newcommand{\rhoCrit}{\hbox{$\rho_{{\rm crit}, 0}$}}
\newcommand{\rhoLr}{\hbox{$\rho_{L_r}$}}

\newcommand{\OHI}{\hbox{$\Omega_{\HIsub}$}}
\newcommand{\msun}{\hbox{${\rm M}_{\odot}$}}

\newcommand{\lsun}{\hbox{${\rm L}_{\odot}$}}

\newcommand{\kms}{\hbox{km\,s$^{-1}$}}
\newcommand{\askapsoft}{ASKAP{\sc soft}}
\newcommand{\CII}{\hbox{[C\,{\sevensize II}]}}

\title[ASKAP DINGO Early Science]{Deep Investigation of Neutral Gas Origins (DINGO): {\HI} stacking experiments with early science data}

\author[J. Rhee et al.]{
Jonghwan Rhee$^{1,2}$\thanks{E-mail: jonghwan.rhee@uwa.edu.au}, Martin Meyer$^{1,2}$, Attila Popping$^{1,3}$, Sabine Bellstedt$^{1}$, Simon P. Driver$^{1}$, 
\newauthor Aaron S. G. Robotham$^{1}$, Matthew Whiting$^{4}$, Ivan K. Baldry$^{5}$, Sarah Brough$^{6,2}$, Michael J. I. Brown$^{7}$, 
\newauthor John D. Bunton$^{4}$, Richard Dodson$^{1}$, Benne W. Holwerda$^{8}$, Andrew M. Hopkins$^{9}$, B\"arbel S. Koribalski$^{4,10}$, 
\newauthor Karen Lee-Waddell$^{1,11}$, \'Angel R. L\'opez-S\'anchez$^{9,12,13,2}$, Jon Loveday$^{14}$, Elizabeth Mahony$^{4}$, 
\newauthor Sambit Roychowdhury$^{1,2}$, Krist\'of Rozgonyi$^{1,2,15}$ and Lister Staveley-Smith$^{1,2}$\\
\\
$^{1}$International Centre for Radio Astronomy Research (ICRAR), University of Western Australia, 35 Stirling Hwy, Crawley, WA 6009, Australia \\
$^{2}$ARC Centre of Excellence for All Sky Astrophysics in 3 Dimensions (ASTRO 3D), Australia \\
$^{3}$ARC Centre of Excellence for All-sky Astrophysics (CAASTRO), Australia \\
$^{4}$Australia Telescope National Facility, CSIRO Space \& Astronomy, P.O. Box 76, Epping, NSW 1710, Australia \\
$^{5}$Astrophysics Research Institute, Liverpool John Moores University, IC2, Liverpool Science Park, 146 Brownlow Hill, Liverpool, L3 5RF \\
$^{6}$School of Physics, University of New South Wales, NSW 2052, Australia \\
$^{7}$School of Physics and Astronomy, Monash Centre for Astrophysics (MoCA), Monash University, Clayton, Victoria 3800, Australia \\
$^{8}$University of Louisville, Department of Physics and Astronomy, 102 Natural Science Building, 40292 KY Louisville, USA \\
$^{9}$Australian Astronomical Optics, Macquarie University, 105 Delhi Rd, North Ryde, NSW 2113, Australia \\
$^{10}$Western Sydney University, Locked Bag 1797, Penrith South DC, NSW 2751, Australia \\
$^{11}$CSIRO Space \& Astronomy, PO Box 1130, Bentley WA 6102, Australia \\
$^{12}$Department of Physics and Astronomy, Macquarie University, NSW 2109, Australia \\
$^{13}$Macquarie University Research Centre for Astronomy, Astrophysics \& Astrophotonics, Sydney, NSW 2109, Australia \\
$^{14}$Astronomy Centre, University of Sussex, Falmer, Brighton BN1 9QH, UK\\
$^{15}$Faculty of Physics, Ludwig-Maximilians-Universit\"at, Scheinerstr. 1, 81679 Munich, Germany
}

\pubyear{2021}

\begin{document}
\label{firstpage}
\pagerange{\pageref{firstpage}--\pageref{lastpage}}
\maketitle

\begin{abstract}
  We present early science results from Deep Investigation of Neutral Gas Origins (DINGO), an {\HI} survey using the Australian Square Kilometre Array Pathfinder (ASKAP). Using ASKAP sub-arrays available during its commissioning phase, DINGO early science data were taken over $\sim$~60~deg$^{2}$ of the Galaxy And Mass Assembly (GAMA) 23~h region with 35.5 hr integration time. We make direct detections of six known and one new sources at $z < 0.01$. Using {\HI} spectral stacking, we investigate the {\HI} gas content of galaxies at $0.04 < z< 0.09$ for different galaxy colours. The results show that galaxy morphology based on optical colour is strongly linked to {\HI} gas properties. To examine environmental impacts on the {\HI} gas content of galaxies, three sub-samples are made based on the GAMA group catalogue. The average {\HI} mass of group central galaxies is larger than those of satellite and isolated galaxies, but with a lower {\HI} gas fraction. 
We derive a variety of {\HI} scaling relations for physical properties of our sample, including stellar mass, stellar mass surface density, $NUV-r$ colour, specific star formation rate, and halo mass.
We find that the derived {\HI} scaling relations are comparable to other published results, with consistent trends also observed to $\sim$0.5 dex lower limits in stellar mass and stellar surface density.
The cosmic {\HI} densities derived from our data are consistent with other published values at similar redshifts.
DINGO early science highlights the power of {\HI} spectral stacking techniques with ASKAP.
\end{abstract}




\section{Introduction}

The 21~cm line emission of neutral atomic hydrogen ({\HI}) has the potential to revolutionise our understanding of galaxy evolution and 
cosmology as the Square Kilometre Array (SKA) and its pathfinders become available \citep{Abdalla:2015,Blyth:2015,Giovanelli:2015,Staveley-Smith:2015,Koribalski:2020}.
Before the first stars and galaxies formed (the so-called dark ages), the Universe was full of neutral hydrogen gas ({\HI}), a significant fraction of which has since been transformed into stars, resulting in the \mbox{Universe} that we observe today.
Neutral hydrogen gas is a good tracer of the large-scale galaxy distribution, and a good probe of the interstellar medium (ISM) within galaxies. 
Kinematic information derived from {\HI} maps of galaxies has been extensively used for studies of the dynamics and structure of the ISM, leading to the recognition of dark matter and the measurements of its distribution \citep[e.g.][]{Bosma:1978, Bosma:1981, Bosma:1981a}. 
Outside galaxies, in the intergalactic medium (IGM), most hydrogen is in the warm-hot phase and has been ionised since the epoch of reionisation (EoR).
{\HI} in galaxies also plays a vital role in understanding environmental effects on galaxy transformation and evolution \citep{Haynes:1984}.
The majority of galaxies in the local Universe live in group-like environments \citep{Eke:2006, Robotham:2011} where galaxies can be easily affected
by a variety of \mbox{mechanisms} such as tidal interaction, harassment, ram pressure stripping and evaporation
\citep[e.g.][]{Gunn:1972, Moore:1996,Chung:2009,Denes:2016,Cortese:2021}.
These environmental effects can be easily seen around the outskirts of a galaxy where {\HI} can extend to larger distances than the optical radii. 
In this sense, {\HI} gas, the raw fuel supply to form stars, is indispensable in understanding galaxy formation and evolution. 

As the sensitivity, field-of-view (FoV) and spectral bandwidth of radio telescopes have dramatically improved over the last few decades, the survey area and depth (that is, survey speed) at 21 cm have correspondingly increased.  From a small number of optically-selected galaxies in the past, current-day research is able to benefit from large-area blind surveys \citep[e.g.][]{Zwaan:1997,Rosenberg:2000,Meyer:2004,Giovanelli:2005}.
However, recent {\HI} surveys have still been confined to gas-rich galaxies in the local Universe
($z < 0.1$) due to the inherent weakness of 21~cm emission and the limited capability of the existing observing facilities. Only a few hundred galaxies have been detected in {\HI} beyond the local Universe at the expense of substantial observing time \citep[e.g.][]{Fernandez:2013, Jaffe:2013, Catinella:2015, Fernandez:2016, Hess:2019,Xi:2021}. 
This situation is changing through the improved sensitivity and larger FoV of the SKA precursors now coming online  and providing faster survey speeds.

The Australian Square Kilometre Array Pathfinder \citep[ASKAP,][]{Johnston:2007, Johnston:2008,Hotan:2021}
radio telescope is one of the SKA precursors and will open up a new window for large extragalactic {\HI} surveys beyond the local Universe due to its wide spectral bandwidth and large instantaneous FoV.
ASKAP consists of 36 dishes, each of diameter 12 m and equipped with phased array feeds (PAFs) forming multiple receiving beams electronically \citep{DeBoer:2009, Hampson:2012}. The phased array feed technology allows ASKAP to have a large $30~{\rm deg}^{2}$ FoV \citep{Bunton:2010} and a wide bandwidth of $288~{\rm MHz}$, which makes it an optimal survey instrument, enabling it to conduct both wide and deep surveys in a comparatively short period of time \citep{Hotan:2021}. The ASKAP prototype, the Boolardy Engineering Test Array \citep[BETA,][]{Hotan:2014,McConnell:2016}, has also demonstrated the capability and performance of ASKAP as a survey instrument \citep[e.g.][]{Serra:2015b, Allison:2015, Harvey-Smith:2016, Heywood:2016, Hobbs:2016, Allison:2017, Moss:2017}.

Previous deep {\HI} surveys have also been limited by small volumes, implying that 
measurements were subject to the effects of cosmic variance \citep[e.g. $\sim$~29 per~cent for AUDS\footnote{Arecibo Ultra Deep Survey},][]{Hoppmann:2015,Xi:2021}. 
The wide-field imaging capability of ASKAP is beneficial for both wide and deep {\HI} surveys in order to obtain the largest volumes of the Universe ever explored in {\HI}, thereby helping us to obtain cosmologically representative {\HI} datasets where the impact of cosmic variance is minimised.

\subsection{DINGO}
Deep Investigation of Neutral Gas Origins \citep[DINGO,][]{Meyer:2009a, Meyer:2009b} is an ASKAP deep {\HI} survey project aiming to provide a cosmologically representative dataset for {\HI} emission, 
enabling studies of the {\HI} gas content of galaxies out to $z\sim 0.43$, corresponding to the past 4 billion years. 
The sky coverage of the DINGO survey is wider than deep {\HI} surveys previously conducted 
and the ongoing deep {\HI} surveys being carried out with other telescopes such as  the VLA\footnote{The Karl G. Jansky Very Large Array}, MeerKAT\footnote{The Meer-Karoo Array Telescope} and FAST\footnote{Five-hundred-meter Aperture Spherical Telescope} (CHILES\footnote{The COSMOS {\HI} Large Extra-galactic Survey \citep{Fernandez:2013}}, LADUMA\footnote{Looking At the Distant Universe with the MeerKAT Array \citep{Holwerda:2012,Blyth:2016}}/MIGHTEE-HI\footnote{The {\HI} emission project of the MeerKAT International GigaHertz Tiered Extragalactic Exploration survey \citep{Maddox:2021}}, and FUDS\footnote{FAST Ultra Deep Survey \citep{Xi:2022}} respectively).

To maximise science returns from the DINGO survey it is optimal to target optical galaxy redshift survey fields because they allow comprehensive studies of the complex physical processes that drive galaxy formation and evolution through the combination of {\HI} and multi-wavelength data \citep{Meyer:2015}.  
The Galaxy and Mass Assembly \citep[GAMA,][]{Driver:2011, Hopkins:2013, Liske:2015, Baldry:2018, Driver:2022} and Wide-Area VISTA Extragalactic Survey \citep[WAVES,][]{Driver:2019} regions are hence excellent target fields for DINGO observations. 

DINGO will explore the {\HI} universe beyond the regime probed by previous {\HI} surveys in conjunction with the physical properties of galaxies obtained from the multi-wavelength GAMA data available.
This can be achieved by constraining {\HI} properties accurately and reliably with direct and stacked detections of the deep {\HI} data of DINGO.
The scientific goals of the DINGO survey address three central questions: (1) How has the {\HI} content of the Universe evolved and what factors have driven its distribution as a function of halo mass? (2) What role has {\HI} played in the baryonic cycle of galaxies and how has this changed as a function of environment and time? (3) What factors have driven the assembly of angular momentum in galaxies and its relationship to {\HI} content?

\subsection{DINGO early science}
\label{sec:dingo}

\begin{figure*}
\centering
\includegraphics[width=0.95\textwidth]{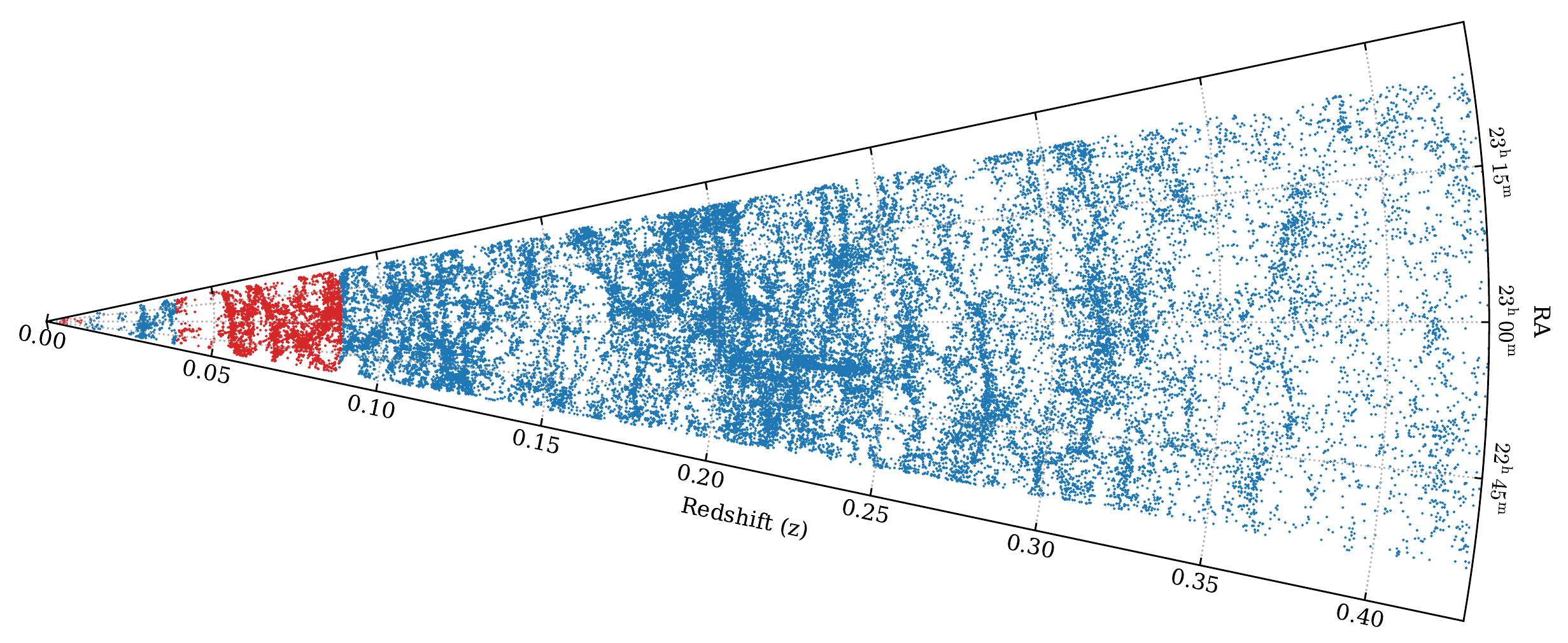} 
\caption[G23 redshift cone diagram]{The redshift cone diagram of the GAMA 23~h field in RA out to $z < 0.43$. The red dots denote the galaxies in the redshift range of this DINGO early science study.}
\label{fig:redshift_cone_diagram}
\end{figure*}

\begin{table*}
\caption[ASKAP DINGO observations]{The summary of ASKAP DINGO early science observations.}
\centering
\label{tab:askap_obs}
\begin{tabular}{@{}cccccccccc}
\hline
     SBID  &  Date  &  Obs. Time  &  Data Volume  &  Num of Antennas & Tile ID & Band ID &  Bandwidth & Centre Frequency & Redshift range\\
               &           &  [h]    &             [TB]        &                               &   &      &  [MHz]        &  [MHz]     & \\
     \hline
     4191  &  2017 Sep 7   &   8.6  &  2.6  &  12  &  0  & 2 &  192 &  1344.5  &   $z < 0.14$ \\
     4208  &  2017 Sep 8   &   8.8  &  2.6  &  12  &  0  & 2 &  192 &   1344.5  &  $z< 0.14$  \\
     4216  &  2017 Sep 9   &   8.7  &  2.6  &  12  &  1  & 2 &  192 &   1344.5  & $z< 0.14$  \\
     4240  &  2017 Sep 11 &   9.4  &  2.8  &  12  &  1  & 2 &  192 &   1344.5  &  $z< 0.14$  \\
      \hline  
\end{tabular}
\end{table*}

During the ASKAP early science phase between 2016 and 2018, each ASKAP survey project was granted limited observing time, with reduced bandwidth and numbers of antennas (see Table~\ref{tab:askap_obs}). We therefore focussed on DINGO science that can use the GAMA Data Release 4 and its value-added catalogues \citep{Driver:2022}.
{\HI} stacking is a powerful tool to conduct statistical analysis as well as to extend survey limits. 
It has contributed to many science cases related to {\HI} since the first pioneering work by \citet{Zwaan:2001} and \citet{Chengalur:2001} including: the evolution of {\HI} gas content of galaxies \citep{Lah:2007, Lah:2009, Delhaize:2013, Rhee:2013, Rhee:2016, Kanekar:2016,
  Rhee:2018, Bera:2019, Hu:2019, Chowdhury:2020, Chen:2021b, Chowdhury:2021}; {\HI} scaling relations and environment effects \citep{Fabello:2011, Fabello:2011a, Gereb:2013, Gereb:2015, Brown:2015, Meyer:2016, Brown:2017, Kleiner:2017, Healy:2019, Hu:2020, Guo:2020,Hu:2021,Roychowdhury:2022,Sinigaglia:2022}; and {\HI} absorption studies \citep{Gereb:2014}.
Recently, a new technique based on a Bayesian approach has been proposed to measure the {\HI} mass function and Tully-Fisher relation \citep{Pan:2020, Pan:2021}.
Despite only having 36 hr of observing time with fewer than half the full set of ASKAP antennas, a small fraction of the final DINGO allocation of 3200 hr on the full ASKAP array, the early science data nicely demonstrates the potential of ASKAP for {\HI} stacking, and the power of complementary wide-area, deep optical surveys such as GAMA.

This paper is structured as follows. In Section~2, we describe the target field, observations, data reduction and data quality assessment procedure. In Section~3, we describe the {\HI} stacking methodology for the DINGO early science data analysis. Section~4 presents the direct detection results. Section~5 and 6 present the results of {\HI} stacking  for measurements of the {\HI} gas content of galaxies and {\HI} scaling relations, respectively.  In Section~7 we focus on the cosmic {\HI} mass density measurements ({\OHI}), followed by our summary and conclusions in Section~8.
All stellar masses quoted in this paper use a \citet{Chabrier:2003} initial mass function (IMF), and all magnitudes are in the AB system. 
We adopt the concordance cosmological parameters of $\Omega_{\Lambda}=$~0.7, $\Omega_{M}=$~0.3 and $H_{0}=$~70~km~s$^{-1}$~Mpc$^{-1}$  throughout this paper.

\section{DINGO Early Science Data}

\begin{figure*}
\centering
\includegraphics[width=0.9\textwidth]{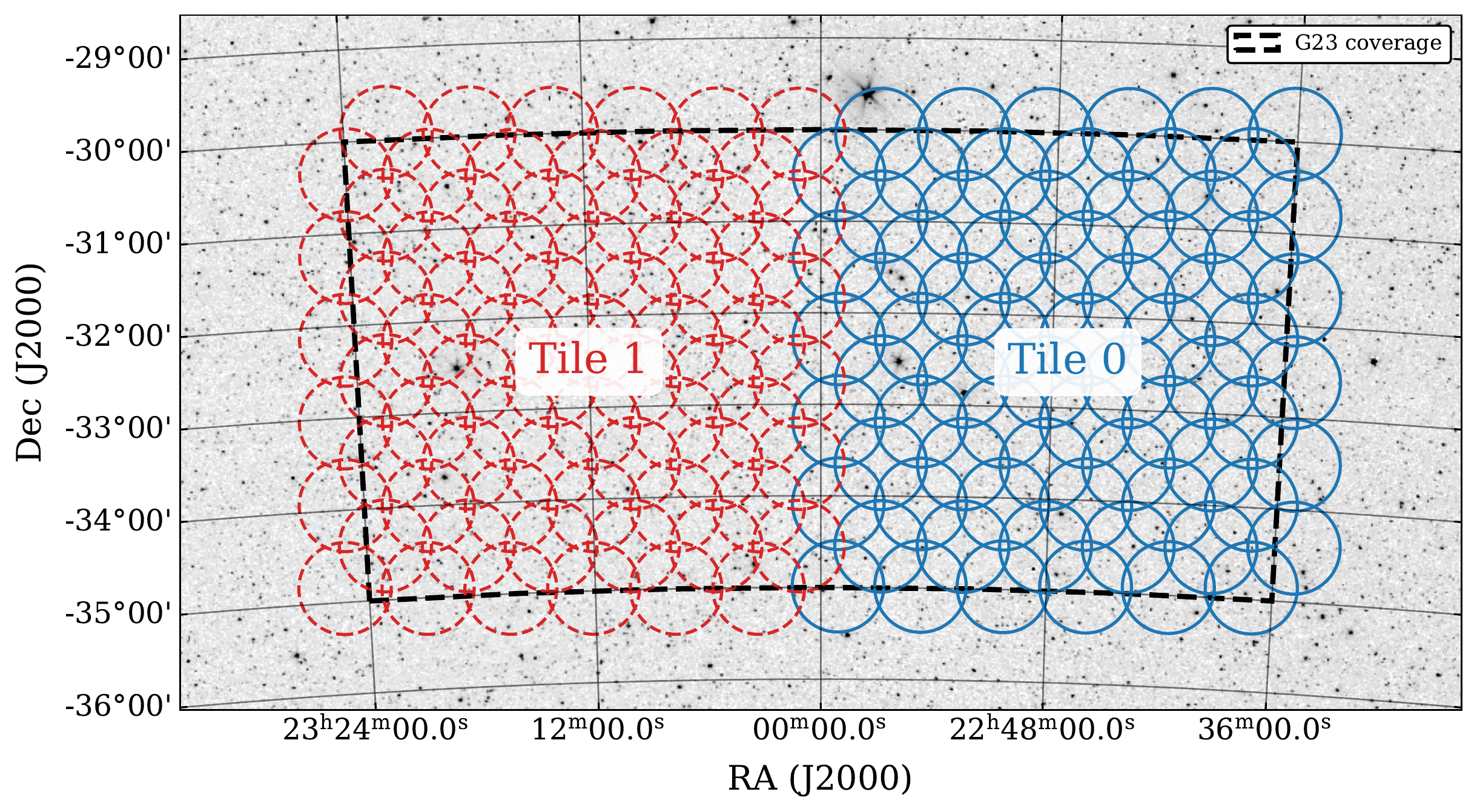} 
\caption[ASKAP 36 beams footprint]{ASKAP beam footprints used for the G23 field observations. Blue (right half) and red (left half) circles each of 1.0\degr in diameter are beams of Tile 0 and 1, respectively. Each tile consists of two interleaving 36-beam patterns of each 6-by-6 footprint generated by the ASKAP PAFs. The black dashed lines denote the nominal G23 survey coverage. The background image is from the optical all-sky image of \citet{Mellinger:2009}.}
\label{fig:beamFootprints}
\end{figure*}

\subsection{Target Field}
For DINGO early science, complementary data from multi-wavelength datasets are essential (e.g. photometry, spectroscopic redshifts and derived values) to facilitate {\HI} spectral stacking experiments. Due to the limited observing time given to each ASKAP survey science project during the early science phase, direct detections are only possible at low redshifts. 
In particular, complete spectroscopic redshift surveys are the key product that enables {\HI} spectral stacking analyses.
GAMA survey fields are optimal for this purpose and also for maximising science returns 
due to the availability of highly-complete AAT spectroscopy \citep{Liske:2015} as well as a variety of imaging data taken using ground and space-based telescopes (e.g. {\it GALEX}, VST, VISTA, UKIRT, {\it WISE}, {\it Herschel}), spanning wavelengths from the ultraviolet (UV) to the far infrared \citep{Driver:2016}.

The GAMA 23~h field (hereafter G23) is a good target as its coordinates lie below the celestial equator (better ASKAP point spread function) and a large number of redshifts ($\sim 40~{\rm k}$) available out to $z\sim 0.43$ (the upper limit of the DINGO survey) from the G23 catalogue \citep{Liske:2015, Bellstedt:2020} as seen in Fig.~\ref{fig:redshift_cone_diagram}. As well as being a target for the DINGO survey, the G23 field has been selected for other ASKAP early science commissioning observations \citep[e.g.][]{Leahy:2019, Allison:2020,Gurkan:2022}.

\subsection{Observations}
\label{sec:obs}
We observed the G23 field centred at  $\alpha, \delta$ (J2000) = $22^{\rm h}59^{\rm m} 00 \rm \fs00, -32\degr18\arcmin00\farcs0$ for DINGO early science commissioning, which was conducted in September 2017.
For the early science observations, 12 of ASKAP's 36 antennas (so called ASKAP-12) were used with bandwidths of 192~MHz (10368 channels) in the ASKAP receiver band~2 \citep{Hotan:2021} and a channel width of 18.5~kHz (equivalent to a velocity resolution of $\sim 4~\kms$ at $z = 0$). The baseline lengths of the sub-arrays were from 22~m to 2.3~km.

The target field of 12 $\times$ 5 $\rm deg^{2}$ was divided into two tiles, each having a large instantaneous FoV of $\sim 30~\rm deg^{2}$. 
Each tile was then covered by two interleaving pointings, each of which forms a square 6~$\times$~6 beam footprint generated by the ASKAP PAFs (see Fig.~\ref{fig:beamFootprints}). For DINGO observations, the two interleaving pointings were changed every 15~min to have one observation block, called a scheduling block, containing two footprints to cover an entire tile at a time. Each beam is 1~deg in diameter (FWHP) with a pitch between two beams of 0.9~deg.
As seen in Fig.~\ref{fig:beamFootprints}, the tiling and beamforming strategy resulted in forming a total of 144 beams across the entire target field. This beam configuration is designed to deliver uniform sensitivity across the field.

The total integration time for data obtained for DINGO early science is about 36 hr, which correspond to  $\sim 11$~TB of data.
For flux and bandpass calibration, we observed PKS~B1934-638, which was placed at the centre of each beam with a 3~min integration time, giving a total of 2 hr for each calibrator observation before or after the science observation. These calibrator observations amount to an additional $\sim 2.6$~TB of data. 
Full details are given in Table~\ref{tab:askap_obs}.

\subsection{Data Reduction}

\begin{table*}
\caption[Spectral windows for DINGO data processing]{The spectral windows of the ASKAP DINGO band 2 data split for processing.}
\centering
\label{tab:spws}
\begin{tabular}{@{}cccccc}
\hline
     Spectral Window ID  &  Bandwidth  &  Frequency Range  &  Redshift range &  Num of channels & Status of processing \\
             (SPW)                &   [MHz]         &  [MHz]                   &                          &   &                           \\
     \hline
     SPW0  &  16   &   1404.5 - 1420.5  & $ z < 0.011$  & 864  & Done     \\
     SPW1  &  16   &   1388.5 - 1404.5  &  $0.011 < z < 0.023$  &  864  & -  \\
     SPW2  &  22   &   1366.5 - 1388.5  &  $0.023 < z < 0.040$  &  1188  & - \\
     SPW3  &  32   &   1335.0 - 1367.0  &  $0.039 < z < 0.064$  &  1728  & Done\\
     SPW4  &  32   &   1305.0 - 1337.0  &  $0.062 < z < 0.088$  &  1728  & Done \\  
     \hline  
\end{tabular}
\end{table*}

\begin{figure}
\centering
\includegraphics[width=0.45\textwidth]{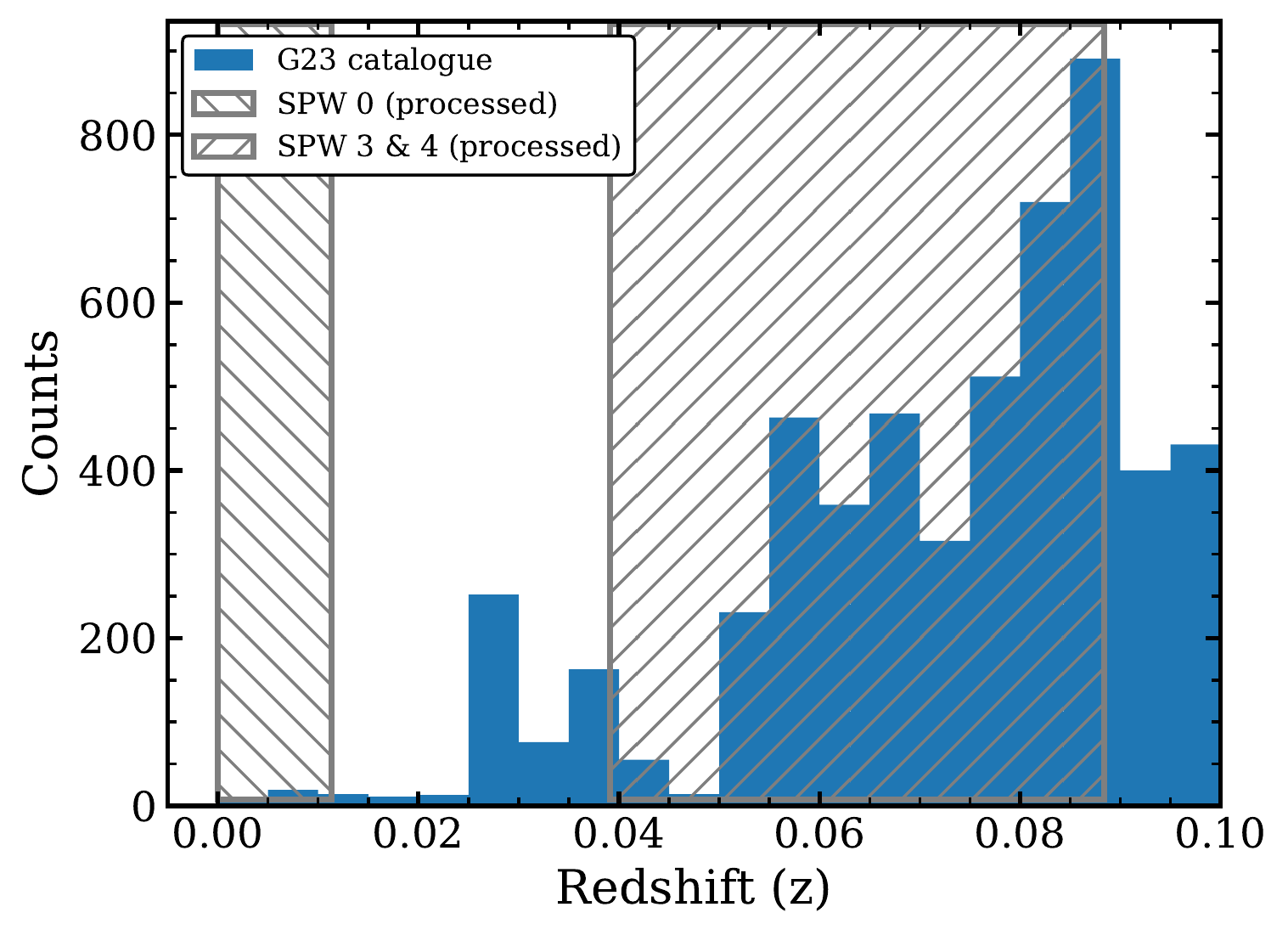}
\caption{The redshift distribution of galaxies in the G23 catalogue out to $z \sim 0.1$. The hatched areas indicate the processed DINGO early science bands.} 
\label{fig:z_dist_processed}
\end{figure}

DINGO early science data were processed using the dedicated ASKAP data processing software,  {\askapsoft} \citep{Guzman:2019} designed to calibrate the majority of ASKAP data (e.g. continuum and spectral line) and produce science images in a high-performance computing environment. {\askapsoft} consists of a number of modules integrated into a data processing pipeline \footnote{\url{http://www.atnf.csiro.au/computing/software/askapsoft/sdp/docs/current/index.html}} which, due to the large data rate and volume generated by ASKAP, is automated and parallelised.
All ASKAP data processing was conducted on the 
{\it Galaxy}\footnote{\url{https://pawsey.org.au/systems/galaxy/}} cluster in the Pawsey supercomputing centre.
A comprehensive description of the {\askapsoft} pipeline is available in \citet{Cornwell:2011, Whiting:2017, Wieringa:2020}.
We provide a brief summary here, focusing on DINGO data processing. 

The processing of DINGO data follows standard {\HI} data reduction procedures, such as editing (or flagging), calibration, self-calibration, continuum subtraction and spectral line imaging. 
A DINGO data set for each night (i.e. one scheduling block) consists of data for 72 separate beams from two interleaving pointings (two 6$\times$6 footprints A and B, respectively) as described in Section~\ref{sec:obs}. For processing, each beam is split from the raw visibility data and calibrated on a per-beam basis.

The primary calibrator, PKS~B1934-638, is used for flux scaling and bandpass calibration, based on the flux model of \citet{Reynolds:1994}.
After splitting, calibrator data sets are inspected to discard data contaminated by radio frequency interference (RFI) using an automated {\askapsoft} flagging module (dynamic flagging). 
Bandpass solutions are obtained by {\askapsoft}  by excluding baselines shorter than 200~m which are vulnerable to solar interference due to calibration observations often being scheduled during daytime.

Science data are also flagged to remove bad data after the bandpass calibration is  applied.
To this end, the amplitude and Stokes~V thresholding methods of the {\askapsoft} flagging utility are applied in the time and spectral domain. 
The data are then copied and averaged into 1~MHz-wide channels for continuum imaging and self-calibration to correct for time-dependent phase errors. Two phase self-calibration loops are applied to obtain a gain solution table, which is subsequently transferred to the full spectral resolution data (18.5~kHz channel width) for spectral line imaging.

Before spectral-line imaging, the continuum model components of continuum sources are subtracted from the calibrated spectral line visibility data. The spectral-line data is imaged with a robust weighting parameter \citep{Briggs:1995} of 0.5 and Gaussian tapering of 18 or 20 arcsec, resulting in $\sim 40 \times 27$ restoring beam sizes (see Table~\ref{tab:processing_results}). 
These parameters provide a good compromise between improving sensitivity and reducing sidelobe levels given the maximum baseline length of $\sim$ 2.3~km for the ASKAP-12 array used in this work.  
To deal with continuum emission residuals in the spectral line data cubes, an additional continuum subtraction step is carried out by subtracting a linear fit to the spectrum at each pixel in the data cubes.

After the second step of continuum subtraction, the independently processed beams are combined using a linear mosaicking algorithm \citep{Serra:2015b}. In this mosaicking process, the primary beam (PB) response pattern is corrected using the ASKAP primary beam models.
As a result of mosaicking, one data set yields a 72-beam (i.e. two 36-beam footprints) combined continuum image and spectral line cube. These images and cubes are then joined together with the combined data products from the other tile to cover the full FoV of the G23 field as seen in Fig.~\ref{fig:beamFootprints} (144 beams in total). For multi-epoch data, we combine the calibrated visibility data for each beam before imaging in order to improve deconvolution of directly-detected sources. 

For this paper, the observed data were split into several small chunks (spectral windows) with narrow bandwidths (e.g. 16~MHz, 22~MHz or 32~MHz) to facilitate faster data processing. Table~\ref{tab:spws} lists the details of the spectral windows.
We processed only the three spectral windows, labelled SPW0, SPW3 $\&$ SPW4 for direct detection and stacking, respectively.
Fig.~\ref{fig:z_dist_processed} shows the redshift distribution of the G23 galaxies out to $z \sim 0.1$ with the processed spectral windows overlaid.

\subsection{Data Quality Assessment}
\label{subsec:qa}

\begin{figure*}
\centering
\includegraphics[width=0.9\textwidth]{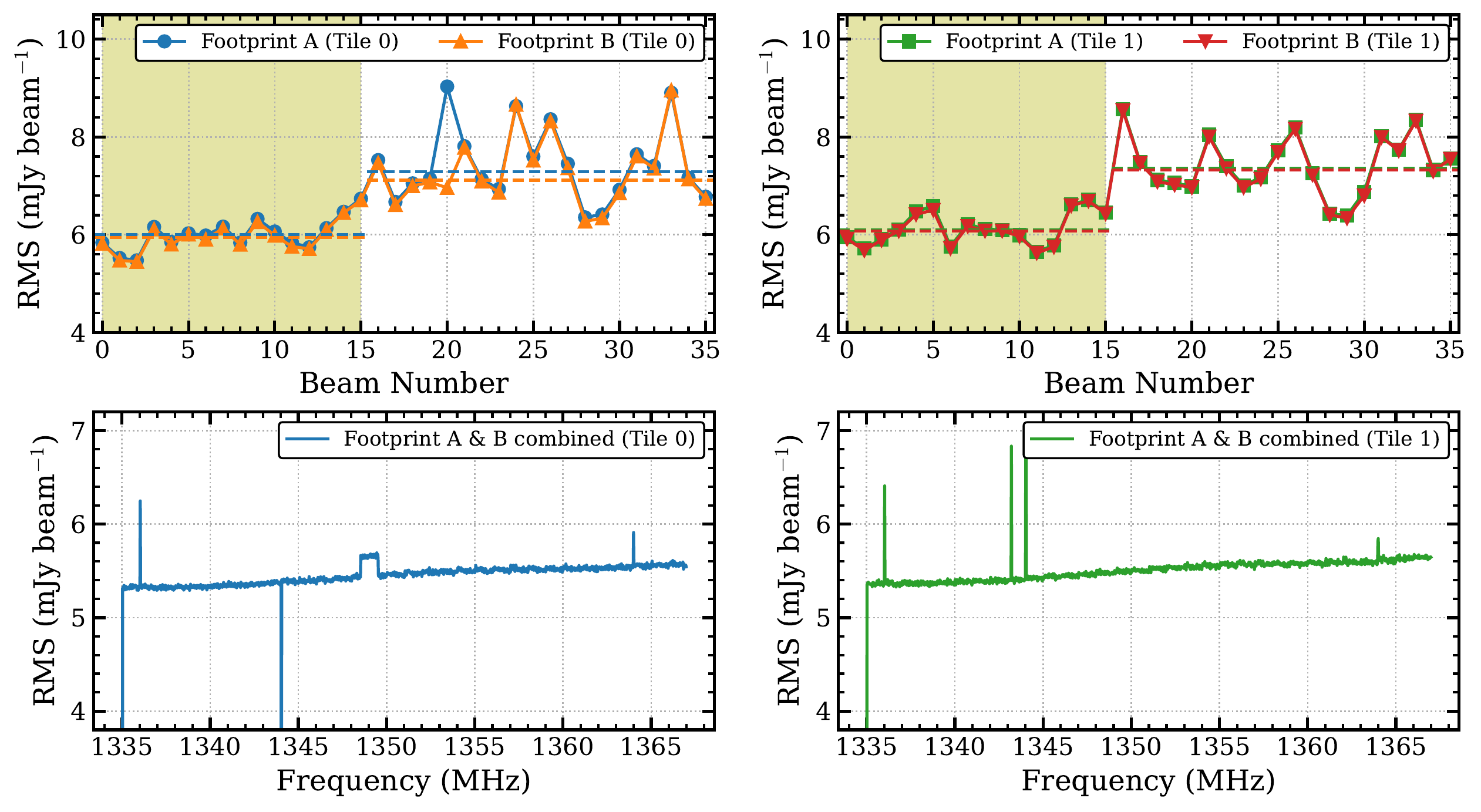}
\caption[rms_variation]{The median RMS noise of each beam for each footprint of a tile ({\it upper}) and the RMS noise measurements with frequency after combining all beams for two footprints and for each tile ({\it bottom}). {\it Left} and {\it Right} columns denote tile 0 and 1, respectively.
The shaded areas in the upper panels indicate the inner 16 beams of each 36-beam footprint. The horizontal dashed lines in the upper panels denote the median RMS of the inner 16 and outer 20 beams, respectively.} 
\label{fig:rms_variation}
\end{figure*}

\begin{figure*}
\centering
\includegraphics[width=0.9\textwidth]{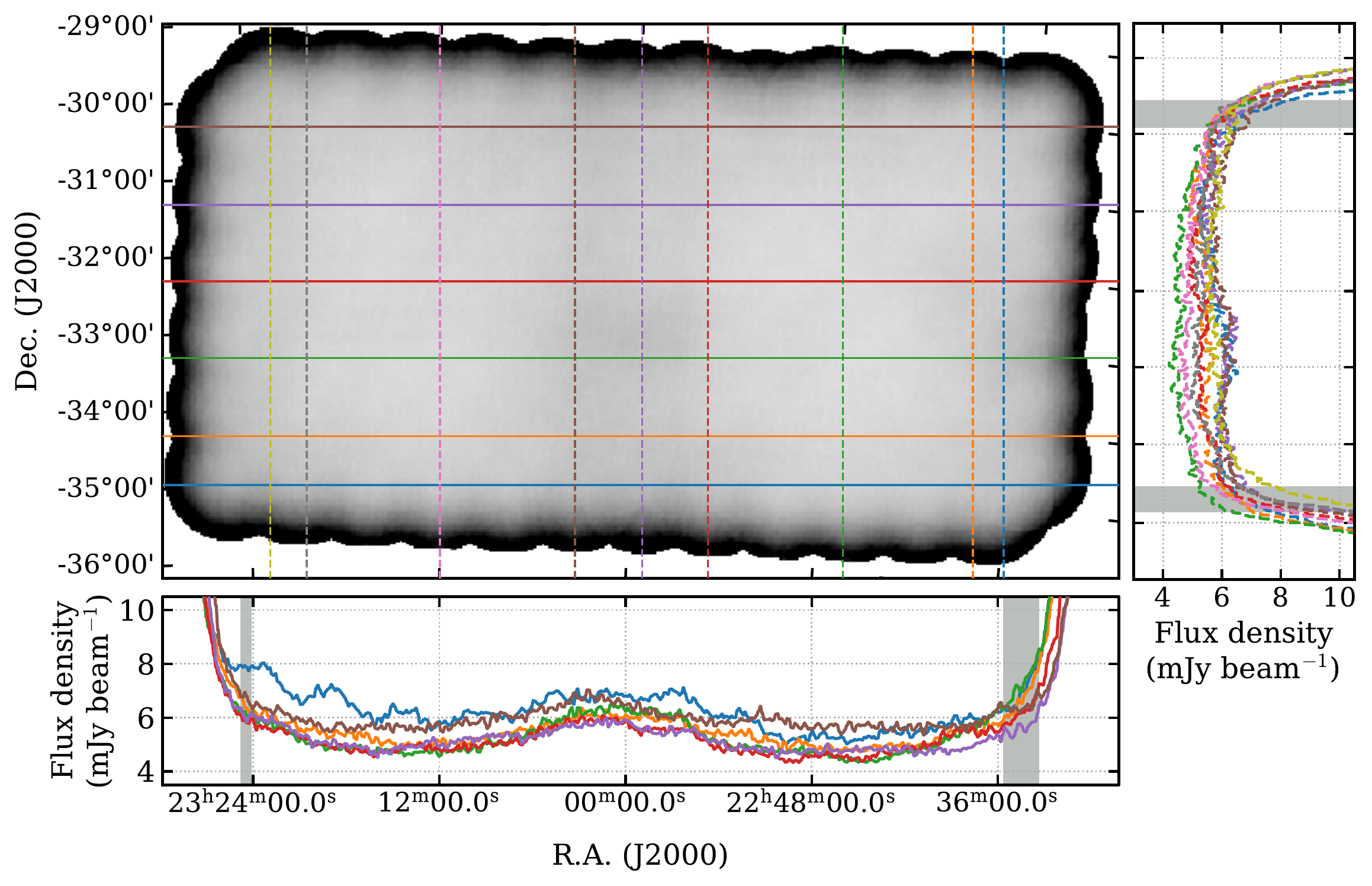}
\caption[rms_variation]{The noise profiles in RA ({\it bottom}) and Dec ({\it right}) directions, respectively.
  The {\it left upper} panel shows the noise map of the central frequency channel where the noise profiles were measured along the horizontal solid and vertical dashed lines. The shaded areas in each noise profile correspond to the outer boundary of the G23 optical survey.}
\label{fig:noise_profile}
\end{figure*}

\begin{table}
\caption{The results of DINGO early science data processing. The expected RMS is a robust-weighted channel noise calculated, accounting for flagged data fraction.}
\centering
\label{tab:processing_results}
\begin{tabular}{@{}cccc}
\hline
     SPW ID  &  Data fraction  &  RMS noise   &  Beam size  \\
             &  flagged       &  expected/measured  &  major $\times$ minor  \\
             &   [$\%$]       &  [mJy channel$^{-1}$]      &  [arcsec $\times$ arcsec] \\
     \hline
     SPW0  &  18.1  &  4.19 / 5.45  &  36.3 $\times$ 28.8  \\
     SPW3  &  24.1  &  4.02 / 5.32  &  37.9 $\times$ 28.0  \\
     SPW4  &  37.5  &  4.29 / 5.60  &  40.8 $\times$ 27.2  \\  
     \hline  
\end{tabular}
\end{table}

DINGO early science data provides a good test bed for examining the performance of ASKAP and its processing pipeline. 
There is a significant fraction of data affected by RFI even though ASKAP is located at the Murchison Radio-astronomy Observatory (MRO) which is a radio-quiet zone. This is mainly due to Global Navigation Satellite Systems and unexpected RFI in the early ASKAP system (e.g. caused by on-dish calibrators, subsequently fixed following ASKAP pilot phase I observations).
The processed SPW0, 3 and 4 data have about 18--38~per~cent of data flagged on average due to RFI as shown in Table~\ref{tab:processing_results}.

Based on the performance specification of ASKAP \citep{Hotan:2021,ASKAP_ACES:2019}, we calculated an expected root-mean-square (RMS) noise per frequency channel taking into account the fraction of data flagged and compared with the measured values for the processed data in Table~\ref{tab:processing_results}. The measured noise is about 30~per~cent higher than expected. 
There is also a beam-wise variation of sensitivity as seen in Fig.~\ref{fig:rms_variation}.
The inner 16 of the 36 beams in each footprint have a lower RMS noise by $\sim 20$~per~cent than the outermost 20 beams for all scheduling blocks.
This difference reflects the expected decrease in sensitivity with radius, most evident for the corner beams \citep{Hotan:2021}. This variation may also explain the higher global noise level, compared with expectation, seen in Table~\ref{tab:processing_results}. 
After combining all beams and scheduling blocks, the variation of RMS noise with frequency for both tiles is shown in the lower panels of Fig.~\ref{fig:rms_variation}. The smooth increase with frequency reflects the measured system temperature trend for ASKAP \citep[Fig.~22 in][]{Hotan:2021}. 

We also measured spatial noise profiles for one channel of a noise cube for SPW3 in both in RA and Dec directions as shown in Fig.~\ref{fig:noise_profile}. The horizontal solid and the vertical dashed lines denote the locations where noise profiles were extracted. The profiles
cover the whole G23 field.
The bottom and right panels show the noise profiles in the RA and Dec directions, respectively.
The grey shaded bars indicate the boundaries of the optical G23 survey field.
Within the optical survey area, there is no significant noise variation (less than 19~per~cent)
as the DINGO observation coverage is larger than that of the optical survey.
In Fig.~\ref{fig:noise_profile}, the central vertical line (dashed purple) coincides with the overlapping areas of both tiles. The noise level here is similar to other regions, with only slightly higher noise indicated by the RA profiles. These measurements vindicate our tiling strategy.

\section{${\HI}$ stacking}
\begin{figure}
\centering
\includegraphics[width=0.45\textwidth]{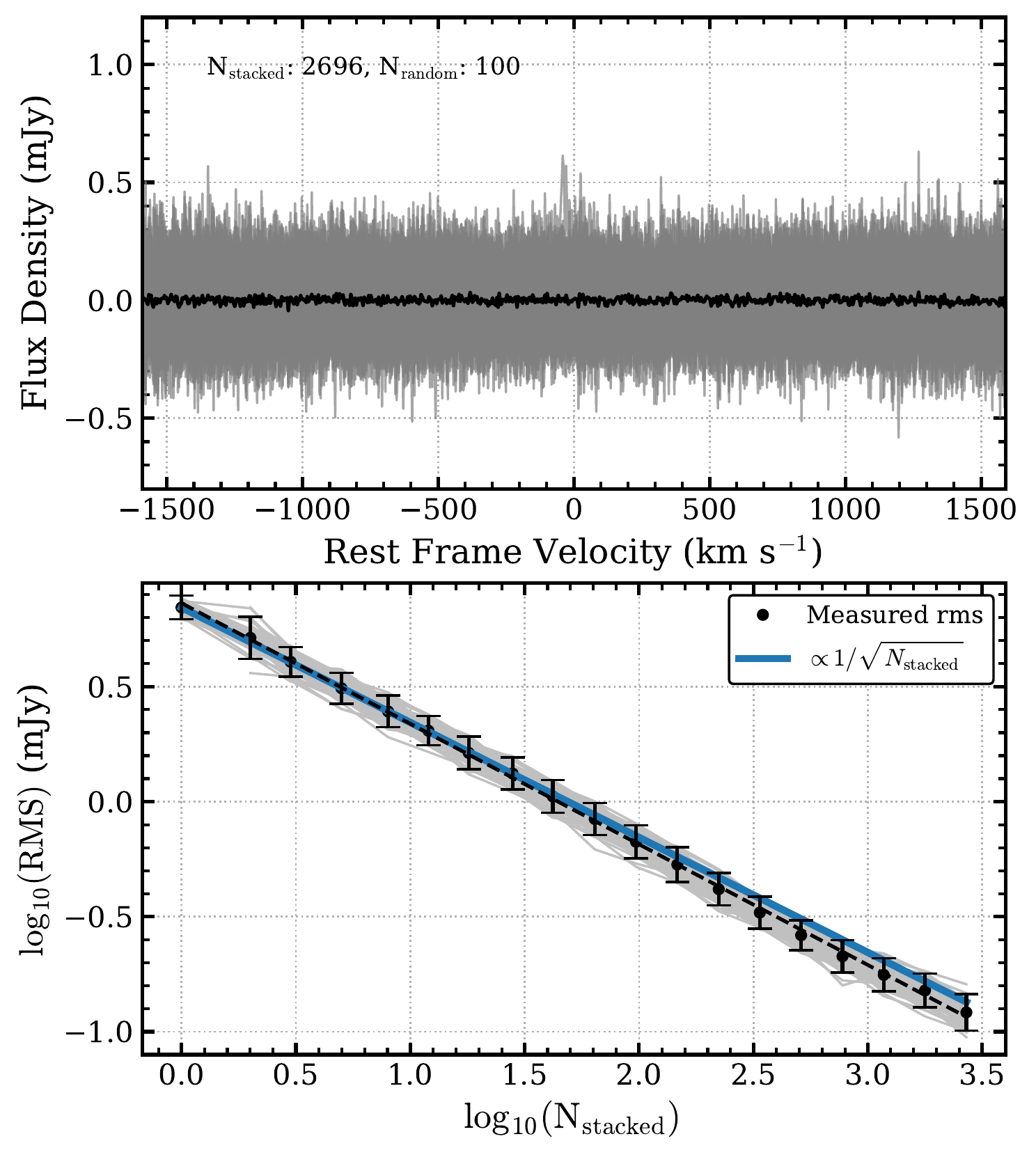} 
\caption[stacked spectra]{This shows the results of {\HI} stacking using DINGO early science data and 100 random catalogues generated based on the original GAMA catalogue of 2696 galaxies with spatial positions fixed and redshifts randomised. The upper panel shows spectra co-added using the 100 random catalogues in grey and their median spectrum in black. The grey solid lines in the lower panel are RMS measurements made when increasing the number of galaxies to be stacked ($N_{\rm stacked}$) for each random catalogue. The black dots and dashed line are the mean values of the RMS measurements and a linear fit to them, respectively, while the blue solid line indicates the expected RMS value proportional to $\sqrt{N_{\rm stacked}}$.}
\label{fig:random_stack}
\end{figure}

\subsection{Noise Characteristics of DINGO Data}
As described in Section~\ref{subsec:qa}, the processed DINGO data has uniform noise characteristics across all the beams and channels. {\HI} stacking can also be used as a diagnostic tool to check the data for noise behaviour that is more likely to impact an {\HI} stacking analysis.
Using the SPW4 data cube and the GAMA catalogue corresponding to the SPW4 redshift range,
we generated 100 random galaxy catalogues with each galaxy's spatial positions fixed but redshifts randomised.
Then, {\HI} stacking was carried out with all the random catalogues.
The upper panel of Fig.~\ref{fig:random_stack} shows 100 co-added random spectra with their median values.
In stacking spectra using each random catalogue, we measured the RMS in the central 300~\kms of the co-added spectra with increasing the number of stacked galaxies ($N_{\rm stacked}$). The measurements of the RMS noise continue to fall as $\sqrt{N_{\rm stacked}}$, as shown in the lower panel of Fig.~\ref{fig:random_stack}. The noise trend of our data follows Gaussian behaviour and this means that the data processing pipeline deals well with non-Gaussian components, for instance continuum residuals which can significantly affect {\HI} stacking analyses, thereby minimising any systematic effect on {\HI} stacking measurements.

\subsection{Stacking Methods}
\label{sec:stacking_method}

\begin{figure}
\centering
\includegraphics[width=0.45\textwidth]{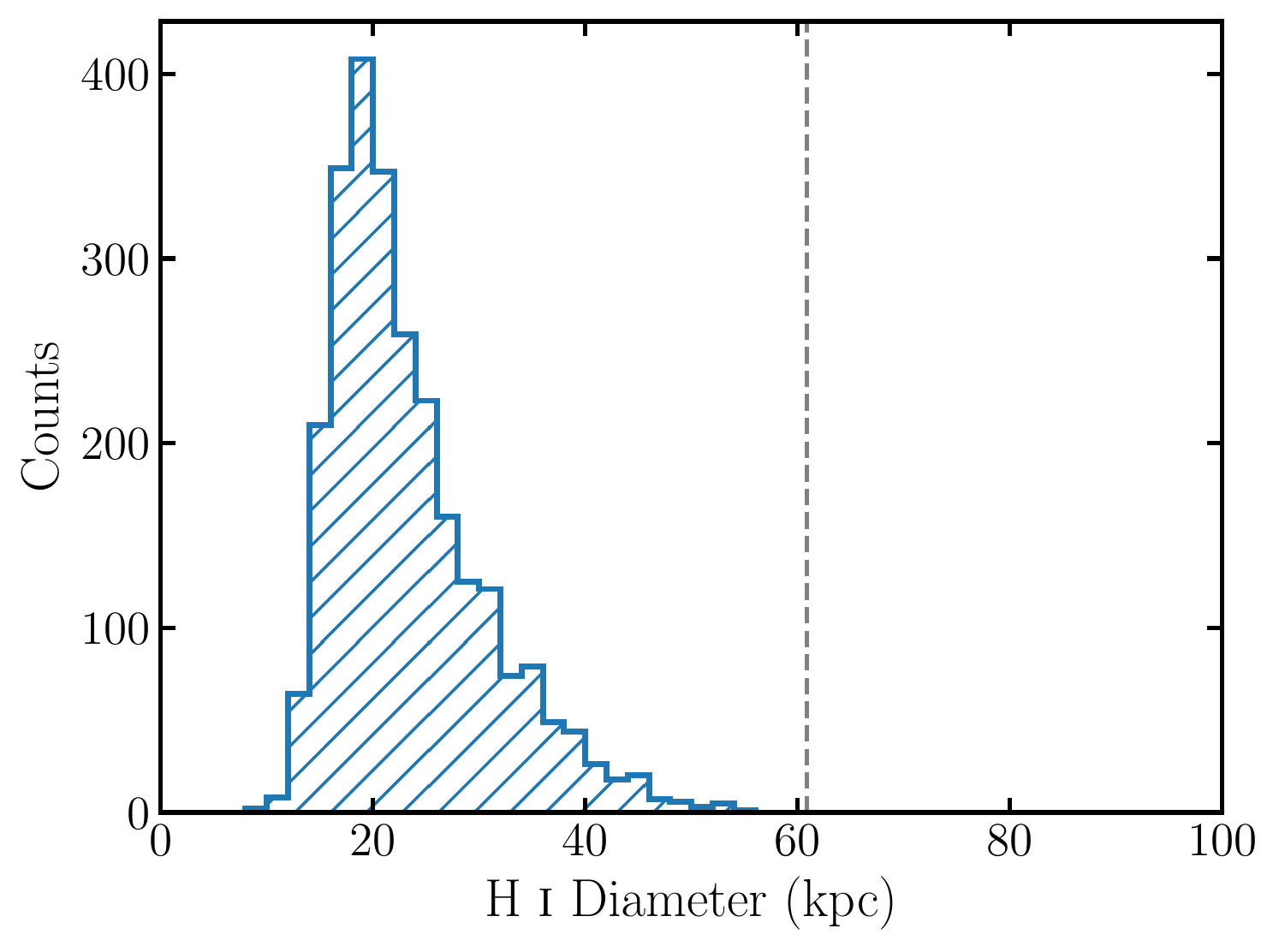}
\includegraphics[width=0.45\textwidth]{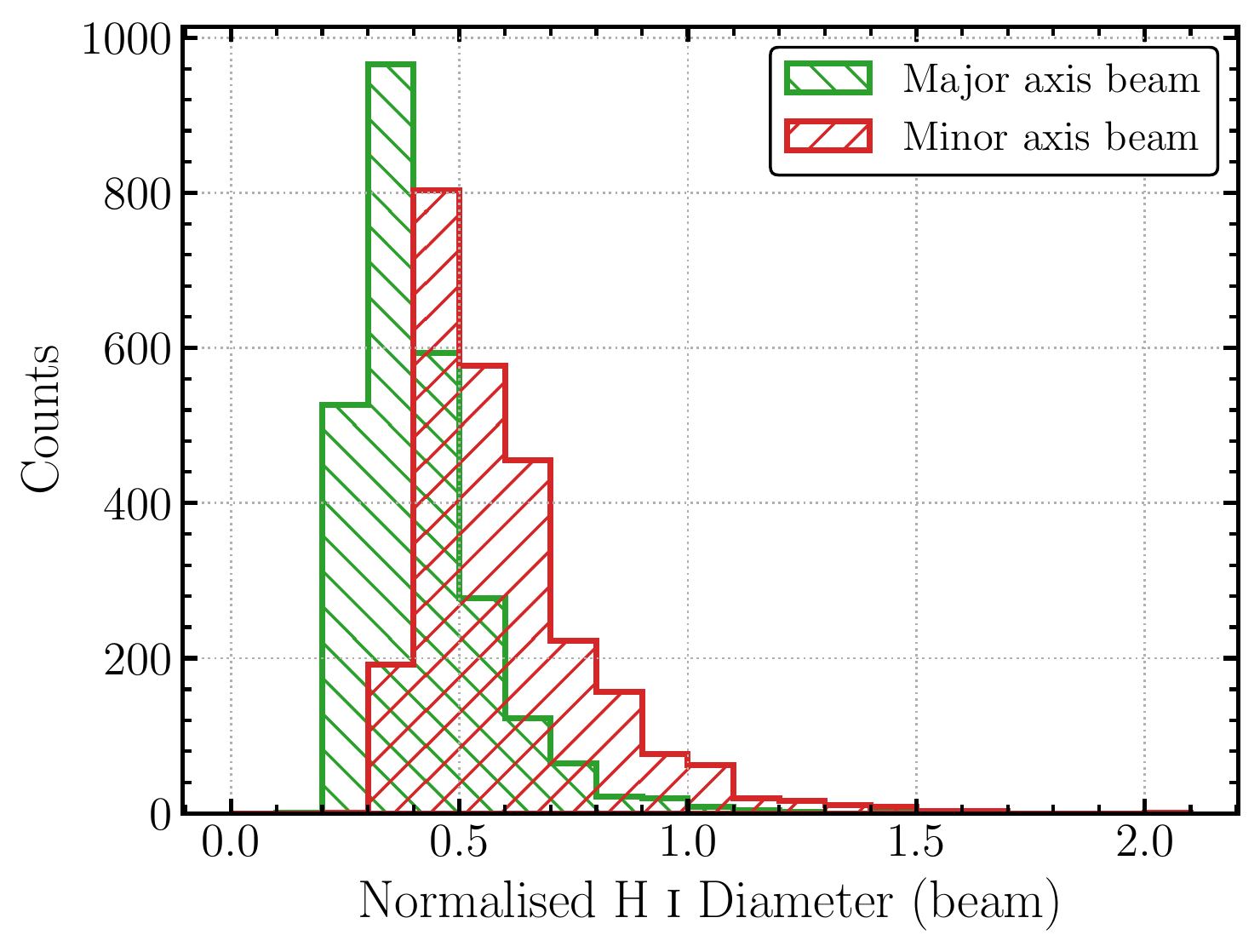} 
\caption{The top panel shows the expected {\HI} size distribution of our sample galaxies in SPW3 and 4 ($0.039 < z < 0.088$) used for stacking. The x-axis indicates {\HI} diameter in kpc units. The vertical dashed line denotes the ASKAP restoring beam size (major axis size of $\sim$~61~kpc) at the median redshift of the SPWs. In the bottom panel, the expected {\HI} sizes are normalised with the major-axis (green) and minor-axis (red) sizes of the ASKAP restoring beam. Only 1 and 5 per~cent respectively of the sample galaxies have a larger predicted {\HI} diameter compared to the major-axis and minor-axis beam sizes.}
\label{fig:HI_size_relation}
\end{figure}

Before stacking {\HI} spectra, we first checked whether galaxies in our sample are likely to be resolved with the ASKAP restoring beam of 
$\sim 40\arcsec \times 27 \arcsec$, corresponding to $\sim$61~kpc~$\times$~41~kpc at the median redshift ($z \sim 0.07$) of the processed data (SPW3 and 4). 
To do this test, we estimated the expected {\HI} size of galaxies in our sample using a scaling relation between {\HI} size ($D_{\HIsub}$) and optical $B$-band luminosity (${\rm M}_{B}$). \citet{Wang:2016} have explored scaling relations related to {\HI} disk size over the wide range of {\HI} mass and optical luminosity with publicly available data.
However,  the data from \citet{Broeils:1997} were only included for the {\HI} size-mass relation, and not the {\HI} size-optical luminosity relation. 
We therefore re-derived the relation by combining the data from both \citet{Broeils:1997} and \citet{Wang:2016}, obtaining:
\begin{equation}
 \label{eq:HI_size_lum_relation}
  {\rm log}_{10}D_{\HIsub} = -0.123~M_{B} - 0.927, 
\end{equation}
where $D_{\HIsub}$ is the major axis diameter of the {\HI} disc (in kpc) at a surface density $\Sigma_{\HIsub} =$ 1~{\msun}~pc$^{-2}$, and $M_{B}$ is the $B$-band absolute magnitude. Using the newly derived relation, we estimated the {\HI} sizes of our sample from the $B$-band luminosity transformed from $g$-band magnitudes and $g-r$ colours of the sample in the GAMA catalogue, following the transformation equation in \citet{Jester:2005}.
Fig.~\ref{fig:HI_size_relation} shows that our sample galaxies are unlikely to be resolved in {\HI} with the ASKAP point spread function (PSF). 

To extract {\HI} spectra from the final data cubes,
we first convert barycentric redshifts of galaxies from the G23 catalogue into the corresponding {\HI} frequencies.
An {\HI} spectrum is extracted over the spatial pixels covering a square sky region of 49~kpc aperture size centred at a galaxy position. The aperture size is selected to 
take into account the predicted {\HI} sizes ($\sim$~23 and 55~kpc mean and maximum, respectively) as shown in Fig.~\ref{fig:HI_size_relation}.
Since our galaxy sample is likely unresolved with the ASKAP PSF as seen in Fig.~\ref{fig:HI_size_relation}, we then calculate the spatially-integrated flux density for each spectral channel following Eq.~8 in \citet{Shostak:1980}:
\begin{equation}
 \label{eq:Flux_int}
  S_v = \frac{\Sigma_x \Sigma_y S_v(x, y)}{\Sigma_x \Sigma_y B(x, y)},
\end{equation}
where $x, y$ are the sky coordinate of a galaxy (RA, Dec) and $B(x, y)$ is the normalised beam response of the ASKAP PSF at the position $(x, y)$, which can be approximated as 2D elliptical Gaussian.

${\HI}$ spectra extracted based on sky positions and redshifts of galaxies are shifted and aligned to the rest frame before co-adding them. 
The stacking process uses the RMS noise (${\sigma}_{\rm rms}$) of each frequency channel as a weight ($w = {\sigma}_{\rm rms}^{-2}$) in stacking {\HI} spectra. The co-added {\HI} spectra are converted to {\HI} mass using the following relation \citep[see][]{Meyer:2017} after baseline correction (excluding the central velocity range where {\HI} signal is expected): 
\begin{equation}
 \label{eq:HImass}
  \frac{{\MHI}}{\msun} = \frac{235.6}{(1+z)} \left( \frac{D_L}{{\rm Mpc}} \right)^2 
  \bigg( \frac{\int S_V dV}{{\rm mJy~\kms}}\bigg),
\end{equation}
where $z$ is redshift, $D_{L}$ is the luminosity distance in units of Mpc, and $\int S_V dV$ is the integrated {\HI} emission flux within a velocity width of $dV$ in units of $\rm {mJy~\kms}$. We selected a velocity width to calculate average {\HI} mass for our sample, based on $w_{50}$ ({\HI} line width at 50 per~cent of the peak flux) derived from the Tully-Fisher relation \citep[TFR, ][]{Tully:1977}, accounting for the uncertainty of redshift measurement for our sample. Following \citet{Meyer:2016} who exploited $K$-band absolute magnitude and $w_{50}$ measurements from HIPASS, sample galaxies in SPW3 and 4 will have 50- and 95-percentile $w_{50}$ of 145.4 and 240.8~{\kms}, respectively. Based on the latter, and a GAMA redshift uncertainty of $2\sigma \sim$~60~\kms\ \citep{Liske:2015}, 
a velocity width of 300~$\kms$ was chosen for all subsequent stacking measurements.

Other than average {\HI} mass $\langle \MHI \rangle$, we also measured average {\HI} mass-to-light ratio in the $r$-band $\langle \MHI / L_{r} \rangle$ and average {\HI}-to-stellar mass ratio (i.e. {\HI} gas fraction, $\langle \MHI / M_{\star} \rangle$). We first converted {\HI} flux spectra extracted from galaxy positions into {\HI} mass spectra using Eq.~\ref{eq:HImass} above. An individual {\HI} mass spectrum is then divided by the $r$-band luminosity and stellar mass of the corresponding galaxy to give $\MHI / L_{r}$ and $\MHI / M_{\star}$ spectra, respectively. Luminosity in the $r$-band is based on updated photometric data in the G23 catalogue \citep{Bellstedt:2020}. Stellar masses are derived using {\sc ProSpect}, a generative galaxy spectral energy distribution (SED) package, which fits SED models with the 20-band photometry in the G23 catalogue to obtain stellar masses of G23 galaxies as well as other galaxy properties such as star formation history and metallicity history \citep[see][for more details]{Robotham:2020, Bellstedt:2020a}. 
These spectra are then stacked to obtain average values of $\MHI / L_{r}$ and $\MHI / M_{\star}$ for our sample. 

To make a robust estimate of the uncertainty of the stacked measurements, 
we use  jackknife resampling. 
One spectrum at a time is excluded from the original sample to generate the same number of jackknife samples as the number of the spectra. 
The error of a stacked measurement is estimated following \citet{Rhee:2013}. These uncertainty measurements are then propagated into the subsequent calculations.

\subsection{Confusion}
\begin{figure}
\centering
\includegraphics[width=0.45\textwidth]{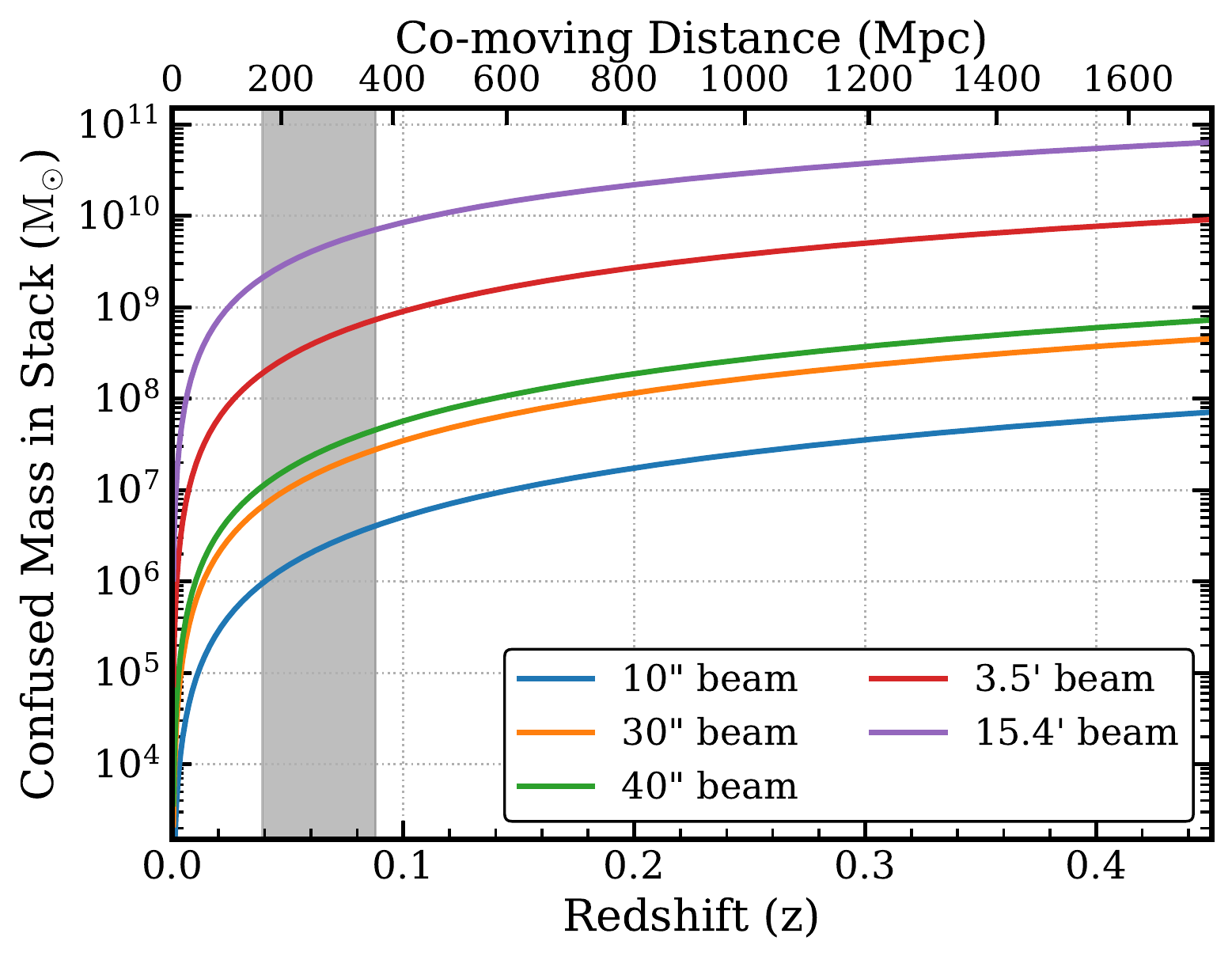} 
\caption{The estimate of confused mass in the stacked {\HI} mass as a function of redshift (or distance) with different synthesized beam sizes: 10$\arcsec$, 30$\arcsec$, 40$\arcsec$, 3.5$\arcmin$, and 15.4$\arcmin$. The grey shaded area denotes the redshift interval where the stacking experiments of this work were conducted (SPW3 and 4).}
\label{fig:confusion}
\end{figure}

One of the main concerns about {\HI} spectral stacking is confusion in the stacked {\HI} mass 
due to unknown or known nearby companions of the stacked galaxies
within a beam of a radio telescope and a velocity line width \citep{Maddox:2013, Jones:2016, Elson:2019}.
However, stacking experiments with interferometers are less vulnerable to this effect than those of single dish telescopes
thanks to a smaller synthesized beam \citep[e.g.][]{Rhee:2013, Rhee:2016, Rhee:2018, Bera:2019, Hu:2019, Chowdhury:2020, Chowdhury:2021}. 
However, deep interferometric data can be influenced at higher redshifts
as the physical size corresponding to the synthesized beam increases. 
We have investigated the impact of confusion on the stacked {\HI} mass 
due to neighbouring galaxies that may or may not be identified in optical surveys. 

First, we checked whether any galaxy pairs had a projected separation less than the ASKAP restoring beam size seen in Table~\ref{tab:processing_results}, and whether the velocity separation was less than 300~{\kms}.
Only $\sim 8$ and $\sim 10$ per~cent of SPW3 and SPW4 sample galaxies are affected, respectively. 

We also tried to estimate the amount of {\HI} mass added to the {\HI} stack due to confused galaxies based on an analytic model from \citet{Jones:2016}.
Using Eq.~2 in \citet{Jones:2016}, we calculated and compared confused {\HI} masses with increasing redshift for five
different synthesized beam sizes: 10$\arcsec$, 30$\arcsec$, 40$\arcsec$, 3.5$\arcmin$ and 15.4$\arcmin$ in Fig.~\ref{fig:confusion}, representing the beam resolutions of DINGO high-$z$, low-$z$ and early science, ALFALFA and HIPASS, respectively.
In the redshift range of DINGO early science, the confused {\HI} mass in stacked {\HI} is below 10$^{7.5}$ {\msun} which is well under the uncertainties (10$^{8}$-10$^{8.1}$ {\msun}) of the stacked measurements  (see Table~\ref{tab:stacking_measurements}).
Since the {\HI} stacking results in this work are unlikely to be subject to  significant confusion,
no confusion correction is applied.

\begin{figure}
\centering
\includegraphics[width=0.45\textwidth]{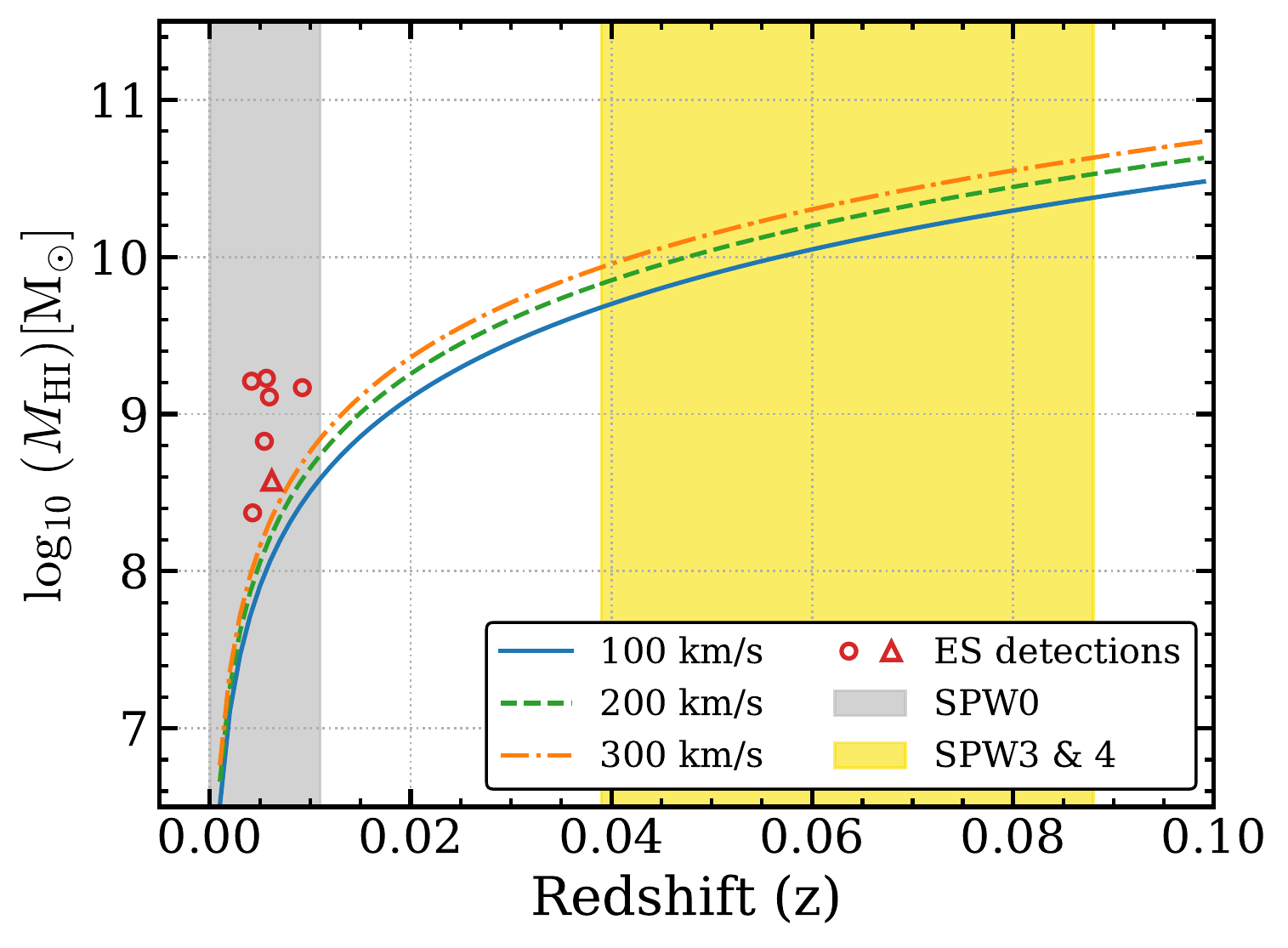}
\caption{{\HI} mass sensitivity prediction as a function of redshift and the distribution of detected {\HI} sources in SPW0 for the DINGO ES data. Three sensitivity curves are calculated with different velocity profile widths of 100, 200, and 300 {\kms}, assuming unresolved 7$\sigma$ detection sources, noise statistics in Table~\ref{tab:processing_results}. The red open circles and triangles are known and new detections in SPW0, respectively. The grey and yellow shaded areas denote the redshift range for SPW0 and SPW3 and 4, respectively. }
\label{fig:hi_sensitivity}
\end{figure}

\begin{figure*}
\centering
\includegraphics[width=0.9\textwidth]{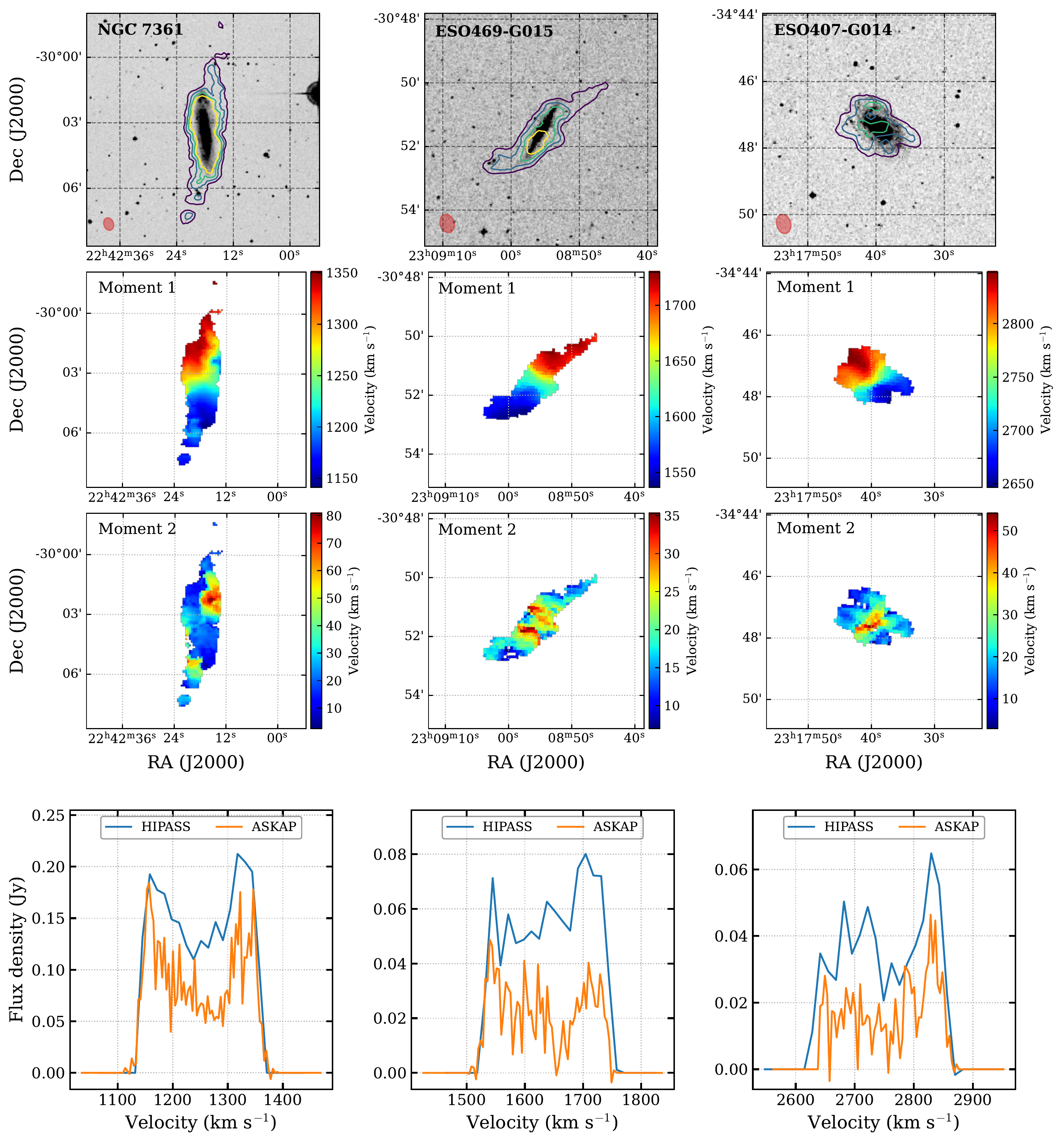} 
\caption[direct detection]{The directly detected {\HI} sources. The first row shows integrated {\HI} column density maps of individual sources overlaid on optical images from the Digitized Sky Survey (DSS). The contour levels are 3, 5, 7, 9 ${\times}$ sensitivity limits of column density for those detections ($\sigma = 8.7, 7.5, 7.8~{\times}~10^{19}$ cm$^{-2}$ for NGC~7361, ESO469-G015, and ESO407-G014, respectively). The red ellipse in the left corner of each panel denotes the ASKAP synthesised beam. The second and third rows are the velocity field and velocity dispersion maps of each detected object, where only pixels above 3$\sigma$ of the column density sensitivity are plotted as the first row panels show. In the last row, the ASKAP spectra are compared with the HIPASS spectra for the three detected sources. The blue spectra are taken from the HIPASS data cubes using SoFiA while the ASKAP spectra are the orange ones. The ASKAP spectra do not seem to recover the HIPASS fluxes. The missing flux is likely due to the lack of short baselines of ASKAP-12 used for early science observations.}
\label{fig:detection_maps}
\end{figure*}

\begin{figure*}
\centering
\includegraphics[width=0.98\textwidth]{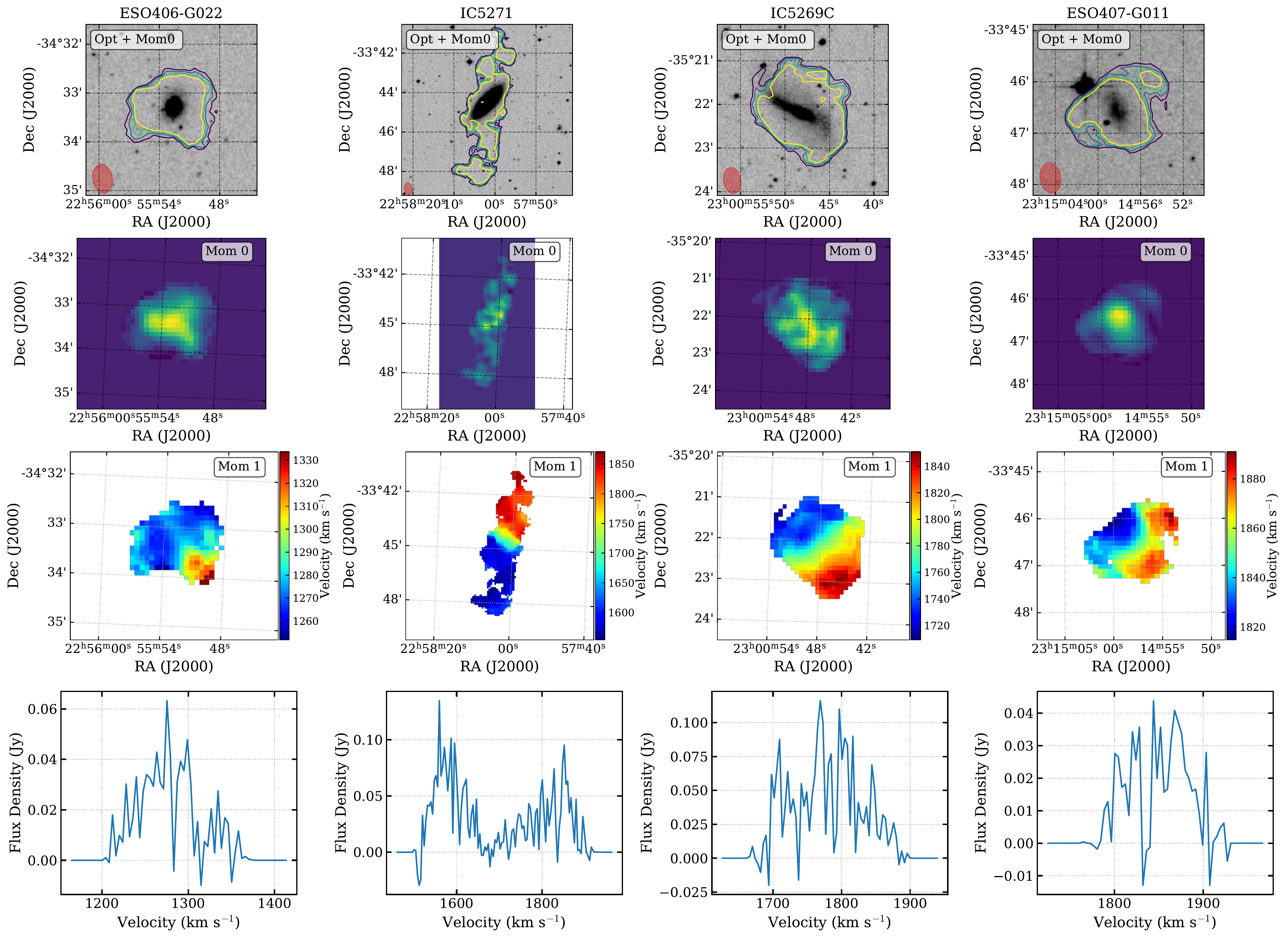}
\caption{The moment maps and spectra of {\HI} detections--ESO406-G022, IC~5271, IC~5269C, ESO407-G011 from left to right column. The first row shows integrated {\HI} column density contours (moment 0) for individual sources overlaid on optical DSS images. The contour levels are 3, 5, 7, 9 ${\times}$ sensitivity limits of column density for those detections. The red ellipse in the left corner of each panel indicates the ASKAP restoring beam. The second and third rows show the moment 0 and 1 maps, respectively. {\HI} spectra of each source are plotted in the last row.}
\label{fig:new_detections}
\end{figure*}

\section{Direct Detection}

\begin{table*}
\caption{The detected source catalogue and derived properties.}
\label{tab:detections}
\centering
\begin{tabular}{@{}llcccccccccc}
\hline
  ID & Name & R.A. & Dec. & $\nu_{\rm obs}$ & $z$ & $V_{\rm opt}$ & $S_{\rm int}$ & $w_{50}$ & log {\MHI}$_{\rm (ASKAP)}$ & log {\MHI}$_{\rm (Parkes)}$ \\
     & & [J2000] & [J2000] & [MHz] & & [{\kms}] & [Jy~Hz] & [{\kms}] & [{\msun}] & [{\msun}] \\
  (1) & (2) & (3) & (4) & (5) & (6) & (7) & (8) & (9) & (10) & (11)\\
\hline
1 & NGC~7361 & 22:42:17.84 & -30:03:26.8 & 1414.53 & 0.0042 & 1245 & 102243 & 208.2 & 9.21 & 9.41$^a$ \\
2 & ESO469-G015 & 23:08:55.39 & -30:51:33.0 & 1412.73 & 0.0054 & 1629 & 24722 & 204.5 & 8.83 & 9.22$^a$ \\
3 & ESO407-G014 & 23:17:40.01 & -34:47:21.1 & 1407.44 & 0.0092 & 2762 & 18766 & 202.6 & 9.17 & 9.50$^a$ \\
4 & ESO406-G022 & 22:55:52.74 & -34:33:20.9 & 1414.37 & 0.0043 & 1278 & 14091 & 62.9 & 8.37 & 8.85$^b$\\
5 & IC~5271            & 22:58:01.86 & -33:44:52.7 & 1412.45 & 0.0056 & 1689 & 57870 & 302.5 & 9.23 & 9.20$^b$\\
6 & IC~5269C          & 23:00:48.00 & -35:22:07.8 & 1412.03 & 0.0059 & 1778 & 39705 & 148.0 & 9.11 & 9.17$^b$\\
7 & ESO407-G011 & 23:14:58.01 & -33:46:33.0 & 1411.70 & 0.0062 & 1850 & 10766 & 103.7 & 8.58 & - \\
\hline
\multicolumn{11}{p{0.94\textwidth}}{\textit{Note}. Cols~(1) and (2): identification and name of the source. Cols~(3) and (4): R.A. and Dec. coordinates of the {\HI} detection. Col (5): the measured central frequency of the source. Col~(6): redshift based on the detected {\HI}. Col~(7): velocity in the optical convention ($cz$). Col~(8): integrated flux. Col~(9): the line width of the integrated profile at 50 per~cent of the peak flux density. Col~(10): logarithmic {\HI} mass from ASKAP. Col~(11): logarithmic {\HI} mass from Parkes.}\\
\multicolumn{11}{p{0.94\textwidth}}{$^a$Direct measurements from HIPASS data cubes using SoFiA}\\
\multicolumn{11}{p{0.94\textwidth}}{$^b$Measurements from Table A4 in \citet{Kilborn:2009}}
\end{tabular}
\end{table*}

Before conducting an {\HI} stacking analysis, we searched for any direct {\HI} detections
using the automated 3D source finder software, Source Finding Application \citep[SoFiA\footnote{\url{https://github.com/SoFiA-Admin/SoFiA-2}},][]{Serra:2015a,Westmeier:2021} 
which was developed for ASKAP {\HI} surveys as well as for future {\HI} surveys.
Since the early science band 2 data have a short integration time compared to the DINGO full survey, we did not expect to directly detect 
galaxies with ${\MHI} < 10^{9.7}$ in the redshift range of SPW3 and 4 ($ 0.039 < z < 0.088$) 
(see Fig.~\ref{fig:hi_sensitivity}), and SoFiA indeed made no such detections.

In contrast, in SPW0 ($z < 0.011$), we detected six resolved sources already identified in HIPASS \citep{Koribalski:2004,Meyer:2004}: NGC~7361, ESO469-G015, ESO407-G014, ESO406-G022, IC~5271 and IC~5269C.
In addition, we made a new {\HI} detection of ESO407-G011. Fig.~\ref{fig:hi_sensitivity} shows that all detections are above the predicted {\HI} mass sensitivity limit calculated for velocity widths of 100, 200, and 300 {\kms}, assuming 7$\sigma$ detection of point sources and the noise statistics listed in Table~\ref{tab:processing_results}. The properties of each source from the SoFiA source finder are given in Table~\ref{tab:detections}. Fig.~\ref{fig:detection_maps} shows integrated flux (1st row) maps overlaid onto optical counterpart images, velocity field maps (2nd row) and velocity dispersion maps (3rd row) for the first three detected sources, respectively. The moment maps and spectra of the latter three detections and a new detection are presented in Fig.~\ref{fig:new_detections}.

We compare the flux from ASKAP with that from HIPASS for the three detected galaxies in the bottom row of Fig.~\ref{fig:detection_maps}. 
37 to 59 per~cent of HIPASS flux is missing in the ASKAP data. This discrepancy between ASKAP and HIPASS fluxes can be attributed to the difference between the two datasets (interferometer vs. single dish) and the lack of short baselines of ASKAP-12 used for DINGO early science observation, which causes missing diffuse {\HI} emission. The rest of the detected sources also have the same flux deficit issue except for IC~5271 when compared with the {\HI} mass measurements from \citet{Kilborn:2009}, as seen in Table~\ref{tab:detections}. The ASKAP BETA array (ASKAP-6) observations \citep{Serra:2015b} have also detected IC~5269C with $\sim$ 35 per~cent flux missing compared to the Parkes data from \citet{Kilborn:2009}.

However, the missing flux issue does not affect the subsequent stacking analysis because our sample galaxies are likely unresolved with the ASKAP restoring beam as seen in Fig~\ref{fig:HI_size_relation}.
This missing flux issue for these galaxies will be re-visited in future work, including the DINGO pilot survey which uses the full ASKAP  array.

\section{{\HI} Gas Content of Galaxies}

\subsection{Blue vs. Red Galaxies}
\label{sec:blue_red}

\begin{figure}
\centering
\includegraphics[width=0.45\textwidth]{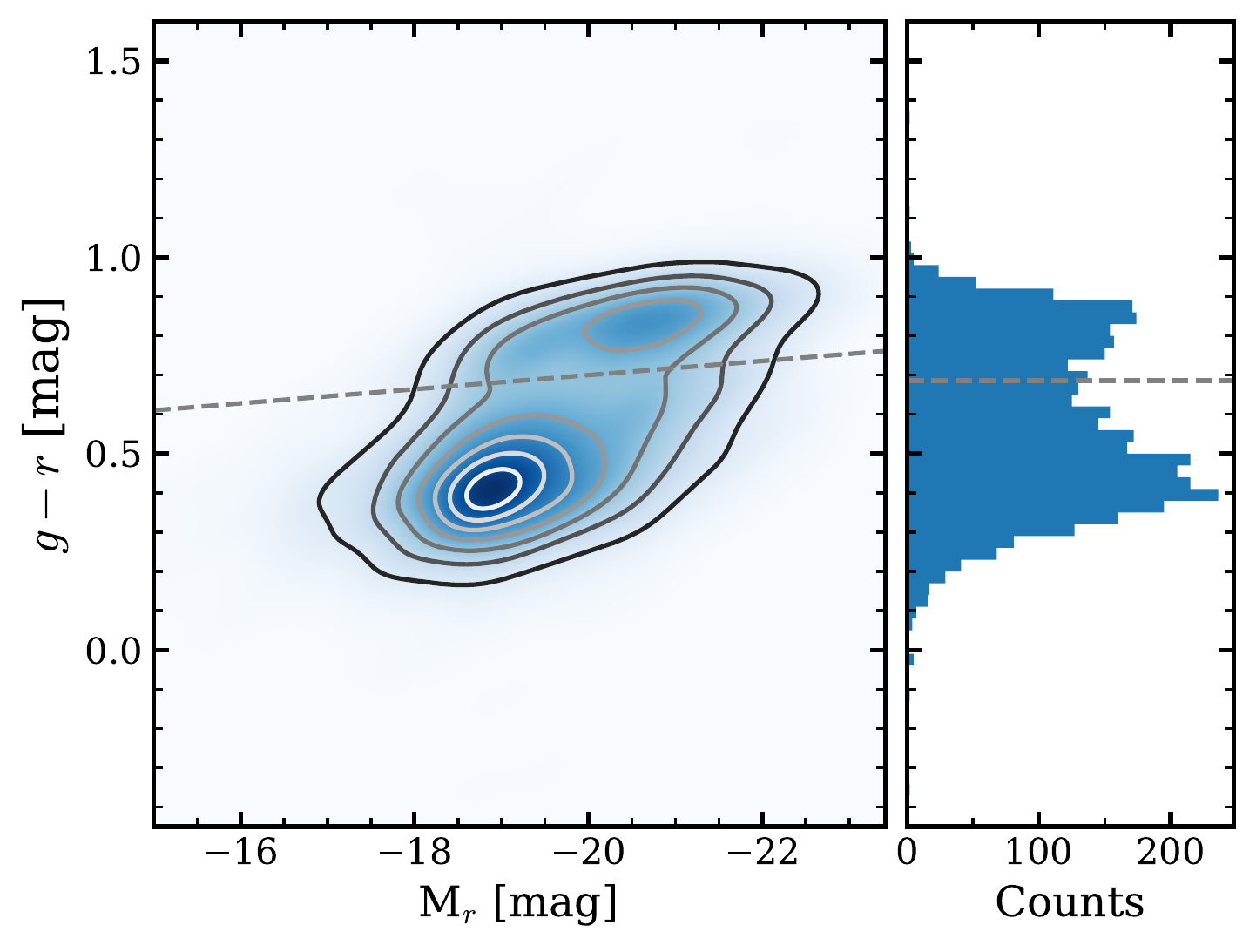} 
\caption{The colour-magnitude diagram (CMD) contour of galaxies in SPW3 and SPW4. The dashed line denotes the colour cut to distinguish between blue and red galaxies. The right panel shows the histogram of $g-r$ colour, indicating the bi-modality of the sample.}
\label{fig:cmd}
\end{figure}

\begin{figure*}
\centering
\includegraphics[width=0.9\textwidth]{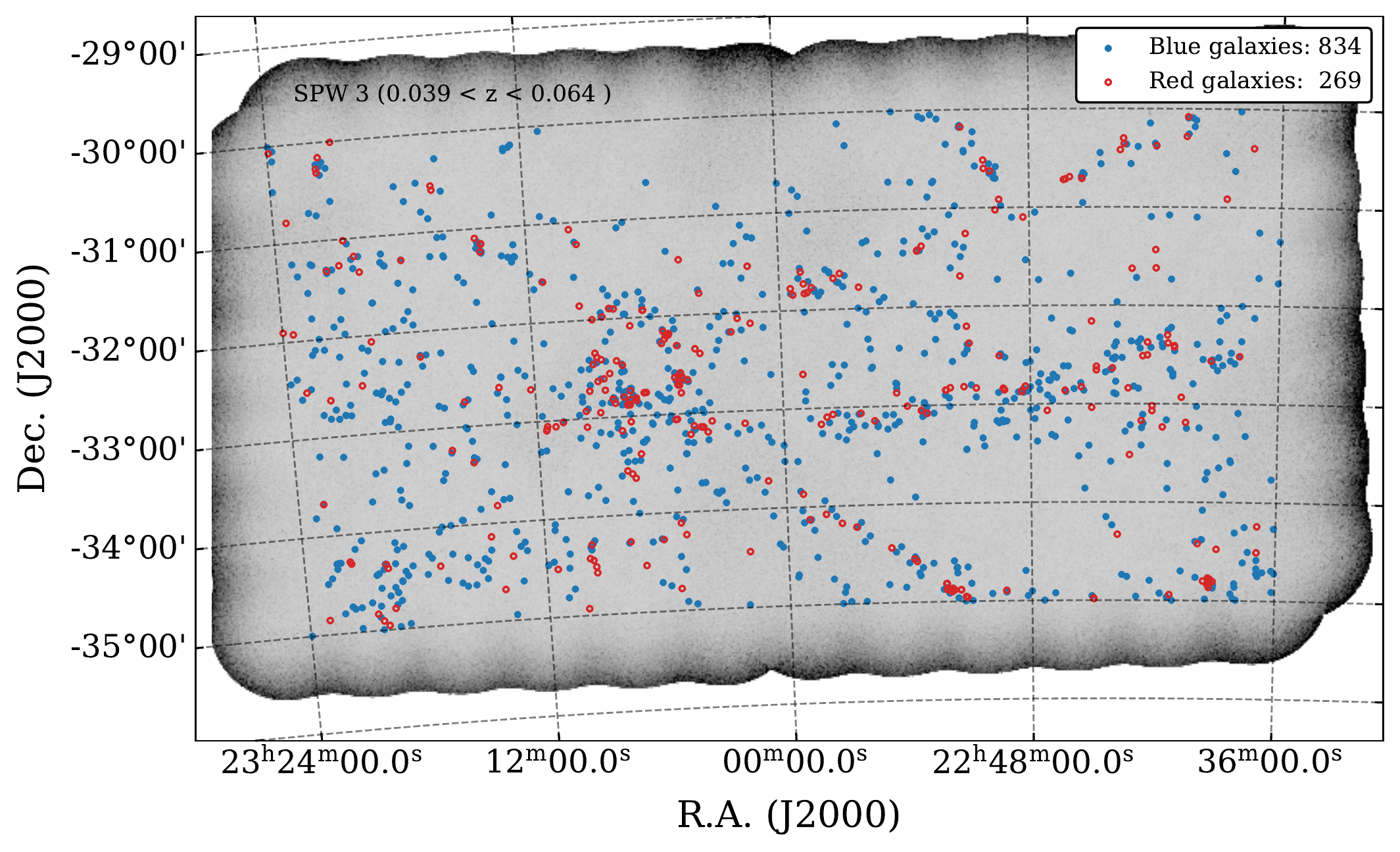} 
\includegraphics[width=0.9\textwidth]{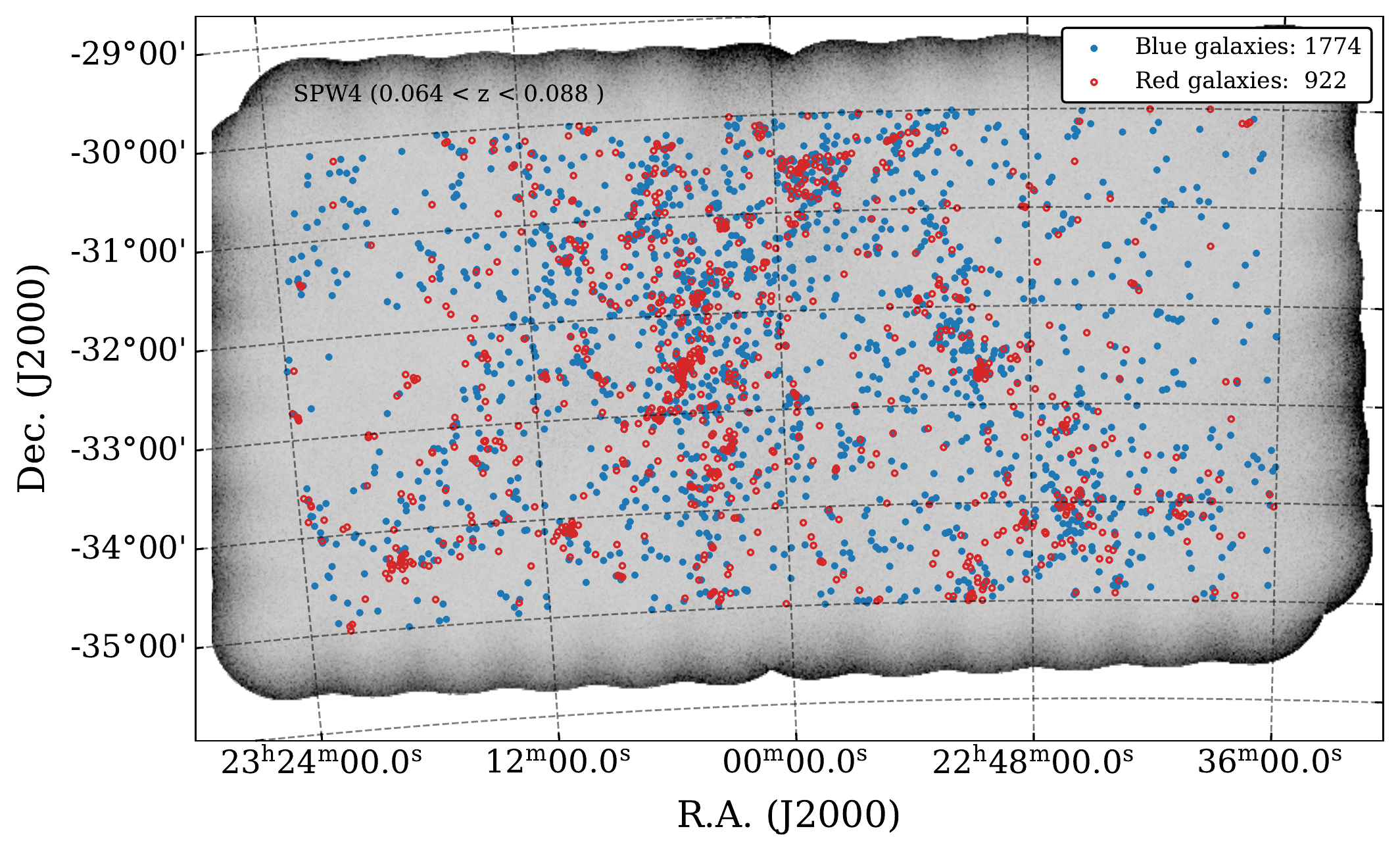} 
\caption{The distribution of blue and red galaxies in the GAMA 23 catalogue for SPW3 ({\it upper}) and SPW4 ({\it lower}), respectively. The background image is a peak flux density map showing the beam coverage of the DINGO early science data.}
\label{fig:galaxy_distribution}
\end{figure*}

\begin{figure*}
\centering
\includegraphics[width=0.9\textwidth]{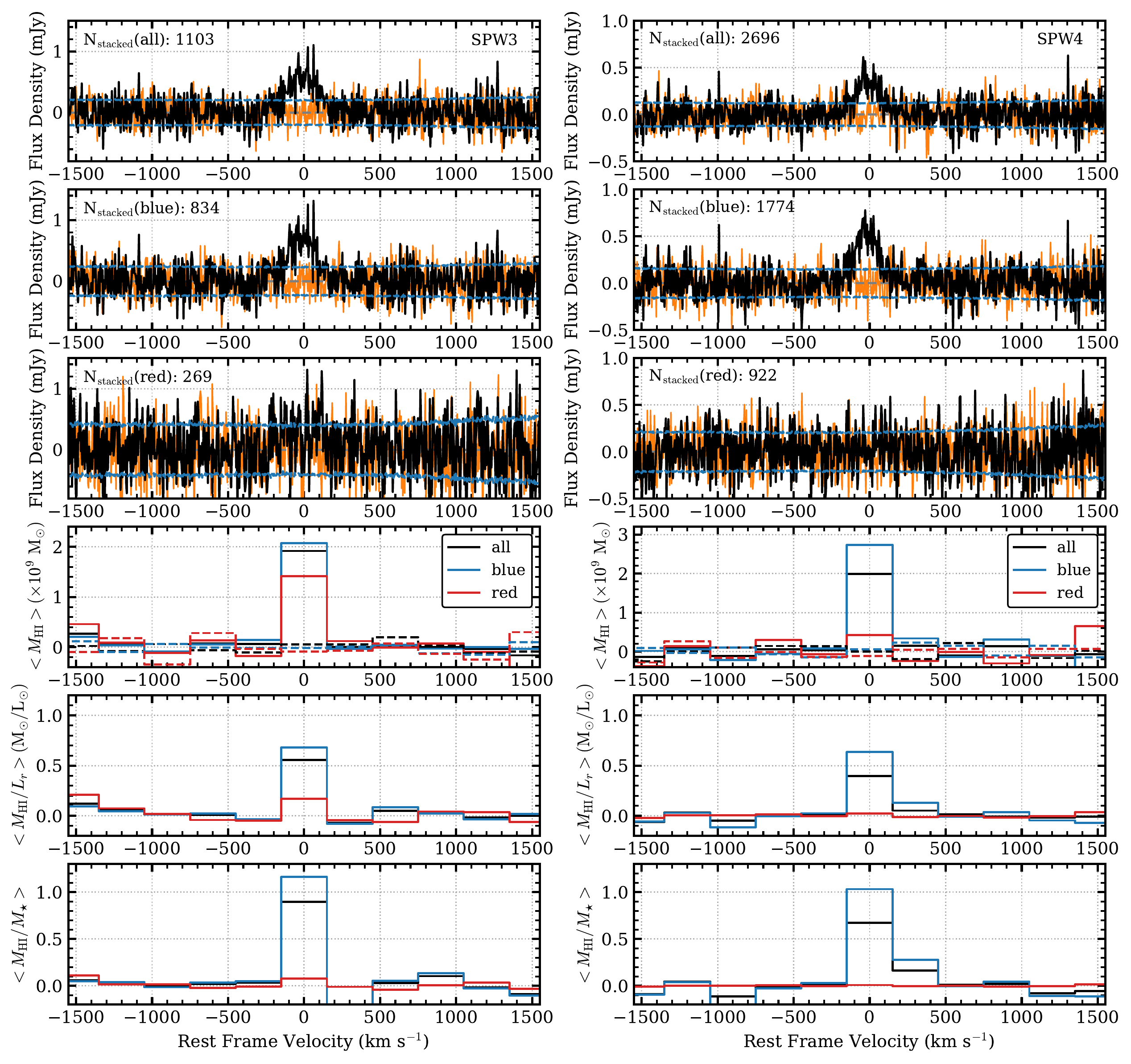} 
\caption{The upper three rows show co-added {\HI} spectra for each galaxy class (all, blue and red) along with the corresponding reference spectra (in orange) in SPW3 and SPW4 ({\it left} and {\it right}), respectively. The reference spectra were extracted at positions offset from each galaxy position and then stacked. The dashed horizontal lines indicate 1~$\sigma$ uncertainties of the stacked spectra. In the lower three rows, the co-added {\HI} mass and {\HI} mass-to-light ratio, and {\HI} mass-to-stellar mass spectra for the sub-classes are re-binned with the velocity width of 300~{\kms} to obtain average {\MHI}, $\MHI / L_{r}$ and $\MHI / M_{\star}$. In the fourth row, the dashed lines denote the re-binned {\HI} mass spectra of the reference spectra for the corresponding galaxy classes.}
\label{fig:stacked_spectra_blue_red}
\end{figure*}

\begin{table*}
\caption{The stacked {\HI} measurements: the number of stacked galaxies, average {\HI} mass, {\HI} mass-to-light ratio in the  $r-$band, and {\HI} to stellar mass fraction for different sub-samples based on colour and group environments in each SPW.}
\label{tab:stacking_measurements}
\centering
\begin{tabular}{lccccccccc}
\hline
   & \multicolumn{4}{c}{SPW3 ($z\sim0.057$)} & & \multicolumn{4}{c}{SPW4 ($z\sim0.080$)}  \\
   Sample & $N_{\rm gal}$ & $\langle \MHI \rangle$ & $\langle \MHI / L_{r} \rangle$ & $\langle \MHI / M_{\star} \rangle$ &  & $N_{\rm gal}$ & $\langle \MHI \rangle$ & $\langle \MHI / L_{r} \rangle$ & $\langle \MHI / M_{\star} \rangle$ \\
   & & [$10^{9}\ \msun$] & [$\msun/\lsun$] &  &  &  & [$10^{9}\ \msun$] & [$\msun/\lsun$] &  \\
   \hline
   All           & 1103 & 1.92~$\pm$~0.10 & 0.53~$\pm$~0.06 & 0.90~$\pm$~0.13 &  & 2696 & 1.99~$\pm$~0.13 & 0.38~$\pm$~0.04  & 0.67~$\pm$~0.06 \\
   Blue         & 834 & 2.08~$\pm$~0.12 & 0.65~$\pm$~0.07 & 1.16~$\pm$~0.17 &  & 1774 & 2.74~$\pm$~0.15 & 0.61~$\pm$~0.05 & 1.03~$\pm$~0.09 \\
   Red          & 269 & 1.42~$\pm$~0.21 & 0.16~$\pm$~0.06 & 0.08~$\pm$~0.03 &  &  922 & 0.42~$\pm$~0.22 & 0.02~$\pm$~0.01  & 0.01~$\pm$~0.01 \\
   Centrals   & 156 & 2.89~$\pm$~0.27 & 0.34~$\pm$~0.06 & 0.45~$\pm$~0.09 &  &  345 & 2.85~$\pm$~0.35 & 0.19~$\pm$~0.04 & 0.32~$\pm$~0.06\\
   Satellites & 331 & 1.50~$\pm$~0.18 & 0.43~$\pm$~0.10 & 0.67~$\pm$~0.15 &  &  874 & 1.55~$\pm$~0.23 & 0.29~$\pm$~0.07 & 0.53~$\pm$~0.08 \\
   Isolated   & 616 & 1.90~$\pm$~0.14 & 0.68~$\pm$~0.09 & 1.13~$\pm$~0.22 &  & 1477 & 2.03~$\pm$~0.16 & 0.51~$\pm$~0.06 & 0.84~$\pm$~0.10 \\
   \hline
\end{tabular}
\end{table*}

To investigate the dependence of {\HI} content in galaxies on galaxy colour, we sub-divided our G23 sample into blue and red galaxies for both SPW3 and SPW4 based on a simple colour cut in the colour-magnitude diagram (CMD) in Fig.~\ref{fig:cmd}. The two sub-samples are well separated in the CMD. The distribution of two galaxy classes in Fig.~\ref{fig:galaxy_distribution} shows that red galaxies seem to be more spatially clustered compared to the blue ones as expected from the well-known colour-density relation \citep[e.g.][]{Balogh:2004,Kauffmann:2004}.

We then stacked {\HI} spectra separately for these colour-based samples as well as the entire sample following the stacking procedure described in Section~\ref{sec:stacking_method}.
Using the stacked spectra, we measured average {\HI} mass, {\HI} mass-to-light ratio and {\HI} mass fraction over stellar mass for each sample in SPW3 and SPW4, respectively, as seen in Fig.~\ref{fig:stacked_spectra_blue_red} and Table~\ref{tab:stacking_measurements}. 
All stacked measurements except for the red sample of SPW4 show significant detections ($> 6~\sigma$).
In particular, the stacked {\HI} measurements for all and blue galaxy samples in both redshift bins have a significance above $15~\sigma$.

In the redshift range ($0.039 < z < 0.088$) of the G23 field covered by both SPWs, most of the {\HI} gas seems to reside in blue galaxies. The comparison of the stacked {\HI} mass between sub-samples shows that blue galaxies in SPW3 have $\sim 46 \%$ more {\HI} gas on average than the red sample, while the amount of {\HI} gas in blue galaxies in SPW4 is 6.5 times larger than that of the red ones. Blue galaxies are also more gas-rich in both SPWs than red galaxies as seen in the {\HI} mass-to-light ratio and {\HI} gas fraction measurements. In particular, the average {\HI} mass of blue galaxies is comparable to their average stellar mass.

\begin{figure}
\centering
\includegraphics[width=0.45\textwidth]{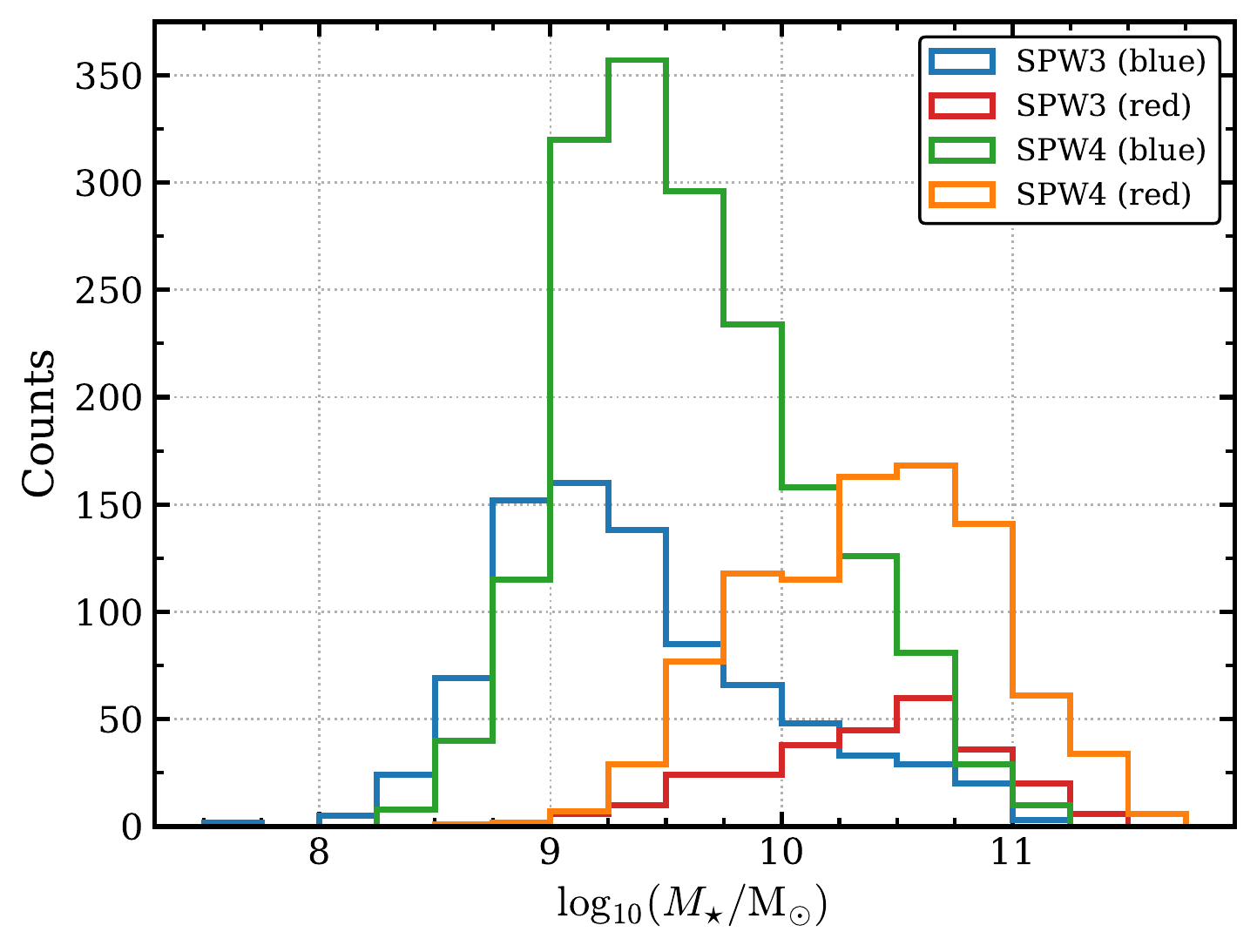} 
\caption{The stellar mass distribution of blue and red galaxies at SPW3 and SPW4.}
\label{fig:hist_smass_spw}
\end{figure}

When comparing the same galaxy populations between different redshift ranges, blue galaxies in SPW4 have $32 \%$ more {\HI} gas than in SPW3. The {\HI} gas fraction of blue galaxies over $r$-band luminosity and stellar mass is consistent between the two redshift bins within the uncertainties, albeit slightly lower in SPW4. However, the {\HI} gas mass in red galaxies in SPW4 is three times less than for SPW3. Moreover, red galaxies in SPW4 are likely to be extremely gas-poor compared to those in SPW3. The difference in {\HI} gas content for red galaxies between the two redshift bins seems to be a selection bias in GAMA rather than an evolutionary trend. For SPW4, more high-mass red galaxies ($M_{\star} > 10^{11}$ M$_{\sun}$) were selected than for SPW3, as seen in Fig.~\ref{fig:hist_smass_spw}. While the mass distributions of red galaxies in SPW3 and SPW4 are similar, the lower redshift (SPW3) sample includes more group central galaxies (see further discussion below), which have higher {\HI} mass than the satellites or isolated centrals. This is likely to explain the higher {\HI} masses we see in SPW3 overall. The trends seen in the blue and red samples demonstrate that the distinct bimodality between blue late-type and red early-type galaxy morphology based on optical colour is strongly linked to {\HI} gas properties \citep[e.g.][]{Kannappan:2004, West:2009, Healy:2019}. In future work, further investigation undertaken with new GAMA morphological classifications \citep{Driver:2022} may reveal more detail regarding the correlation between galaxy morphology and {\HI} gas content.

\subsection{Group vs. Isolated Galaxies}

\begin{figure*}
\centering
\includegraphics[width=0.9\textwidth]{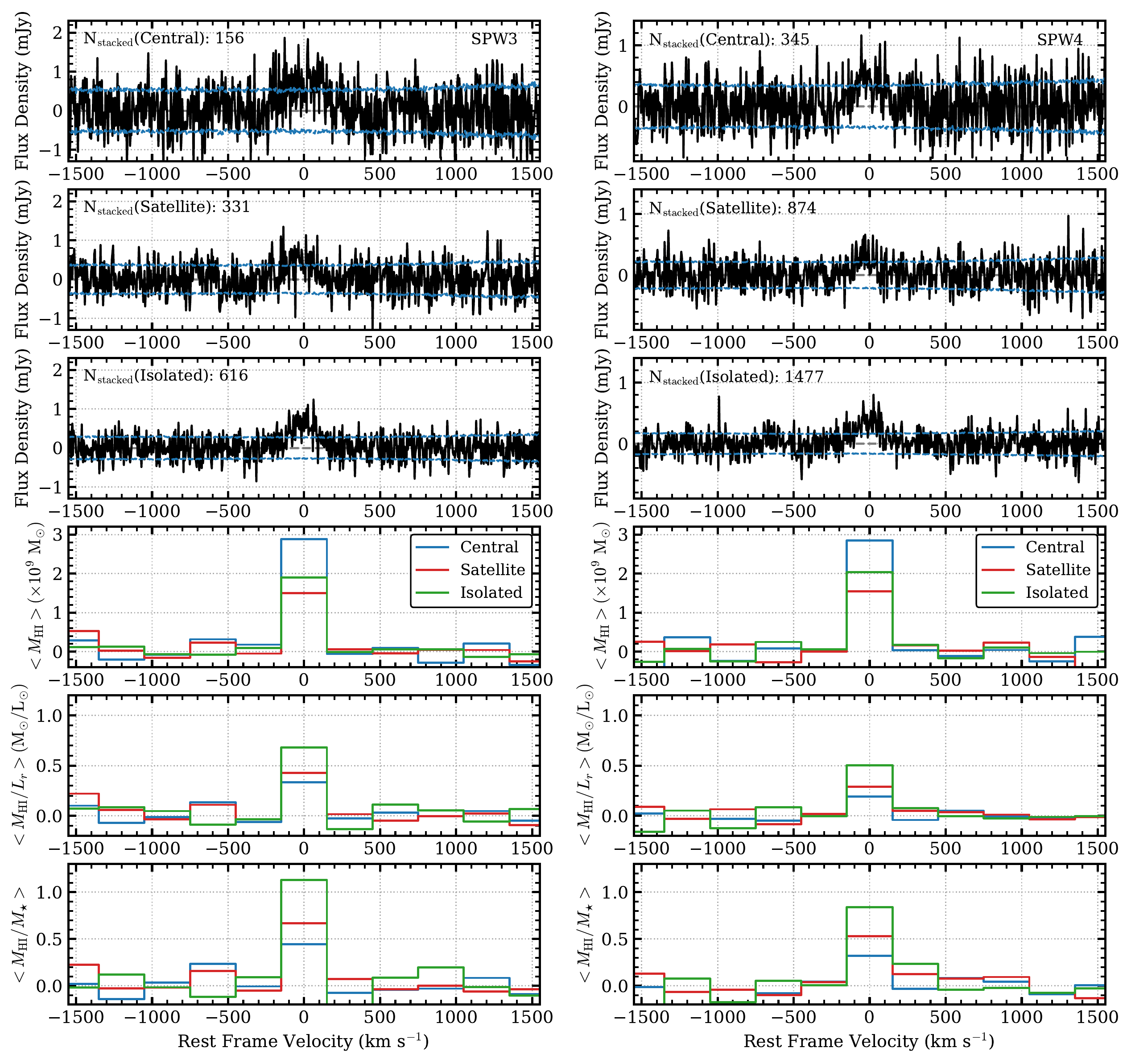} 
\caption{The upper three rows show co-added {\HI} spectra for each galaxy class (group central, satellite and isolated) in SPW3 and SPW4  ({\it left} and {\it right}), respectively. The dashed horizontal lines indicate 1~$\sigma$ uncertainties of the stacked spectra. In the rest of the rows, the co-added {\HI} mass and {\HI} mass-to-light ratio, and {\HI} mass-to-stellar mass spectra for the sub-classes are re-binned with the velocity width of 300~{\kms} to obtain average {\MHI}, $\MHI / L_{r}$ and $\MHI / M_{\star}$.}
\label{fig:stacked_spectra_group}
\end{figure*}

\begin{figure*}
\centering
\includegraphics[width=0.95\textwidth]{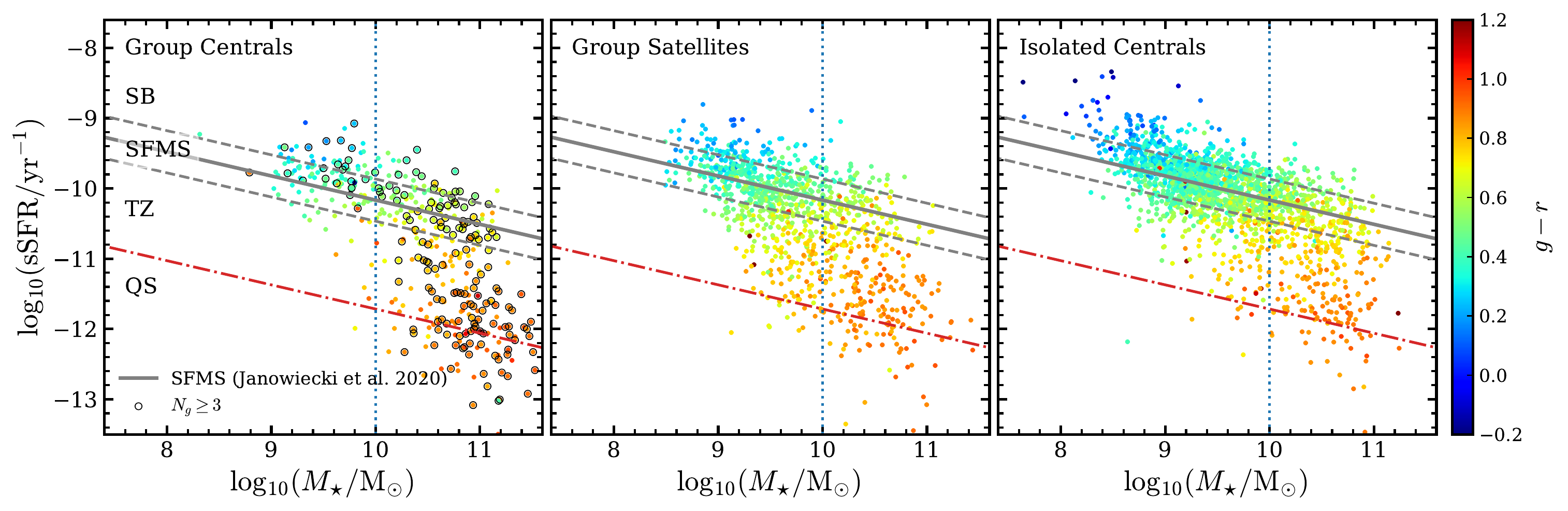} 
\caption{This shows the distribution of three sub-samples for different group environments on the specific SFR and stellar mass plane, colour-coded by the $g-r$ colour of each galaxy. In the panels, the solid and the dashed lines indicate a fit of the star-forming main sequence based on xGASS \citep{Janowiecki:2020} and its $\pm$0.3 dex offsets ($\Delta {\rm SFMS}=\pm0.3$), respectively. The red dash-dotted lines show a limit to separate quiescent galaxies. The blue vertical dotted lines are the stellar mass cut used to divide the sub-samples of starburst (SB) and star-forming main sequence (SFMS) by stellar mass. The black open circles in the first panel indicate group central galaxies with more than three group members of $N_{g} \geq 3$.}
\label{fig:sfms_group}
\end{figure*}

\begin{table}
\caption{The stacked {\HI} measurements of group centrals, group satellites and isolated centrals for different status of star formation: the number of stacked galaxies, average stellar mass, average {\HI} mass, and {\HI} gas fraction for starbursts (SB), star-forming main sequence (SFMS), transition zone (TZ) and quiescent (QS) galaxies, respectively. The additional {\HI} stacking measurements are made for the sub-samples of SB and SFMS galaxies split by their stellar mass--$M_{\star} < 10^{10}~\msun$ for low and $M_{\star} \ge 10^{10}~\msun$ for high, respectively.}
\label{tab:stacking_group}
\centering
\begin{tabular}{lcccc}
\hline
   Sample & $N_{\rm gal}$ & $\langle M_{\star} \rangle$ & $\langle \MHI \rangle$ & $\langle \MHI / M_{\star} \rangle$ \\
   & & [$10^{9}\ \msun$] & [$10^{9}\ \msun$] &  \\
  \hline
  \multicolumn{5}{l}{Group Centrals} \\
  All Centrals  & 501 &  59.90  & 3.21~$\pm$~0.25  &  0.35~$\pm$~0.05 \\
  SB       & 67  & 20.94 & 6.41~$\pm$~0.66 & 1.30~$\pm$~0.26 \\
   SB (low) & 35 & 4.50 & 7.09~$\pm$~0.96 & 2.34~$\pm$~0.49 \\
   SB (high) & 32 & 38.93 & 6.08~$\pm$~0.94 & 0.18~$\pm$~0.05 \\  
  SFMS  & 172 & 28.52 & 4.33~$\pm$~0.40 & 0.51~$\pm$~0.10 \\
   SFMS (low) & 71 & 4.26 & 3.29~$\pm$~0.58 & 1.03~$\pm$~0.24 \\
   SFMS (high) & 101 & 45.58 & 5.39~$\pm$~0.58 & 0.14~$\pm$~0.03 \\
  TZ      &  131  & 72.30  & 2.13~$\pm$~0.52  & 0.004~$\pm$~0.046 \\
  QS      & 131 & 108.61 & 0.91~$\pm$~0.49 & 0.02~$\pm$~0.02 \\
  & & & & \\
  \multicolumn{5}{l}{Group Satellites} \\
  All Satellites &  1205  & 16.75  &  1.78~$\pm$~0.17  &  0.58~$\pm$~0.07 \\
  SB       & 128 & 7.64 &  4.95~$\pm$~0.55 & 2.09~$\pm$~0.35 \\ 
   SB (low) & 97 & 2.76 & 4.69~$\pm$~0.63 & 2.66~$\pm$~0.46 \\
   SB (high) & 31 & 22.90 & 5.88~$\pm$~1.16 & 0.30~$\pm$~0.06 \\  
  SFMS  & 483 & 8.16 & 2.06~$\pm$~0.25 & 0.75~$\pm$~0.13 \\
   SFMS (low) & 382 & 3.13 & 1.78~$\pm$~0.27 & 0.92~$\pm$~0.17 \\
  SFMS (high) & 101 & 27.21 & 3.08~$\pm$~0.64 & 0.10~$\pm$~0.04 \\
  TZ      &  360 &  20.31  & 0.95~$\pm$~0.33  &  0.17~$\pm$~0.07 \\
  QS      & 234  &  33.98  & 0.49~$\pm$~0.41 & 0.04~$\pm$~0.06 \\
  & & & & \\
  \multicolumn{5}{l}{Isolated Centrals} \\
  All Isolated &  2093  &  12.28  &  2.22~$\pm$~0.12  &  0.93~$\pm$~0.10 \\
  SB       & 292  & 6.60 & 4.00~$\pm$~0.35 & 2.40~$\pm$~0.51 \\
   SB (low) & 241 & 2.41 & 3.48~$\pm$~0.37 & 2.85~$\pm$~0.61 \\
   SB (high) & 51 & 26.38 & 6.80~$\pm$~0.90 & 0.25~$\pm$~0.05 \\    
  SFMS  & 1208 & 7.88 & 2.40~$\pm$~0.15 & 0.98~$\pm$~0.11 \\
   SFMS (low) & 974 & 2.77 & 2.06~$\pm$~0.16 & 1.18~$\pm$~0.13 \\
   SFMS (high) & 234 & 29.16 & 4.15~$\pm$~0.41 & 0.17~$\pm$~0.02 \\
  TZ      &  413  &  19.08  &  1.07~$\pm$~0.27  &  0.18~$\pm$~0.10 \\
  QS      &  180  & 35.43 & 0.32~$\pm$~0.44 & 0.01~$\pm$~0.01 \\
  \hline
\end{tabular}
\end{table}

The study of the environmental impact on {\HI} gas content is one of the scientific goals for the DINGO survey.
The GAMA survey allows us to conduct this study due to its well-defined group catalogue generated based on the high spectroscopic completeness (98 per~cent to $r < 19.8$) of the survey \citep{Robotham:2011}.
We used version 8 of the group catalogue for G23, which identifies groups using a friends-of-friends algorithm taking into account redshift space distortions.
Based on the group catalogue, we divided our galaxies into three sub-samples for each redshift bin (SPW): group central, group satellite, and isolated central galaxies.
For these sub-samples, we conducted {\HI} stacking to measure the average {\HI} mass, {\HI} mass-to-light ratio, and {\HI} gas fraction of each sample, enabling us to examine the variation of {\HI} gas content with different environments statistically and thereby providing observational evidence for physical mechanisms driving the environmental effects.

In both SPWs, group central galaxies contain more {\HI} gas than satellite and isolated central galaxies while the former are more {\HI}-deficient than the latter two when comparing {\HI} mass-to-light ratio and {\HI} gas fraction between them (see Table~\ref{tab:stacking_measurements} and Fig.~\ref{fig:stacked_spectra_group}).
The lower {\HI} gas fraction of group centrals, albeit containing larger {\HI} mass, likely originates from these galaxies having much higher stellar mass than satellites and isolated centrals. 

To explore this tendency in more detail, we look into how the {\HI} content of galaxies in different environments varies with physical properties, namely star formation activity and stellar mass. To this end, the two redshift bins of SPW3 ($z \sim 0.057$) and SPW4 ($z \sim 0.080$) are combined to increase the sample size for stacking measurements. 
This step is reasonable because of the lack of any significant evolution of the {\HI} gas content in the different environments between the two redshift intervals. 

Fig.~\ref{fig:sfms_group} shows the distribution of each sample in the specific star formation rate (sSFR $\equiv {\rm SFR}/M_{\star}$)--stellar mass plane, colour-coded by $g-r$ colour.
In all three sub-samples, galaxies with bluer colours are populated on and above the star-forming main sequence \citep[solid and dashed lines, e.g.][]{Brinchmann:2004, Noeske:2007, Whitaker:2012}, while galaxies with low sSFR ($< 10^{-11}$ yr$^{-1}$) are consistent with the sample categorised as red galaxies in the previous section. 
This highlights the correspondence between the colour and the star formation properties of galaxies \citep[e.g.][]{Schawinski:2014,Davies:2019,Corcho-Caballero:2020}.

Based on the sSFR and stellar mass of our sample galaxies, we divide each environmental sample into three sub-categories in the sSFR-$M_{\star}$ plane.
In each panel of Fig.~\ref{fig:sfms_group}, the grey solid line denotes the star-forming main sequence determined by \citet{Janowiecki:2020} and the dashed lines are $\pm$0.3~dex ($\sim \pm 1 \sigma$) offsets from the star-forming main sequence. We define galaxies within this range as star-forming main sequence (SFMS), while galaxies above the main sequence are referred to as starbursts (SB). For quiescent (QS) galaxies, we adopt $\Delta {\rm SFMS}=-1.55$ dex, following \citet{Janowiecki:2020}. We then classify the galaxies between SFMS and QS as the transition zone (TZ) sample. In addition, we sub-divide the SB and SFMS samples to check further dependence of the {\HI} properties of SB and SFMS galaxies on stellar mass: low mass with $M_{\star} < 10^{10}~\msun$ and high mass with $M_{\star} \geq 10^{10}~\msun$.
For all these samples, we make {\HI} stacking measurements to obtain their average {\HI} mass and {\HI} gas fraction.

It is apparent from the results in Table~\ref{tab:stacking_group} that the {\HI} abundance and {\HI} richness of all the sub-samples based on star formation properties in each environment are in line with the global trends found in Table~\ref{tab:stacking_measurements}.  
The SB, SFMS, TZ, and QS samples of the group central galaxies have higher stellar and {\HI} mass on average than the counterparts of both group satellites and isolated centrals, ending up being more dominant {\HI} gas reservoirs. It is likely because central galaxies can obtain extra {\HI} gas from their satellites \citep{Stevens:2019}. 
Indeed, there are many groups having more than three group members, as shown in the first panel of Fig.~\ref{fig:sfms_group}. They are distributed mostly above the star-forming main sequence fit close to the SB regime. However, as a result of their higher stellar mass, the {\HI} fraction of group central samples is relatively low compared to those of the two other environments.
Our findings appear to be contrary to those of other observations and simulations \citep[e.g.][]{Brown:2017, Stevens:2019}, where central galaxies are more gas-rich than satellites. However, this contrast is due to the difference in samples with the previous work. Their central galaxies consist of both group and isolated centrals. If these two groups in our data are combined, the gas fraction trends  are in agreement.

In the comparison to the {\HI} properties of centrals in different environments, central galaxies in isolated environments appear to be less massive in {\HI} but {\HI}-richer than centrals in group environments regardless of their stellar mass.
This trend seems to disagree with that found by \citet{Janowiecki:2017} based on single-dish Arecibo radio telescope observations.
They adopted the same selection strategy to separate central galaxies between group and isolated environments as here. They find that group centrals have a higher {\HI} gas fraction than isolated central galaxies, especially in a low mass range of $M_{\star} < 10^{10.5}~ \msun$. 
However, this disagreement can be explained by the difference between interferometric and single-dish data. Due to the larger beam size of single-dishes, observations of group central galaxies can readily be confused, leading to an overestimation of the {\HI} gas fraction in group central galaxies. This result is also found in other studies comparing {\HI} properties between different environments using both Arecibo and ASKAP data in overlapping target fields \citep{Roychowdhury:2022}. The {\HI} gas fraction of their low stellar mass groups  is larger than for isolated galaxies measured using Arecibo, while this trend is reversed with ASKAP data. In future work, we will look into this discrepancy in more detail using DINGO pilot survey data obtained with the ASKAP full array and an increased sample from other GAMA fields.

\begin{table}
\caption{The average SFR and {\HI} gas depletion time ($\langle t_{\rm dep} \rangle \equiv \langle \MHI \rangle / \langle {\rm SFR} \rangle$) of starbursts (SB) and star-forming main sequence (SFMS) samples in group centrals, group satellites and isolated centrals for different stellar mass bins (low and high) identified by the stellar mass threshold of $M_{\star} = 10^{10}~\msun$.}
\label{tab:t_dep}
\centering
\begin{tabular}{lcc}
\hline
   Sample & $\langle {\rm SFR} \rangle$ & $\langle t_{\rm dep} \rangle$ \\
   & [$\msun {\rm yr}^{-1}$] & [Gyr]  \\
  \hline
  \multicolumn{3}{l}{Group Centrals} \\
   SB (low) & 1.40 & 5.07~$\pm$~0.67 \\
   SB (high) & 4.55 & 1.34~$\pm$~0.21  \\  
   SFMS (low) & 0.48 & 6.86~$\pm$~1.20 \\
   SFMS (high) & 1.83 & 2.95~$\pm$~0.32 \\
  & & \\
  \multicolumn{3}{l}{Group Satellites} \\
   SB (low) & 0.95 & 4.92~$\pm$~0.66 \\
   SB (high) & 3.78 & 1.56~$\pm$~0.31 \\  
   SFMS (low) & 0.33 & 5.47~$\pm$~0.82 \\
   SFMS (high) & 1.29 & 2.40~$\pm$~0.49 \\
  & & \\
  \multicolumn{3}{l}{Isolated Centrals} \\
   SB (low) & 0.75 & 4.62~$\pm$~0.50 \\
   SB (high) & 3.70 & 1.84~$\pm$~0.24 \\    
   SFMS (low) & 0.31 & 6.69~$\pm$~0.53 \\
   SFMS (high) & 1.37 & 3.03~$\pm$~0.30 \\  
  \hline
\end{tabular}
\end{table}

For a given environmental category, SB galaxies contain more {\HI} gas and are {\HI}-richer than SFMS, TZ and QS galaxies. When comparing sub-samples of SB and SFMS galaxies separated by stellar mass, the {\HI} mass of galaxies is correlated with stellar mass while the {\HI} fraction is anti-correlated with stellar mass except for one sub-sample. Interestingly, the SB group centrals with low stellar mass have a large amount of {\HI} gas with higher {\HI} gas fractions. This means that the SB (low) group centrals are likely to consume their {\HI} gas slowly, therefore resulting inefficient star formation. Table~\ref{tab:t_dep} lists the average SFRs and the average {\HI} gas depletion times expressed as $\langle t_{\rm dep} \rangle \equiv \langle \MHI \rangle / \langle {\rm SFR} \rangle$ for all the sub-samples of SB and SFMS across all environments. SB populations tend to have higher SFR and shorter $t_{\rm dep}$ than SFMS within a given environment, and lower stellar mass samples at a given sub-sample are found to be less active in star formation and take a longer time to deplete their {\HI} reservoirs. These features in the lower stellar mass populations might indicate an earlier stage of evolution. Compared to different environments, group satellite samples appear to be rapidly evolving.
This may be driven by the many external and internal mechanisms that can more easily influence satellite galaxies in dense environments \citep{Cortese:2021}.

\section{{\HI} Scaling Relations}
\subsection{{\HI} Gas Fraction Scaling Relations}
\begin{figure}
\centering
\includegraphics[width=0.48\textwidth]{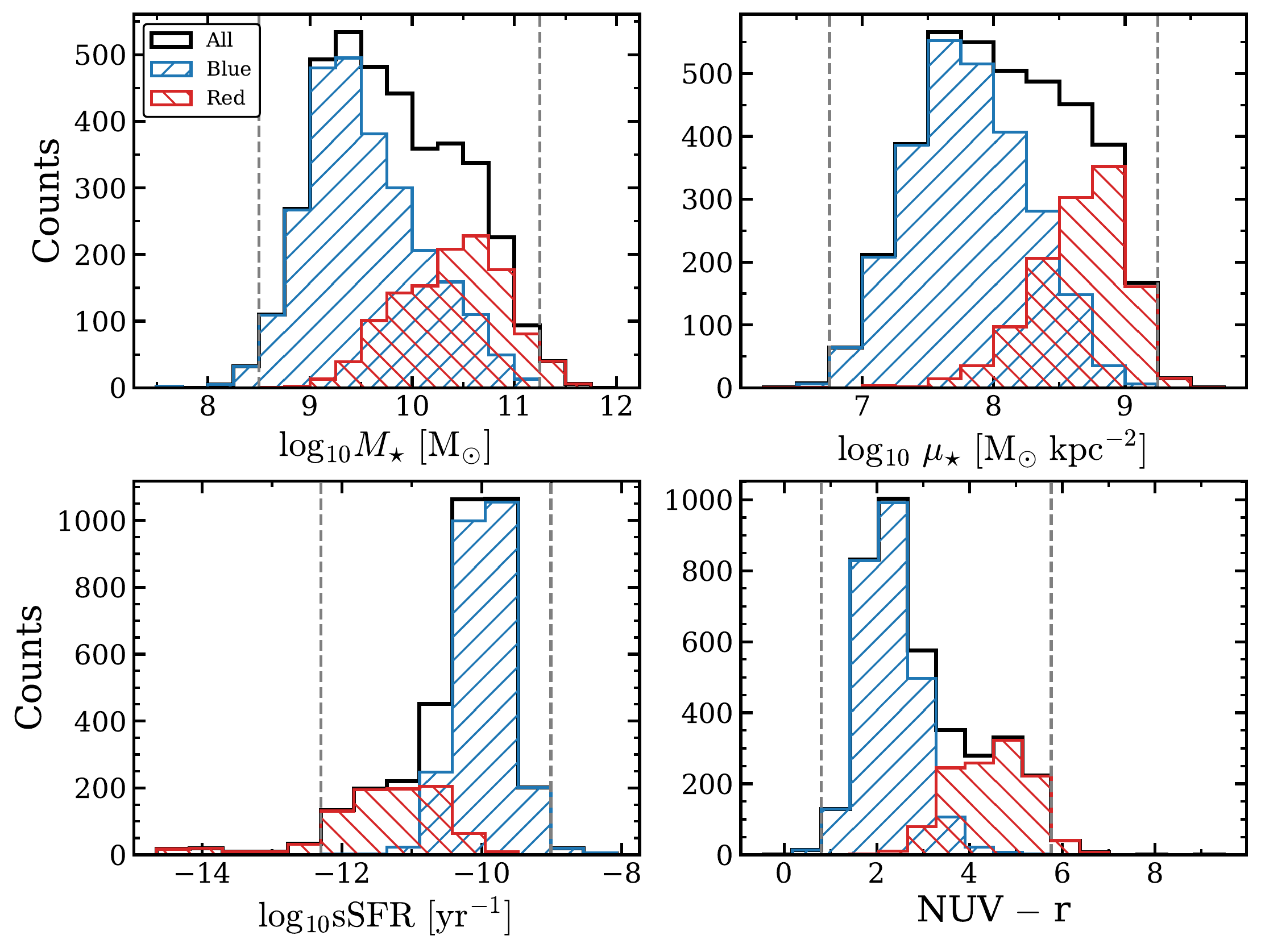} 
\caption{The distributions of stellar mass, stellar surface density, $NUV-r$ and specific SFR  for the sample in the clockwise direction from the top left. The dashed lines in each panel denote the cuts of each physical parameter within which {\HI} scaling relations are derived. In each panel, blue and red histograms are distributions of blue and red galaxies for the corresponding physical property, respectively.}
\label{fig:hist_smass_mu_nuvr_ssfr}
\end{figure}

Studying the relation of {\HI} gas content with other physical properties of galaxies is crucial in understanding the physical processes regulating galaxy evolution \citep[e.g.][]{Catinella:2010,Catinella:2018}.
To investigate a variety of {\HI} scaling relations, we first check the distributions of the physical properties that we want to relate to {\HI}, for instance stellar mass ($M_{\star}$), stellar mass surface density  (${\rm \mu_{\star}}$), star formation rate (SFR) and $NUV-r$ colour. To this end, we combine two redshift bins data into one sample ($0.039 < z < 0.088$), ending up with the sample size of 3799 in total.
For those galaxies, the $R_{50}$, ${\it NUV}$ and ${\it r}$ are all taken from the new GAMA ProFound photometry catalogue \citep{Bellstedt:2020}, and the stellar mass and SFR values are taken from the ProSpect SED fits to photometry \citep{Bellstedt:2020a}, which are part of Data Release 4 in GAMA \citep{Driver:2022}.
The stellar mass surface density is defined as:
\begin{equation}
\label{eq:mu_star}
{\mathbf \mu_{\star}} = \frac{M_{\star}}{2 \pi R_{50}^{2} C_{\rm axrat}},
\end{equation}
where $M_{\star}$ is stellar mass in units of solar mass and $R_{50}$ is the effective half-light radius in kpc, measured using isophotal fitting in a stack of $r$ and $Z$ bands as the elliptical semi-major axis containing 50 per~cent of the flux. $C_{\rm axrat}$ is the axial ratio of semi-major and semi-minor axes of the isophotal fit to compensate for a systematic underestimation of the stellar mass surface density due to the elliptical shape of the fit. Specific star formation rate (sSFR) is derived as SFR divided by stellar mass (${\rm SFR}/M_{\star}$).
In Fig.~\ref{fig:hist_smass_mu_nuvr_ssfr}, the distributions of these physical properties of our sample show that blue galaxies tend to be less massive, more disk-dominated (${\rm log}\ \mu_{\star}\ [{\msun}\ {\rm kpc}^{-2}] < 8.5$), and more star-forming than red ones as expected. In addition, the colour cut used to the divide blue and red samples seems to work well as the $NUV-r$ colour distribution is consistent with the sample selected based on $g-r$ colour.  

Due to low-number statistics at both the low and high end, we applied the cuts for stellar mass, surface density, specific SFR, and $NUV-r$ colour as follows:
\begin{itemize}
  \centering
  \item[] $8.5 < {\rm log}\ M_{\star}\ [{\msun}] < 11.25$, 
  \item[] $6.75 < {\rm log}\ \mu_{\star}\ [{\msun}\ {\rm kpc}^{-2}] < 9.0$,
  \item[] $-12.3 < {\rm log\ sSFR\ [yr^{-1}]} < -9.01$,
  \item[] $0.8 < NUV - r < 5.8$. 
  \end{itemize}

\begin{figure*}
\centering
\includegraphics[width=0.9\textwidth]{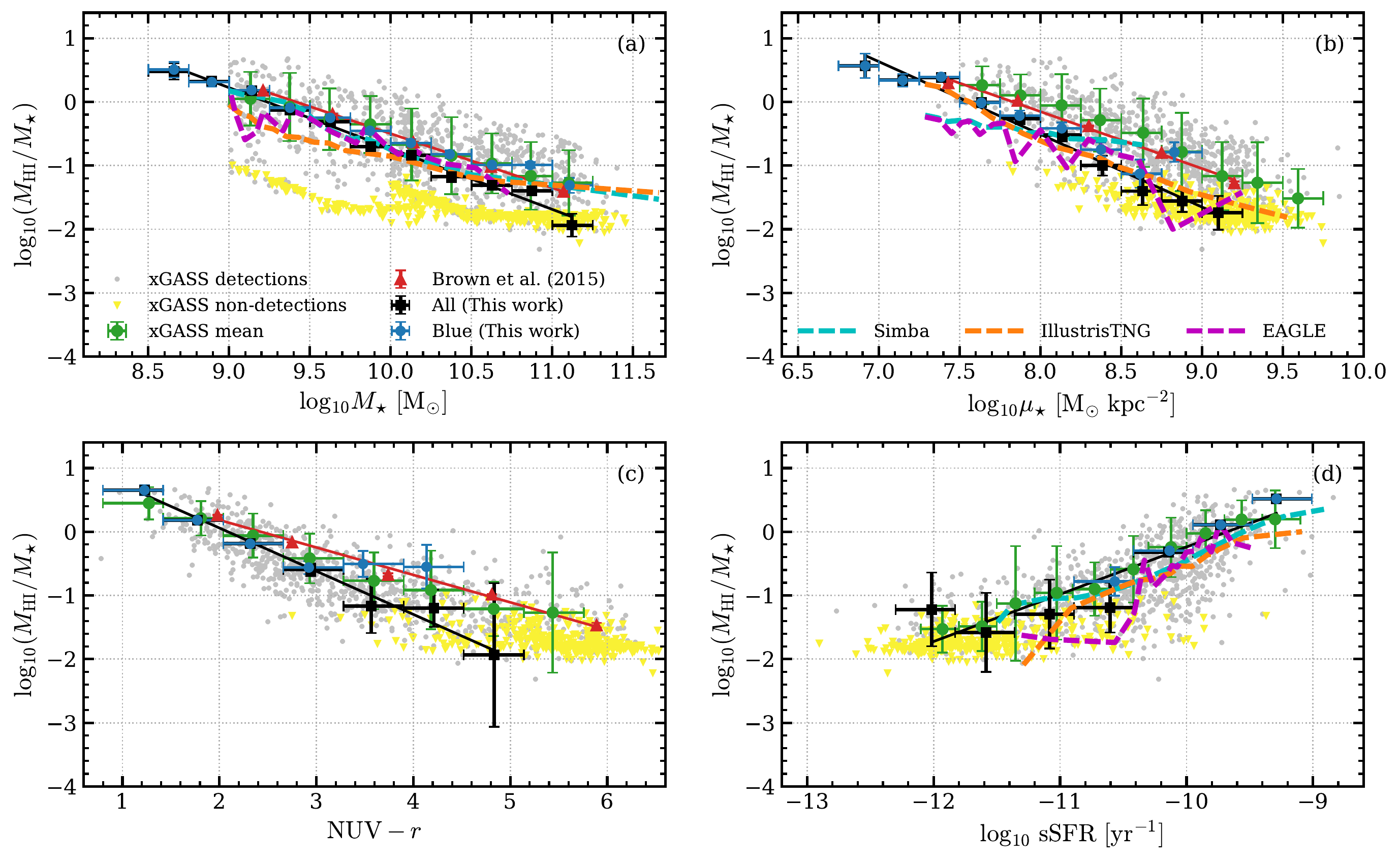} 
\caption{This shows average {\HI} gas fraction as a function of: (a) galaxy stellar mass; (b) stellar surface density; (c)
  ${\rm NUV}-r$ colour; and (d) specific star formation rate for the whole sample and sub-sample of blue galaxies with black squares and blue circles, respectively. The black dashed lines are linear fits to our entire sample. Grey points and yellow downward triangles are the xGASS detections and the $5 \sigma$ upper limits of non-detections, respectively. The green circles denote the mean of linear values of all the xGASS measurements including non-detections. The red triangle and dashed line are measurements from \citet{Brown:2015} and fitted lines. Cyan, orange and purple dashed lines are theoretical model predictions from SIMBA, IllustrisTNG, and EAGLE simulations, respectively.}
\label{fig:scaling_relations}
\end{figure*}

These selection criteria overlap well with the range of physical parameters covered by the extended {\it GALEX} Arecibo SDSS Survey \citep[xGASS,][]{Catinella:2018} that has conducted a comprehensive investigation of {\HI} scaling relations with a large sample in the local Universe ($z < 0.05$). xGASS is an ideal reference sample for our scaling relations with the DINGO early science data. We compare our scaling relations with those of \citet{Brown:2015} who also adopted {\HI} spectral stacking techniques to investigate {\HI} scaling relations using the ALFALFA and SDSS data spanning a similar dynamic range of physical parameters. 
However, our dynamic range extends further into the lower stellar mass and surface density regime by about 0.5~dex. Moreover, the redshift range of the DINGO early science data is higher than those of xGASS and ALFALFA. This extension is beneficial in determining how {\HI} scaling relations extend to the lower mass and more disk-dominated regimes and if there is any evolution in {\HI} scaling relations out to $z \sim 0.09$. Within the selected range, we divide our sample into bins and then make {\HI} stacking measurements for the average {\HI} gas fraction in these bins to examine scaling relations with the corresponding physical parameters.

\begin{table}
  \caption{The average {\HI} gas fractions for different physical properties of the DINGO early science sample shown in Fig.~\ref{fig:scaling_relations}--stellar mass, stellar mass surface density, $NUV-r$ colour, sSFR.
    $N_{\rm gal}$ is the number of galaxies used for the stacked {\HI} gas fraction measurements of each physical property bin.}
\label{tab:scaling_relations}
\centering
\begin{tabular}{lcrr}
 \hline
   $x$ & $\langle x \rangle$ &  $N_{\rm gal}$ & ${\rm log} \langle \MHI / M_{\star} \rangle$ \\
 \hline
$ {\rm log}\ M_{\star}$ & 8.66 & 110 & $0.48~\pm~0.13$ \\ 
                                    & 8.89 & 269 & $0.32~\pm~0.07$ \\
                                    & 9.14 & 493 & $0.18~\pm~0.05$ \\
                                    & 9.38 & 534 & $-0.13~\pm~0.06$ \\
                                    & 9.63 & 482 & $-0.31~\pm~0.06$ \\
                                    & 9.88 & 442 & $-0.70~\pm~0.09$ \\
                                    &10.13 & 359 & $-0.84~\pm~0.07$ \\
                                    &10.38 & 367 & $-1.17~\pm~0.09$ \\
                                    &10.63 & 338 & $-1.31~\pm~0.07$ \\
                                    &10.87 & 226 & $-1.40~\pm~0.06$ \\
                                    &11.12 & 94 & $-1.94~\pm~0.18$ \\
 & & & \\
$ {\rm log}\ \mu_{\star}$ & 6.92 & 64 & $0.57~\pm~ 0.19$ \\
                                        & 7.15 & 211 & $0.34~\pm~0.10$ \\
                                        & 7.39 & 388 & $0.38~\pm~0.05$ \\
                                        & 7.64 & 566 & $-0.01~\pm~0.07$ \\
                                        & 7.88 & 550 & $-0.26~\pm~0.09$ \\
                                        & 8.14 & 504 & $-0.52~\pm~0.10$ \\
                                        & 8.38 & 487 & $-1.00~\pm~0.16$ \\
                                        & 8.63 & 451 & $-1.40~\pm~0.22$ \\
                                        & 8.88 & 387 & $-1.56~\pm~0.17$ \\
                                        & 9.10 & 167 & $-1.74~\pm~0.26$ \\
 & & & \\
$NUV-r$          & 1.23 & 129 & $0.65~\pm~0.08$ \\
                        & 1.77 & 831 & $0.18~\pm~0.05$ \\
                        & 2.32 & 1002 & $-0.19~\pm~0.06$ \\
                        & 2.94 & 576 & $-0.59~\pm~0.10$ \\
                        & 3.56 & 352 & $-1.17~\pm~0.42$  \\
                        & 4.21 & 280 & $-1.20~\pm~0.29$ \\
                        & 4.83 & 330 & $-1.94~\pm~1.23$ \\ 
 & & & \\
$ {\rm log\ sSFR}$ & $-$12.02 & 134 & $-1.22~\pm~0.58$ \\
                              & $-$11.58 & 198 & $-1.58~\pm~0.62$ \\
                              & $-$11.08 & 220 & $-1.29~\pm~0.54$ \\
                              & $-$10.60 & 452 & $-1.19~\pm~0.39$ \\  
                              & $-$10.14 & 1063 & $-0.32~\pm~0.06$ \\   
                              & $-$9.73 & 1064 & $0.11~\pm~0.04$ \\   
                              & $-$9.29 & 201 & $0.52~\pm~ 0.07$ \\   
  \hline
\end{tabular}
\end{table}

Fig.~\ref{fig:scaling_relations} presents {\HI} gas fraction scaling relations derived from the DINGO early science data for: (a) stellar mass; (b) stellar mass surface density; (c)  $NUV-r$ colour; and (d) specific SFR for the whole sample and the sub-sample of blue galaxies, shown by black squares and blue circles, respectively. The black solid line denotes a linear fit to the whole sample in each panel.
Our scaling relations are compared with the previous work using the xGASS representative sample\footnote{\url{https://xgass.icrar.org/}} \citep{Catinella:2018} and the ALFALFA/SDSS data \citep{Brown:2015}. We also compared our scaling relations to the predictions from \citet{Dave:2020}  using three cosmological hydrodynamic simulations, SIMBA \citep{Dave:2019}, EAGLE \citep{Schaye:2015}, and IllustrisTNG \citep{Pillepich:2018}. For EAGLE, we use the results from the EAGLE-Recal simulation (Recal-L25N752) with higher mass and spatial resolutions.

In each panel, grey dots and yellow downward triangles indicate xGASS detections and $5 \sigma$ upper limits of xGASS non-detections, respectively.
The green circles are the average of linear values of the xGASS sample including the upper limits. 
Red triangles and dashed lines are values from Table~1 of \citet{Brown:2015} and linear fits to them.
Cyan, orange and purple dashed lines denote theoretical model predictions from SIMBA, IllustrisTNG, and EAGLE simulations, respectively. 
The overall observed trends of our scaling relations follow well those of xGASS and ALFALFA/SDSS
although our relations become consistently lower with increasing stellar mass, stellar surface density and $NUV-r$ colour. The three model predictions appear to agree well with our scaling relations, although there are some variations among the three models.

In Fig.~\ref{fig:scaling_relations}~(a), the average {\HI} gas fraction for our sample is shown as a function of stellar mass along with other reference data.
The relation for the whole sample is consistent with the xGASS mean values and the \citeauthor{Brown:2015} stacking measurements for the lower mass bins, but diverges at higher stellar mass ($M_{\star} > 10^{10}$ {\msun}). But the blue galaxy sample is in excellent agreement with the reference relations. As seen in Fig.~\ref{fig:hist_smass_mu_nuvr_ssfr}, the high-mass end is dominated by gas-poor red galaxies, which likely contributes to this offset.

SIMBA simulation predictions agree reasonably well with our entire sample at most masses
but show better agreement with the blue galaxy sample at high masses ($M_{\star} > 10^{11}$ {\msun}) where our relation is subject to the impact of the gas poor population. IllustrisTNG shows a lower slope of the relation, resulting in underprediction and overprediction at low and high masses, respectively. The EAGLE simulation is in good agreement with our relations except at lower masses ($M_{\star} < 10^{9.5}$ {\msun}).

Average {\HI} gas fraction with stellar mass surface density is shown in Fig.~\ref{fig:scaling_relations}~(b).
By definition, galaxies having higher stellar surface density ($ {\rm log}\ \mu_{\star} \gtrsim 8.5$) are bulge-dominated while the others are disk-dominated  \citep{Kauffmann:2006}.
The offset of this scaling relation appears to be larger than the relation with stellar mass ($\sim$ 0.2 to 0.6~dex).
As in case of the scaling relation with stellar mass, exclusion of the red, bulge-dominated systems from the sample reduces the offset but for this relation the overall offset remains significant.
This may be due to the difference in the values for $R_{50}$ used for calculating stellar surface density in Eq.~\ref{eq:mu_star}.
The GAMA $R_{50}$ is measured from a stacked image of $r$ and $Z$ band data
while xGASS and \citeauthor{Brown:2015} use the Petrosian half-light radius measured only in the $z$ band.
The GAMA photometry is optimised in order to capture all the flux of an object, and therefore there are dilations of the segments implemented. This may well be producing larger segments, and therefore slightly larger $R_{50}$ values.
If our $R_{50}$ is larger than that used for the two ALFALFA data, this will lead to lower values for stellar surface density, 
thus resulting in a horizontal shift of our scaling relation toward the lower surface density end. 

By contrast, simulations are generally consistent with our results. IllustrisTNG is in excellent agreement with our scaling relation for the entire sample, while SIMBA has a flatter trend with $\mu_{\star}$ than observations, and EAGLE tends to produce lower {\HI} gas fraction at lower $\mu_{\star}$. The different maximum of $\mu_{\star}$ in each simulation is attributed to numerical resolution: IllustrisTNG has the highest resolution, while SIMBA has the lowest resolution.

\begin{figure}
\centering
\includegraphics[width=0.45\textwidth]{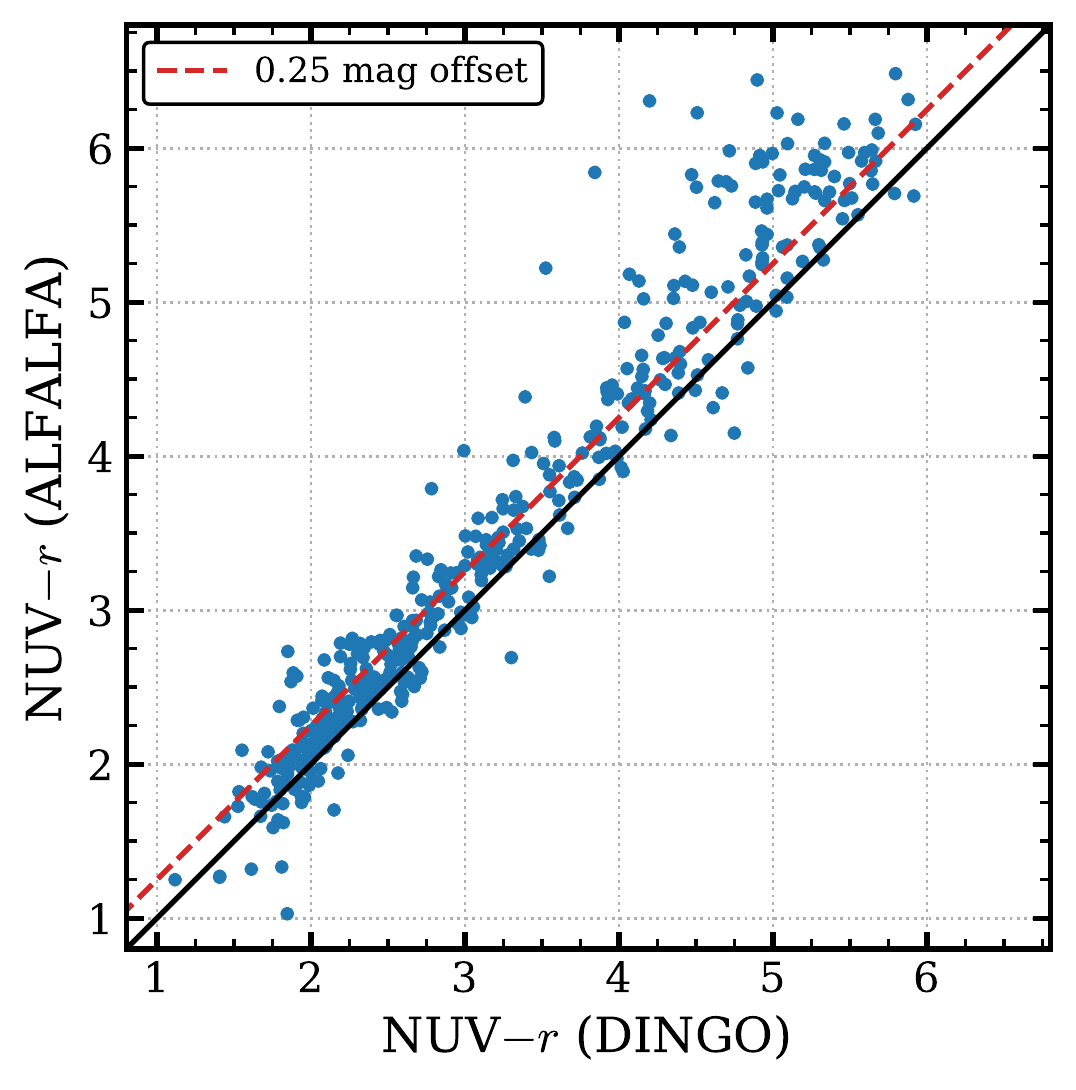} 
\caption{$NUV-r$ colour comparison between \citet{Brown:2015} and this work. The black solid and red dashed lines denote the one-to-one and 0.25 dex offset lines, respectively.}
\label{fig:compare_nuvr}
\end{figure}

The deviation in the {\HI} gas fraction vs. $NUV-r$ relation seen in Fig.~\ref{fig:scaling_relations}~(c) 
is probably due to a systematic difference in $NUV-r$ measurement.
Since other GAMA (equatorial) fields partially overlap with the ALFALFA survey area,
we cross-match the GAMA catalogues for the fields with the sample of \citeauthor{Brown:2015}, resulting in 435 common galaxies between the two samples. 
As seen in Fig.~\ref{fig:compare_nuvr}, the $NUV-r$ values from \citeauthor{Brown:2015} are systematically higher than the GAMA values.
The offset in $NUV-r$ becomes larger as the $NUV-r$ colour becomes redder, particularly for $NUV-r > 4$. The red galaxy sample also reflects the larger offset as seen in Fig.~\ref{fig:scaling_relations}~(a) and (b). 
Correction of the colour offset allows the scaling relations to be shifted horizontally, removing the discrepancy. For this scaling relation, there are no simulation data available in \citet{Dave:2020}.

The scaling relation with sSFR shown in Fig.~\ref{fig:scaling_relations}~(d) are consistent with xGASS. \citeauthor{Brown:2015} did not explore this relation. All simulations also reproduce well the observed trend of increasing {\HI} gas fraction with sSFR, although EAGLE shows a significant drop in {\HI} gas fraction at low sSFR.

\subsection{{\HI}--Stellar Mass Relation}
\begin{figure}
\centering
\includegraphics[width=0.45\textwidth]{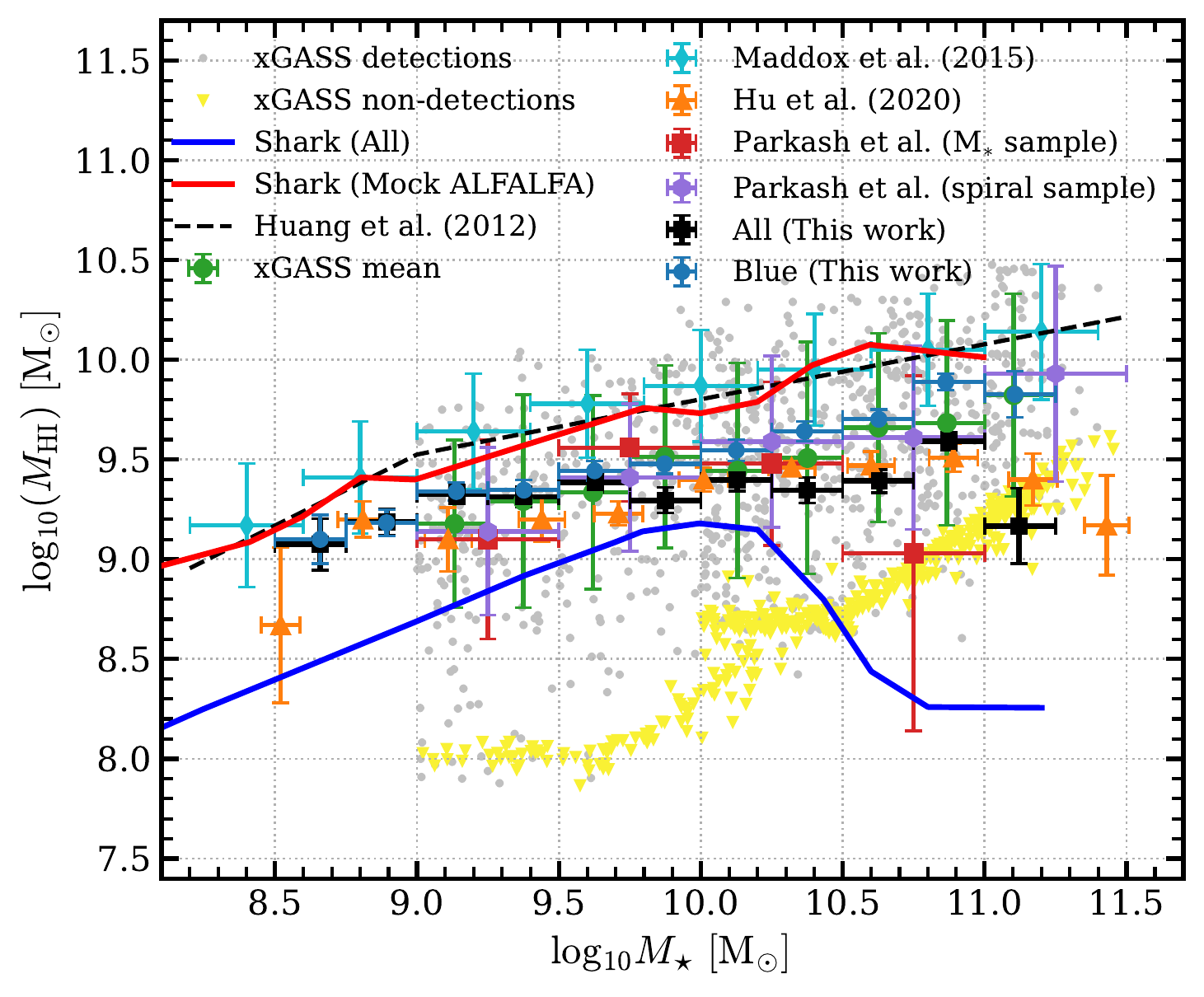} 
\caption{The average {\HI} mass as a function of stellar mass for all our sample (black squares) and a sub-sample of blue galaxies  (blue circles). Grey points and yellow downward triangles are xGASS detections and $5 \sigma$ upper limits for non-detections, respectively. The green circles denote the mean of linear values of all the xGASS sample. WSRT stacking measurements by \citet{Hu:2020} are overlaid with the orange triangles, and ALFALFA measurements \citep{Maddox:2015} are also presented with cyan diamonds for comparison. The red squares and purple circles are from the HIPASS-WISE sample \citep{Parkash:2018}. The black dashed line is the {\HI}-stellar mass relation based on ALFALFA from \citet{Huang:2012}. The red and blue solid lines represent predictions from a semi-analytic galaxy evolution model \citep[{\sc Shark},][]{Chauhan:2019} with ALFALFA survey selection criteria applied and no selection function adopted, respectively.}
\label{fig:hism_relation}
\end{figure}

The average {\HI} mass measured in each stellar mass bin is presented in Fig.~\ref{fig:hism_relation}. The black squares and blue circles indicate the measurements from our entire sample and the blue galaxy sample, respectively. We compare our measurements with observational data, such as the xGASS representative sample catalogue \citep{Catinella:2018}, the ALFALFA 40 per~cent catalogue \citep[$\alpha$.40,][]{Maddox:2015}, the HIPASS-WISE sample \citep{Parkash:2018}, {\HI} stacking measurements using the WSRT-SDSS data, and simulated data using a semi-analytic model \citep{Chauhan:2019}. 

The $\alpha$.40 data is an upper limit of the {\HI}--stellar mass relation because  only directly-detected {\HI} galaxies were used to derive the relation. This relation is in excellent agreement with \citet{Huang:2012} who used the same $\alpha$.40 data.   
By and large, our galaxy sample is in good agreement with the result of \citet{Hu:2020} who also adopted an {\HI} stacking technique, recovering a trend of gradually increasing {\HI} mass as a function of stellar mass. 
Our results also seem to be consistent with the xGASS mean relation within the large uncertainties of the xGASS data, except for the measurement of all the sample galaxies at highest stellar mass bin ($M_{\star} > 10^{11}\ {\msun}$) which is close to the xGASS upper limits (yellow downward triangles). Additionally, the relation for our entire sample starts to deviate from the increasing trend at around $M_{\star} > 10^{10}\ {\msun}$. This is because the sub-sample of gas-poor red galaxies begins to dominate this stellar mass range as seen in the top left panel of Fig.~\ref{fig:hist_smass_mu_nuvr_ssfr}.
This can also explain why our  relation and that of \citet{Hu:2020} both decline at the highest mass bins of  $M_{\star} > 10^{11}\ {\msun}$ where red early-type galaxies are the main population.

\citet{Parkash:2018} use an {\HI}-selected sample from HIPASS to show that the {\HI}--stellar mass relation increases is a similar manner to that shown by \citet{Huang:2012} and \citet{Maddox:2015}, albeit with a slightly different shape. However, their relation from a sample based on stellar mass appears to be flatter due to the significant contribution of early-type galaxies as also seen in our relation.

The simulated data from the {\sc Shark} semi-analytic model show the same phenomenon in its predictions. \citet{Chauhan:2019} constructed a mock ALFALFA survey data (red line) by applying the ALFALFA survey selection function to the original {\sc Shark} data (blue line). The mock data are well consistent with the $\alpha$.40 data, while the original data are systematically lower than the mock ALFALFA. They explained that this difference is due to a selection bias of the ALFALFA survey preferentially selecting gas-rich galaxies. They also ascribed a drop in the original data at $M_{\star} > 10^{10.3}$ to the impact of gas-poor elliptical galaxies being dominant in the mass range, as seen in our entire sample, \citet{Hu:2020} and the stellar mass sample of \citet{Parkash:2018}. 

The differences of the {\HI}-stellar mass relation between different studies therefore appears to be mainly due to sample selection bias. {\HI}-selected samples explicitly show an increasing trend while other samples are dominated by red and gas-poor galaxy populations, leading to a flatter relation at high stellar mass.

\subsection{{\HI}--Halo Mass Relation}
\begin{figure}
\centering
\includegraphics[width=0.45\textwidth]{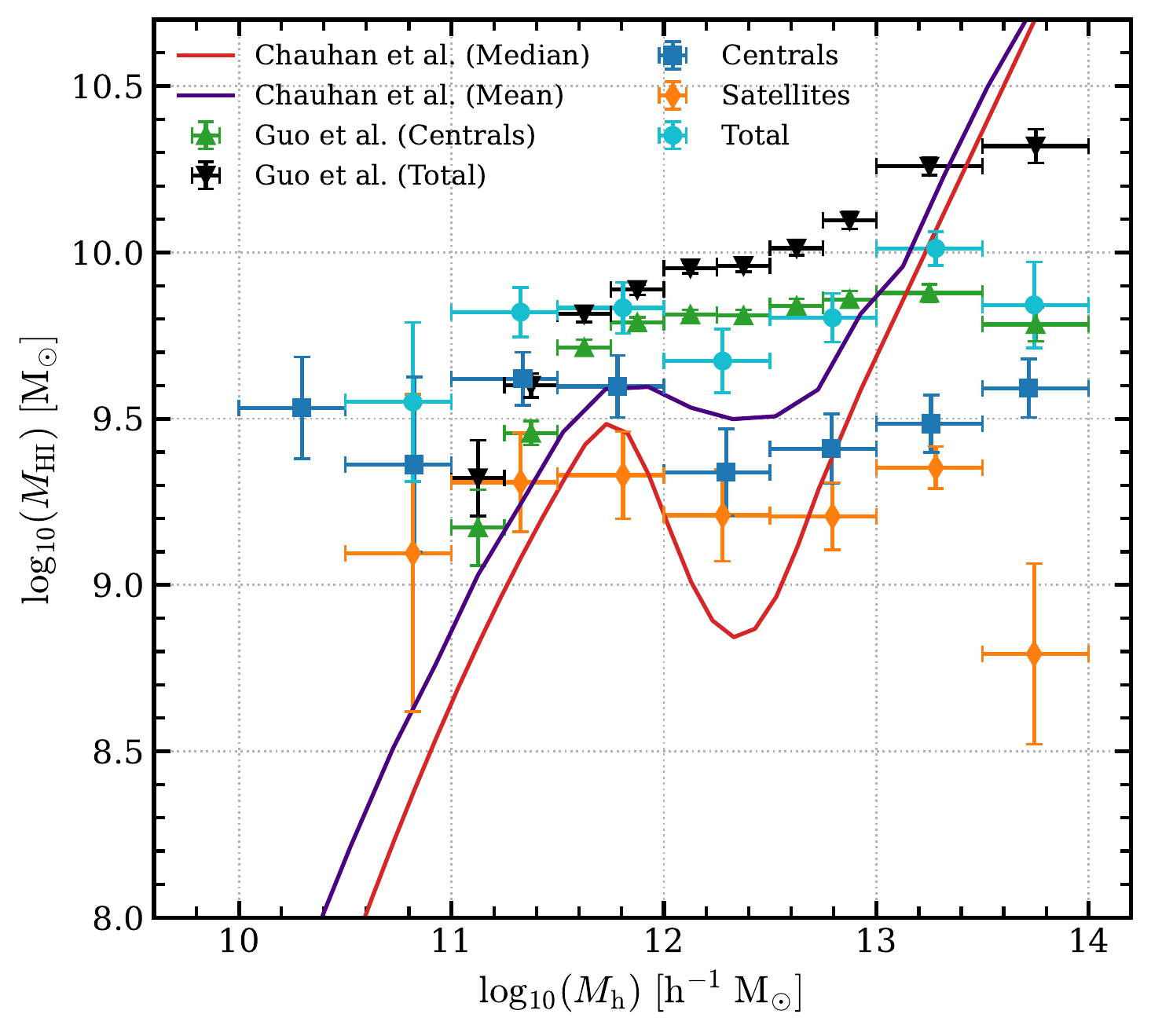} 
\caption{The {\HI}--halo mass relation. Blue squares and orange diamonds indicate the average {\HI} masses of group central and satellite galaxies from the DINGO early science data as a function of halo mass, respectively. The average total {\HI} masses of groups calculated with Eq.~\ref{eq:total_MHI} are overlaid with cyan circles. We compare our relation with the ALFALFA stacking results for galaxy groups having the number of group members of $N_{\rm g} \geq 2$ from Table~A1 and A2 of \citet{Guo:2020}. The measurements from central galaxies only and all entire groups are shown with green triangles and black downward triangles, respectively. Relations predicted from a semi-analytic galaxy evolution model ({\sc Shark}) are denoted with red (median) and purple (mean) lines, respectively \citep{Chauhan:2020}.}
\label{fig:hihm_relation}
\end{figure}

\begin{figure}
\centering
\includegraphics[width=0.45\textwidth]{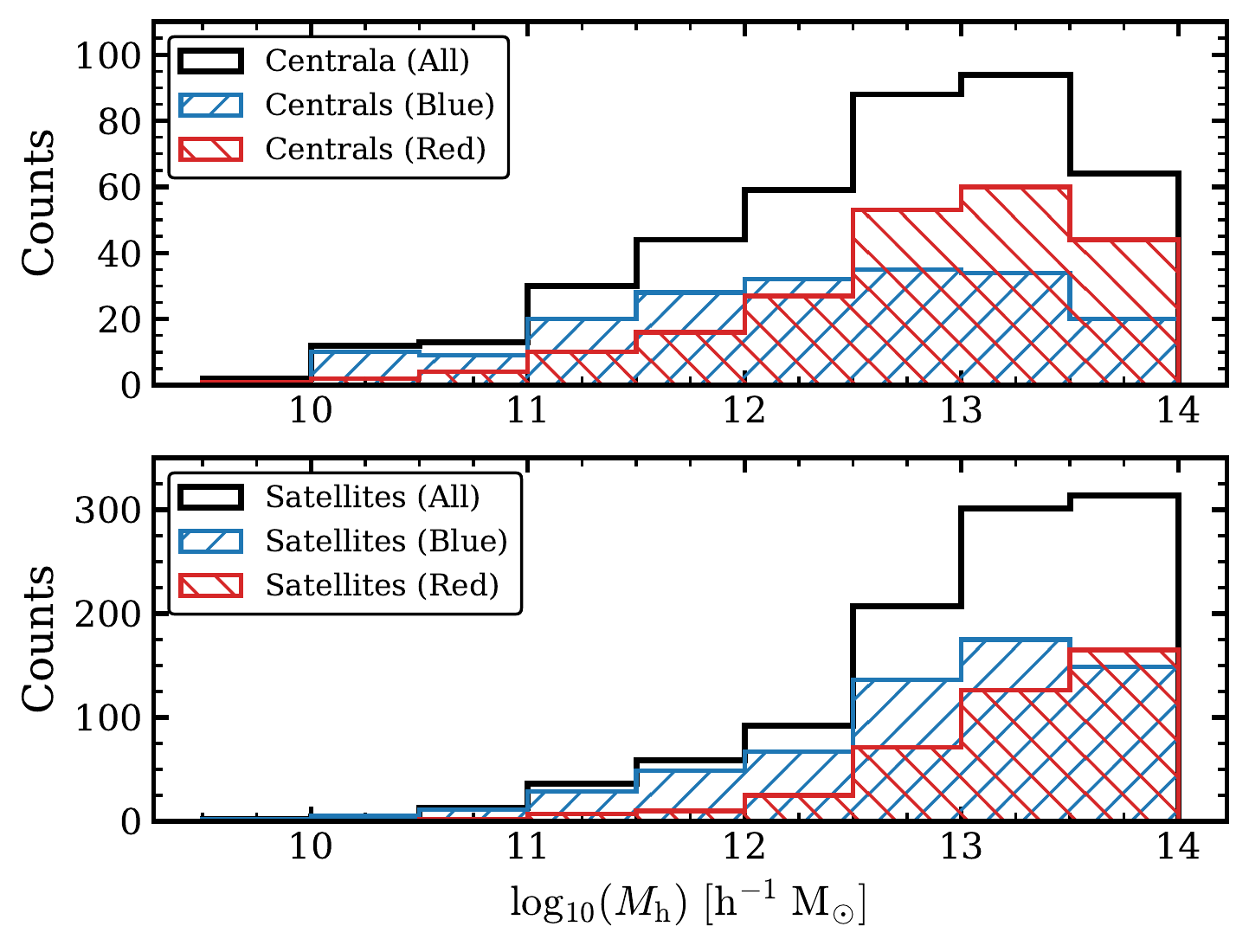} 
\caption{The halo mass distribution of central ({\it upper}) and satellite ({\it lower}) galaxies. In each panel, the sample galaxies are sub-divided into blue and red subsamples based on the $g-r$ colour cut used in Section~\ref{sec:blue_red}.}
\label{fig:hmass_distribution}
\end{figure}

For the {\HI}--halo mass relation, we only selected group central and satellite galaxies, corresponding to galaxy groups with more than one group member ($N_{\rm g} \geq 2$). Average {\HI} masses of centrals and satellites are separately measured using {\HI} stacking for each halo mass bin in the range $10^{10}\ h^{-1} {\msun} < M_{\rm h} < 10^{14}\ h^{-1} {\msun}$. Then, the average total {\HI} mass in individual halo mass bins ($\langle \MHI \rangle_{{\rm total}, M_{\rm h}}$) is calculated as follows:
\begin{equation}
 \label{eq:total_MHI}
 \langle \MHI \rangle_{{\rm total}, M_{\rm h}} = \frac{N_{{\rm c}, M_{\rm h}} \times \langle \MHI \rangle_{{\rm c}, M_{\rm h}} + N_{{\rm s}, M_{\rm h}} \times \langle \MHI \rangle_{{\rm s}, M_{\rm h}}} {N_{{\rm group}, M_{\rm h}}},
\end{equation}
where $N_{{\rm c}, M_{\rm h}}$ and  $N_{{\rm s}, M_{\rm h}}$ are the number of central and satellite galaxies co-added in a halo mass bin, respectively. 
$\langle \MHI \rangle_{{\rm c}, M_{\rm h}}$ and $\langle \MHI \rangle_{{\rm s}, M_{\rm h}}$ correspond to average {\HI} mass measurements for centrals and satellites in each halo mass bin.
$N_{{\rm group}, M_{\rm h}}$ is the number of groups in the respective halo mass bins.
Table~\ref{tab:hihm_relation} lists the average total {\HI} mass of groups from this calculation as well as the stacked {\HI} mass measurements with halo mass of groups for centrals and satellites. 
Since this calculation assumes that contribution of intergalactic {\HI} gas to the total {\HI} gas in halos is negligible, our measurements can be regarded as a lower limit to the {\HI}--halo mass relation.

The {\HI}--halo relation is compared with those derived from other observational and simulated data \citep{Guo:2020, Chauhan:2020} in Fig.~\ref{fig:hihm_relation}. \citet{Guo:2020} inferred their {\HI}--halo mass relation from ALFALFA and SDSS data using an {\HI} stacking method (black and green triangles). But instead of stacking {\HI} spectra of individual galaxies, they co-added {\HI} spectra from entire galaxy
groups to estimate average total {\HI} mass of groups based on the SDSS galaxy group catalogue \citep{Lim:2017}, so-called {\HI} group stacking. In their work, the average total {\HI} mass increases with halo mass except for a plateau in the halo mass range of $10^{11.8}\ h^{-1} \msun< M_{\rm h} < 10^{13}\ h^{-1} \msun$. But it starts to increase again at  higher  halo masses ($ M_{\rm h} > 10^{13}\ h^{-1} \msun$) while the plateau found in central galaxies below this halo mass extends to higher halo masses. They asserted that a three-phase scenario of {\HI}-rich galaxy formation could explain the trend. In the lower halo mass regime, smooth cold gas accretion seems to be associated with the increase of {\HI} gas content. The flat relation in the more massive halo mass range of $10^{11.8}\ h^{-1} \msun< M_{\rm h} < 10^{13}\ h^{-1} \msun$, called the transition mass range hereafter, is likely due to the effect of virial halo shock-heating and Active Galactic Nuclei (AGN) feedback that leads to reduction in {\HI} gas supply.

The {\sc Shark}  semi-analytic model of galaxy formation \citep{Lagos:2018} predicts an {\HI}--halo mass relation  with a dip between $10^{11.8}\ h^{-1} \msun$ and $10^{13}\ h^{-1} \msun$ \citep{Chauhan:2020}. The dip is predicted to occur again because of AGN feedback. There is some evidence for this in our measurements. As seen in Fig.~\ref{fig:hihm_relation}, the shape of our {\HI}--halo mass relation is similar to the predictions of \citet{Chauhan:2020} up to the transition mass range. For $ M_{\rm h} > 10^{13.5}\ h^{-1} \msun$, the average total {\HI} mass decreases due to reduced contribution of satellite galaxies to total {\HI} mass. However, the simulation predicts significant contribution from satellites as large galaxy groups or clusters having a large number of satellites normally span the highest halo mass range (see Table~\ref{tab:hihm_relation}). In this dense environment, satellites tend to be red and {\HI} deficient \citep[e.g.][]{Giovanelli:1985, Marasco:2016} because of environmental processes such as ram pressure stripping \citep{Bravo-Alfaro:2000, Chung:2007, Yoon:2017, Stevens:2017, Stevens:2019}. Indeed, \citet{Brown:2017} conducted an {\HI} stacking analysis to explore environmental effects in satellite galaxies, showing significant depletion of {\HI} gas of satellite galaxies in large group and cluster regimes of $ M_{\rm h} > 10^{13} \msun$ where ram-pressure stripping is suggested as a dominant environmental driver of {\HI} gas removal. Similarly, as shown in Fig.~\ref{fig:hmass_distribution}, our sample in the highest halo mass range are dominated by red galaxies, which likely makes our relation different from the results of simulations and other observations. 

The discrepancy of the {\HI}--halo mass relations between observations and theoretical predictions can stem from a mixture of different stacking techniques (group vs. individual galaxy stacking), different halo mass estimation methods and unknown underlying physics \citep{Chauhan:2021}. 
We have assumed that there is no contribution of {\HI} from intra-group or inter-galactic medium to the average total {\HI} mass of groups. The {\sc Shark} prediction is also based on the same assumption. But these assumptions 
may have to be modified if significant amounts of neutral gas are present in the IGM \citep[e.g.][]{Serra:2015b, Kleiner:2021, Namumba:2021}.
Future work with the DINGO pilot survey data with the ASKAP full 36-array 
will enable us to study larger galaxy samples with selection effects minimised.
We will also be able to scrutinise the discrepant {\HI}--halo mass relations using common DINGO/ALFALFA samples of galaxies 
because of the overlap between the GAMA and ALFALFA survey areas. 

\begin{table}
  \caption{Measurements of the {\HI}--halo mass relation for the different DINGO early science samples shown in Fig.~\ref{fig:hmass_distribution}: groups, centrals and satellites. Note that only groups having more than one group member were used for the {\HI}--halo mass scaling relation.  
    $N_{\rm g}$ is the number of groups or galaxies used for the stacked {\HI} mass measurements of each halo mass bin.}
\label{tab:hihm_relation}
\centering
\begin{tabular}{lcrc}
 \hline
   Sample  &  ${\rm log} M_{\rm h}$ &  $N_{\rm g}$ & ${\rm log} \langle \MHI \rangle$ \\
 \hline
Group       & 10.82 & 13 & $9.55~\pm~0.24$ \\
                 & 11.33 & 30 & $9.82~\pm~0.07$ \\
                 & 11.81 & 44 & $9.83~\pm~0.08$ \\
                 & 12.28 & 59 & $9.67~\pm~0.10$ \\
                 & 12.79 & 88 & $9.80~\pm~0.07$ \\
                 & 13.28 & 94 & $10.01~\pm~0.05$ \\
                 & 13.74 & 64 & $9.84~\pm~0.13$ \\
  & & & \\
 Centrals    & 10.30 & 12 & $9.53~\pm~ 0.15$ \\
                  & 10.83 & 13 & $9.36~\pm~0.26$ \\
                  & 11.34 & 30 & $9.62~\pm~0.08$ \\
                  & 11.78 & 44 & $9.60~\pm~0.09$ \\
                  & 12.29 & 59 & $9.34~\pm~0.13$ \\
                  & 12.79 & 88 & $9.41~\pm~0.10$ \\
                  & 13.26 & 94 & $9.49~\pm~0.09$ \\
                  & 13.72 & 64 & $9.59~\pm~0.09$ \\

  & & & \\
 Satellites   & 10.82 & 13 & $9.10~\pm~0.48$ \\
                  & 11.33 & 36 & $9.31~\pm~0.15$ \\
                  & 11.81 & 59 & $9.33~\pm~0.13$ \\
                  & 12.28 & 92 & $9.21~\pm~0.14$ \\
                  & 12.79 & 207 & $9.21~\pm~0.10$  \\
                  & 13.28 & 301 & $9.35~\pm~0.06$ \\
                  & 13.74 & 314 & $8.79~\pm~0.27$ \\

  \hline
\end{tabular}
\end{table}


\section{Cosmic {\HI} mass density}
The cosmic {\HI} mass density ({\OHI}) measures the co-moving density relative to the present-day critical density \citep[{\rhoCrit}, ][]{Meyer:2017}. Its evolution is an important factor in determining the veracity of cosmological simulations which balance gas accretion with consumption, and it is an important test of feedback models and gas-phase balance. 
To calculate {\rhoHI}, we can make use of $\langle \MHI / L_{r} \rangle$ ratio with the optical luminosity density ({\rhoLr}) derived for GAMA galaxies \citep{Loveday:2015}, as follows:
\begin{align}
\label{eq:omega_HI}
{\rhoHI} & = \bigg \langle \frac{\MHI}{L_{r}} \bigg \rangle \times {\rhoLr}.
\end{align}

Although the GAMA survey overall achieved a very high spectroscopic completeness of 98 per~cent to a limiting magnitude of $r < 19.8$ mag, G23 has a relatively lower completeness of 94 per~cent with a limit of $i < 19.0$ mag \citep[see ][]{Liske:2015, Bellstedt:2020}. We introduce a correction factor to compensate for the small level of incompleteness which may affect faint and gas-rich populations, as previous {\HI} stacking works did, to derive {\OHI} \citep[e.g. ][]{Delhaize:2013, Rhee:2013}. We apply this correction factor only to our blue sample because it is the dominant population subject to the incompleteness effect and could, in principle, be more gas-rich, as seen in Fig.~\ref{fig:hist_smass_mu_nuvr_ssfr} (a) of the scaling relation between {\HI} fraction and stellar mass. But the red galaxy sample is dominant only in the high mass range (i.e. the bright end of the luminosity function) and gas-poor.

To calculate such a correction factor, we follow the method used in \citet{Delhaize:2013} because we measure $\langle \MHI / L_{r} \rangle$ as they did:
\begin{align}
\label{eq:correction}
  f_{\rm corr} & = \bigg \langle \frac{\MHI}{L} \bigg \rangle_{\rm all} \bigg / \bigg \langle \frac{\MHI}{L} \bigg \rangle_{\rm observed}, \nonumber \\
               & = \frac{\int \MHI(L)\, \phi(L)\, dL}{\int L \, \phi(L) \, dL} \bigg / \frac{\int \MHI(L)\, N(L)\, dL}{\int \, N(L)\, dL},
\end{align}
where $\phi(L) $ is the optical luminosity function expressed as a Schechter function form and {\MHI}(L) is a relation of {\HI} mass with luminosity, assuming: 
\begin{align}
\label{eq:MHI_L}
\frac{\MHI}{L} \propto L^{\beta}.
\end{align}

\begin{figure}
\centering
\includegraphics[width=0.45\textwidth]{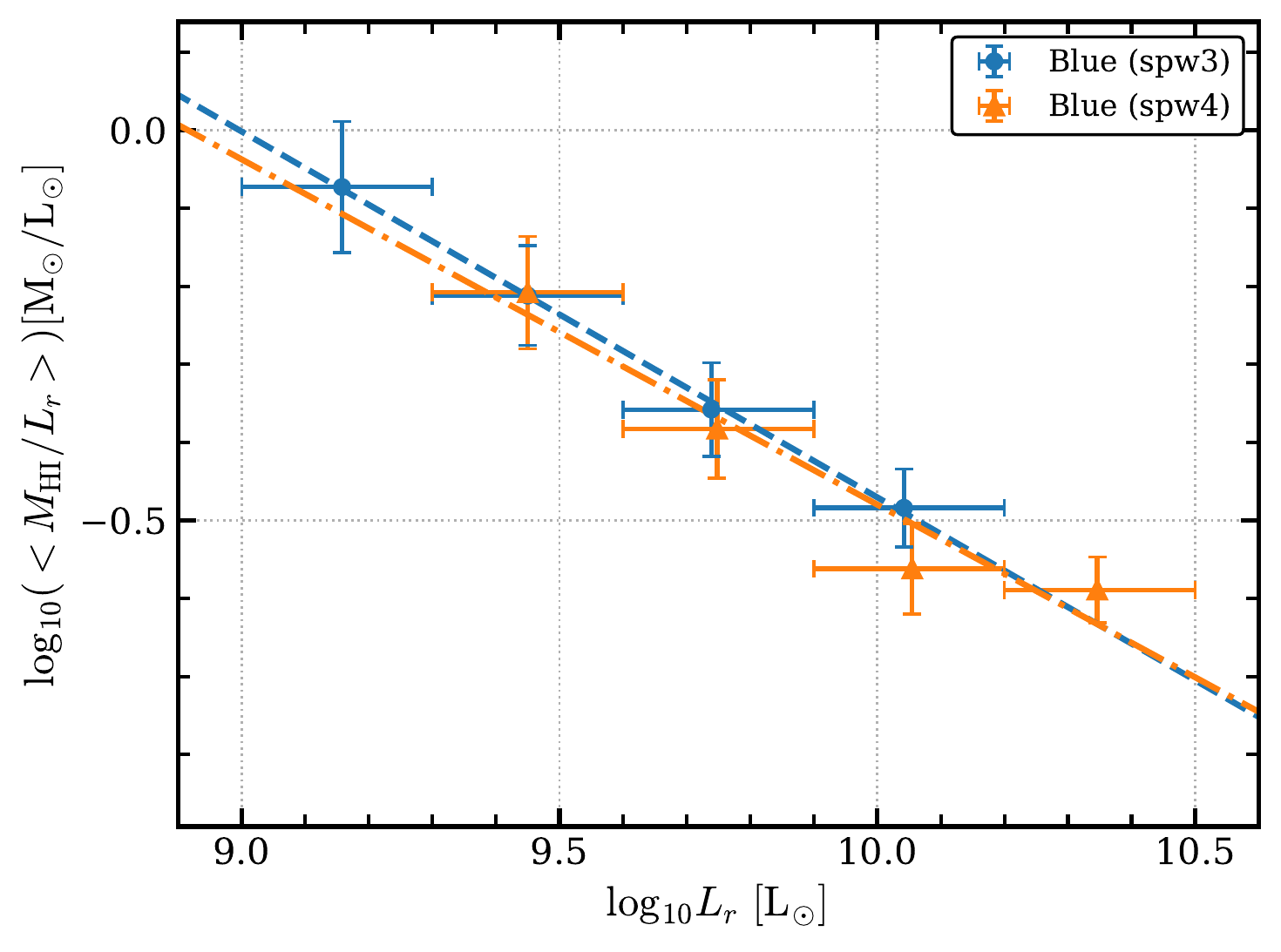} 
\caption{The relation of  $\langle \MHI / L_{r} \rangle$ ratio with luminosity for SPW3 (blue circles) and SPW4 (orange triangles). The dashed and dotted-dashed lines denote linear fits to SPW3 and 4 data, respectively.}
\label{fig:HI_L_relation}
\end{figure}

Fig.~\ref{fig:HI_L_relation} shows the relations derived separately for blue sample in two redshift bins (SPW3 and SPW4), giving us 
$\beta = -0.47, -0.44$ for SPW3 and SPW4, respectively. 
Given the relations, the correction equation above becomes:
\begin{align}
\label{eq:correction2}
  f_{\rm corr} & = \frac{(L^{*})^{\beta} \Gamma (2+\alpha+\beta)}{\Gamma (2+\alpha)} \times \frac{\sum w_{i} }{\sum L_{i}^{\beta} w_{i}},
\end{align}
where $L^{*}$ and $\alpha$ are the characteristic luminosity and the faint-end slope of the optical luminosity function from \citet{Loveday:2015}:
$L^{*} = 2.14 \times 10^{10}$ and $2.09 \times 10^{10}~\lsun$ for blue sample in SPW3 and SPW4, respectively, and $\alpha = -1.37$. 
$\beta$ is the slope of the relation of {\HI} mass-to-light ratio with luminosity shown in Fig.~\ref{fig:HI_L_relation}, $\Gamma$ is the complete gamma function, and 
$w_{i}$ is the weight used in stacking {\HI} spectra.
After applying this correction factor, the {\HI} mass density ({\rhoHI}) is divided by the present critical density ({\rhoCrit}) to obtain {\OHI}:
\begin{align}
\label{eq:omega_hi}
 {\OHI} & = \frac{\rhoHI \times f_{\rm corr}}{\rhoCrit}. 
\end{align}

\begin{table*}
 \centering
 \caption{The {\HI} mass density ({\rhoHI}) and cosmic {\HI} mass density ({\OHI}) values measured with $\langle \MHI / L_{r} \rangle$ ratio, the optical $r-$band luminosity density (${\rhoLr}$) and the correction factor ($f_{\rm corr}$) for two redshift bins.}
\begin{tabular}{lccccccccc}
\hline
  & \multicolumn{4}{c}{SPW3 ($z\sim0.057$)} & & \multicolumn{4}{c}{SPW4 ($z\sim0.080$)}  \\
\cline{2-5} \cline{7-10}
  Sample & $\langle \MHI / L_{r} \rangle$ & ${\rhoLr}$ & $\rhoHI$ & $f_{\rm corr}$ &  & $\langle \MHI / L_{r} \rangle$ & ${\rhoLr}$ & $\rhoHI$ & $f_{\rm corr}$ \\
 & [$\msun/\lsun$] &  [10$^{8}$~{\lsun}~Mpc$^{-3}$] & [10$^{8}$~{\msun}~Mpc$^{-3}$] &  &  & [$\msun/\lsun$] & [10$^{8}$~{\lsun}~Mpc$^{-3}$] & [10$^{8}$~{\msun}~Mpc$^{-3}$] &   \\
   \hline
   Blue         & 0.65~$\pm$~0.07 & 0.58~$\pm$~0.07 &  0.38~$\pm$~0.06 & 1.44 &  & 0.61~$\pm$~0.05 & 0.61~$\pm$~0.08 &  0.37~$\pm$~0.06 & 1.68  \\
   Red          & 0.16~$\pm$~0.06 & 0.18~$\pm$~0.42 &  0.03~$\pm$~0.07 & - &  & 0.02~$\pm$~0.01 & 0.19~$\pm$~0.44 &  0.004~$\pm$~0.010 & - \\
  \hline
  All (Blue+Red)    & \multicolumn{4}{c}{{\OHI} ($z~\sim0.057$) = 0.42~$\pm$~0.08} & & \multicolumn{4}{c}{{\OHI} ($z\sim0.080$) = 0.46~$\pm$~0.07}  \\
  \hline
\end{tabular}
\label{tab:OHI}
\end{table*}

\begin{figure*}
\centering
\includegraphics[width=0.95\textwidth]{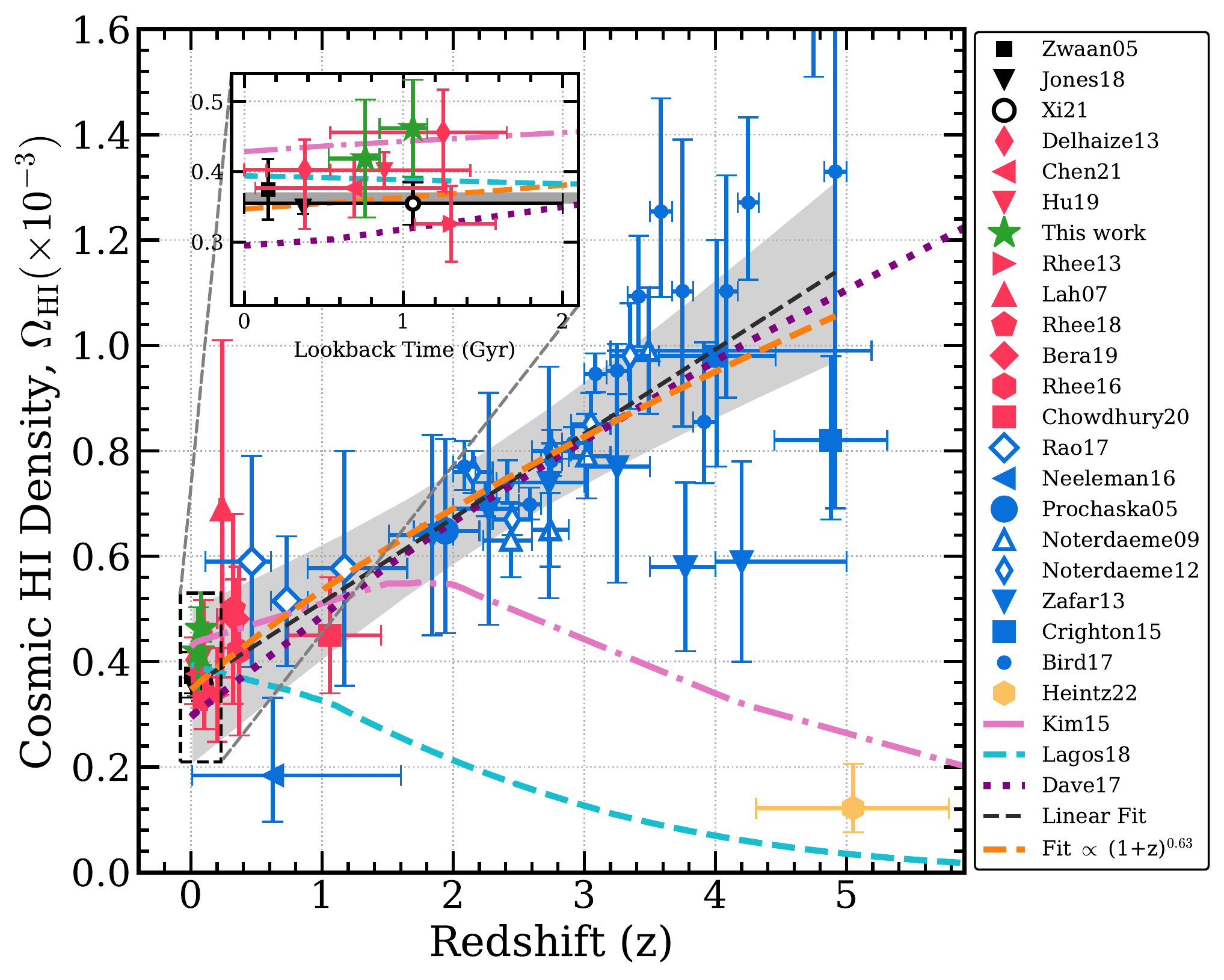}
\caption{Cosmic {\HI} gas density ({\OHI}) measurements as a function of redshift, colour-coded by observing methods: direct detection of {\HI} emission ({\it black}), {\HI} stacking ({\it red}; {\it green} for this work) and Damped Lyman $\alpha$ absorption (DLA, {\it blue}). The inset shows {\OHI} measurements as a function of lookback time over the last 2~Gyr. The green stars shows our {\HI} stacking measurements. All measurements are corrected to the same cosmological parameters used in this work and a consistent definition of {\OHI}. The black square and downward triangle at $z \sim 0$ are the HIPASS and ALFALFA 21-cm emission direct measurements by \citet{Zwaan:2005} and \citet{Jones:2018}, respectively. The empty black circle is the result from the Arecibo Ultra Deep Survey \citep[AUDS,][]{Freudling:2011} 100 per~cent data \citep{Xi:2021}. The red diamonds are the measurements using the Parkes telescope data \citep{Delhaize:2013}. The red left-pointing triangle is from DINGO-VLA data \citep{Chen:2021b}. The red downward and right-pointing triangles are measured using the WSRT \citep{Hu:2019, Rhee:2013}, respectively. All GRMT {\HI} stacking measurements \citep{Lah:2007, Rhee:2018, Bera:2019, Rhee:2016, Chowdhury:2020} are denoted with red triangle, pentagon, diamond, hexagon, and square, respectively. The blue open diamonds, left-pointing triangle, big circle, open triangles, open diamonds and small circles are DLA measurements from {\it HST}, SDSS, VLT, and Gemini by \citet{rao:2017, Neeleman:2016, Prochaska:2005, Noterdaeme:2009, Noterdaeme:2012, Zafar:2013, Crighton:2015, Bird:2017}, respectively. The yellow hexagon shows the recent {\OHI} value derived from {\CII} 158~${\umu}$m
emission of main-sequence star-forming galaxies at $z \approx 5$ \citep{Heintz:2022}. The grey dashed line with a shaded area shows a weighted linear fit of all {\OHI} measurements and its 95 per~cent confidence interval. The orange line is a power-law fit.
The predictions from two semi-analytic medels of galaxy formation and evolution \citep{Kim:2015, Lagos:2018} are denoted with pink dot-dashed and cyan dashed lines, respectively. The purple dotted line indicates {\OHI} predicted using cosmological hydrodynamical simulations of galaxy evolution \citep{Dave:2017}.}
\label{fig:omega_hi}
\end{figure*}

We derive {\OHI} separately for blue and red samples with a correction factor applied only to the blue sample and then obtain the total {\OHI} by summing {\OHI} values from both samples, giving ${\OHI} = (0.42 \pm 0.08) \times 10^{-3}$ at  $z \sim 0.057$ and ${\OHI} = (0.46 \pm 0.07) \times 10^{-3}$ at $z \sim 0.080$. 
Table~\ref{tab:OHI} lists the $\rhoHI$ and {\OHI} values derived for the two redshift bins and the corresponding parameters ($\langle \MHI / L_{r} \rangle$, {\rhoLr} and $f_{\rm corr}$) used for the calculation in Eq.~\ref{eq:omega_HI} and \ref{eq:omega_hi}. 

The {\OHI} values from the blue and red samples in the two redshift bins show that there is no or negligible evolution of the {\HI} gas content of galaxies in this narrow redshift span. 
The two {\OHI} measurements at $z\sim0.057$ and $0.080$ are compared with those from other studies in Fig.~\ref{fig:omega_hi}, where all {\OHI} measurements are colour-coded by observing method. The two large blind {\HI} surveys HIPASS and ALFALFA have constrained accurately
the {\OHI} in the local Universe based on the {\HI} mass functions (HIMF), and the derived {\OHI} values
({\it black} square and downward triangle, respectively) are in excellent agreement \citep{Zwaan:2005, Jones:2018}.
Beyond the local Universe, it becomes difficult to make {\OHI} measurements using directly detected {\HI}
21-cm emission due to the poor sensitivity of existing radio telescopes and inherent faintness of 21-cm emission.
Thus, only one direct measurement ({\it black} empty circle) has been made at $z \sim 0.16$ from a blind {\HI} survey -- the Arecibo Ultra Deep Survey \citep[AUDS, ][]{Xi:2021} which achieved an rms noise of $\sim 75 \mu$Jy per 4.5~{\kms}, but required a  large amount of observing time ($\sim 700$~hrs). From their data, an HIMF was constructed with 247 galaxies detected in {\HI}, thereby deriving {\OHI}. 
The {\OHI} measurement is consistent with those of the other blind surveys in the local Universe,
although it is subject to cosmic variance due to the smaller sky coverage of 1.35~${\rm deg^{2}}$
than the local blind surveys.

In the redshift regime at $z > 0.1$, {\HI} stacking has substantially contributed to measuring {\OHI} with a variety of radio telescopes, such as Parkes, VLA, WSRT, and GMRT \citep[{\it red} markers, ][]{Delhaize:2013, Chen:2021b, Rhee:2013,  Hu:2019, Lah:2007, Rhee:2016, Rhee:2018, Bera:2019, Chowdhury:2020}.
In particular, the recent measurement from the upgraded GMRT (uGMRT) at $z \sim 1.06$ \citep{Chowdhury:2020} is the highest-$z$ {\OHI} ever made with {\HI} stacking. This measurement plays an important role in bridging the gap between low and high redshifts ($0.5 < z < 2.0$). As Fig.~\ref{fig:omega_hi} shows, all {\HI} stacking measurements indicate that there is no conspicuous evolution of the {\HI} gas content of galaxies at $z < 0.5$ (corresponding to the last 4~Gyr). The work of \citet{Chowdhury:2020} extends this trend to 8~Gyr ($z \sim 1.0$).
Our measurements are compared in detail in the inset of Fig.~\ref{fig:omega_hi} where {\OHI} is plotted
as a function of lookback time, especially zooming in the last 2~Gyr. Our measurement at $z \sim 0.057$ is
in accordance with what was measured at $z \sim 0.051$ with the DINGO-VLA data \citep{Chen:2021b}.
They carried out a DINGO precursor project  using the VLA, covering a total sky area of $\sim 40~{\rm deg^{2}}$ (267 VLA pointings) in the GAMA 9 hr (G09) region out to $z < 0.1$ with a total integration time of $\sim 92$~hrs. A new {\HI} stacking technique was developed and applied to their {\HI} stacking analysis, called {\it cubelet} stacking \citep{Chen:2021a}, due to sparse $uv$-coverage in each pointing of the short integration time of $\sim$~28~min. The agreement between the DINGO early science and the DINGO-VLA data would be indicative of the feasibility of the new stacking technique. Now that the G09 area has also been observed as a part of the ASKAP observatory projects during the pilot survey phase, this measurement will be able to be directly compared with the DINGO pilot survey data. 
Our other {\OHI} measurement at $z \sim 0.080$ is also in line with other stacking and direct detection measurements using WSRT and AUDS at low redshifts \citep{Hu:2019, Xi:2021}.   

At higher redshifts, Damped Lyman $\alpha$ absorption (DLA) systems and their proxies ({\it blue} markers) are used to estimate {\OHI} values 
\citep{rao:2017, Neeleman:2016,Prochaska:2005, Noterdaeme:2009, Noterdaeme:2012, Zafar:2013, Crighton:2015, Bird:2017}.
These show a mild evolution in {\OHI}, at least by a factor of two in the range $0.5 < z < 5$.
The grey and orange fitting lines show an overall gradual increase in {\OHI} from $z \sim 0$ to $5$, which can be compared with theoretical predictions (pink dot-dashed and cyan dashed lines) from the semi-analytic models of galaxy formation and evolution \citep{Kim:2015, Lagos:2018}. This comparison reveals that, while the models reproduce the observations well at lower redshifts, there are discrepancies at higher redshifts  ($z >0.5$). This tension is related to semi-analytic simulations only accounting for {\HI} in the interstellar medium (ISM) of galaxies \citep{Lagos:2018}, whereas at higher redshift {\HI} is also distributed in the gaseous haloes around galaxies \citep[e.g.][]{van-de-Voort:2012}.
In contrast, the prediction of cosmological hydrodynamical simulations \citep{Dave:2017} seems to align well with observations, and follow our power-law fit (orange dashed line) to the observational values: ${\OHI}(z) = 3.5 (1 + z)^{0.63} \times 10^{-3}$, presumably because of the better treatment of gas accreting onto halos. 
In the local Universe ($z<0.1$),  the semi-analytic models are in better agreement with the measured values, as seen in the inset of Fig.~\ref{fig:omega_hi}.

However, recent studies by \citet{Heintz:2021, Heintz:2022} used {\CII} 158~${\umu}$m emission data from the ALMA Reionization Era Bright Emission Line Survey \citep[REBELS,][]{Bouwens:2022} to infer an empirical relation enabling one to convert {\CII} luminosity to {\HI} mass in galaxies, thereby deriving {\OHI} at $z > 5$. This {\OHI} value is lower by a factor of 7 to 11 than those of DLAs at similar redshifts \citep{Crighton:2015, Bird:2017}, which is closer to the predictions from two semi-analytic models rather than the hydrodynamical simulation. These observations do appear to provide conclusive evidence for the cause of the difference mentioned above between semi-analytic and hydrodynamical simulations.

\section{Summary and Conclusion}
We have observed the whole GAMA~23 region ($\sim$ 60~deg$^{2}$) with the ASKAP-12 array as part of the DINGO early science project. The quality of the data, which was processed with {\askapsoft}, the ASKAP data processing pipeline, is in a good agreement with the expectation based on the specification of the ASKAP telescope at the time.
The processed data achieve uniform sensitivity across a wide spatial and spectral coverage although there are some variations (20 per~cent) around the outermost edge of the full DINGO coverage which is larger than the nominal G23 survey area as shown in Fig.~\ref{fig:noise_profile}.
The characteristics of the processed ASKAP DINGO data demonstrate the excellent quality of the ASKAP data and the processing pipeline.
Using the high-quality DINGO early science data, we have searched for direct detections at $ z < 0.01$ and conducted {\HI} stacking experiments at $0.04 < z < 0.09$. 
Our main results are as follows:

\begin{itemize}
\item With the low redshift data (SPW0), we make direct detections of six sources $z < 0.01$ already known in HIPASS: NGC~7361, ESO469-G015, ESO407-G014, ESO406-G022, IC5271, and IC5269C. The fluxes of the five detections that ASKAP measured do not reproduce those of HIPASS due to the lack of short baselines in ASKAP-12. Additionally, we make a new direct {\HI} detection of ESO407-G011.

\item We find strong correlations between galaxy morphology based on optical colour and {\HI} gas properties using {\HI} stacking with galaxies at $0.04 < z< 0.09$. Blue galaxies have more {\HI} gas and larger {\HI} gas fractions than red galaxies in general. This trend does not seem to change significantly over the redshift range analysed here. For the red galaxy sample, the {\HI} gas abundance and fraction falls with increasing redshift. This is because the higher redshift bin includes more massive red galaxies, which contribute to the reduction in {\HI} gas mass and fraction compared to the lower redshift bin.

\item We statistically assess the impact of environment on {\HI} gas properties with three sub-samples of a total of 3799 galaxies based on the well-constructed GAMA group catalogue: group centrals, group satellites, and isolated centrals. Group central galaxies tend to have more {\HI} gas mass but lower gas fractions than group satellites and isolated central galaxies. The isolated centrals are the most {\HI} gas-rich of the three samples. We also find a further dependance of {\HI} properties on star formation properties and stellar mass of our sample galaxies sub-divided in the sSFR-stellar mass plane. For a given environment, galaxies with more active star formation tend to be more massive in {\HI} and gas-richer than the other samples. Lower stellar mass samples have higher {\HI} gas fraction and a longer depletion time in all the sub-categories,
indicating that lower stellar mass systems are at an earlier evolution stage.

\item Thanks to the extensive complementary data from GAMA, we derive a variety of {\HI} scaling relations with physical properties of our sample such as stellar mass, stellar mass surface density, $NUV-r$, sSFR, and halo mass. The DINGO early science data reproduce well the scaling relations associated with {\HI} gas fraction, in line with those of other observations \citep{Brown:2015, Catinella:2018} and three  state-of-the-art hydrodynamic simulations \citep{Dave:2020}. What is striking in our results is that the trends of the {\HI} scaling relations, especially with stellar mass and stellar mass surface density, persist through to lower stellar masses, lower stellar surface densities, and out to a higher redshift ($z \sim 0.09$) than previous studies have been able to explore, despite the short amount of observing time available.

\item The {\HI}--stellar mass relation for our blue galaxies is in agreement with the results from the xGASS representative sample and the WSRT stacking experiment by \citet{Catinella:2018} and \citet{Hu:2019}, respectively. 
Combining the blue and red samples reveals that the relation changes from the continual increase of {\HI} mass with stellar mass to a flat trend in the stellar mass range of $10^{9} < M_{\star} < 10^{10.75}$ along with a more complicated shape at higher stellar mass where {\HI}-deficient red galaxies are the dominant population. The impact of massive gas-poor populations is reproduced by a semi-analytic simulation \citep{Chauhan:2019}.  

\item We derive an {\HI}--halo mass relation with stacked {\HI} masses and halo masses measured for our group centrals and satellites for the first time, using interferometric data.
The {\HI}--halo mass relation is compared with the inferred relation using single-dish stacking measurements from ALFALFA by \citet{Guo:2020} and the theoretical relation that \citet{Chauhan:2020} derived using the {\sc Shark} semi-analytic galaxy formation model. Our relation provides some evidence for the `dip' feature that {\sc Shark} predicts in the transition halo mass range between 10$^{11.8}$ and 10$^{13}$ {\msun}, where the AGN feedback comes into play, not seen in the results of \citeauthor{Guo:2020}.

\item Using the {\HI} stacking measurements, we measure the cosmic {\HI} mass density ({\OHI}): ${\OHI} = (0.42~\pm~0.08) \times 10^{-3}$\ at $z \sim 0.057$ and $(0.46~\pm~0.07) \times 10^{-3}$ at $z \sim 0.080$, respectively.
These measurements are consistent with each other as well as other values at similar redshifts, confirming the findings from previous work \citep[e.g. ][]{Rhee:2018} that there is no significant evolution of the {\HI} gas content in galaxies over the last 4~Gyr.
\end{itemize}

In conclusion, this DINGO early science study demonstrates the power of {\HI} stacking techniques to study the {\HI} gas content of galaxies, environmental effects, the {\HI} scaling relations, and  {\HI} gas evolution. This covers many of the key DINGO scientific areas. 
In future work, we will conduct a more detailed investigation to understand better the relation of the {\HI} gas content of galaxies with various galaxy properties such as morphology and determine the systematics causing the differences found in the {\HI} scaling relations, using the DINGO pilot survey data taken with the ASKAP full array, which will increase the sample size and sensitivity of the DINGO early science data presented here. Eventually, the full DINGO survey will extend the current work to further higher redshifts and address the central questions of galaxy evolution, such as the interplay among galaxy constituents (gas, stars, dark matter) in various environments where galaxies reside.

\section*{Acknowledgments}
We are grateful to the anonymous referee for helpful comments that have improved this paper.
JR thanks Michael G.  Jones, Wenkai Hu and Garima Chauhan for providing useful information used in Fig.~\ref{fig:confusion}, Fig.~\ref{fig:hism_relation} and Fig~\ref{fig:hihm_relation}, respectively.
JR is grateful to Barbara Catinella, Luca Cortese and Ivy Wong for helpful comments and discussions. 
We thank Han-Seek Kim and  Claudia del P. Lagos for providing their model predictions used in Fig.~\ref{fig:omega_hi}. KR acknowledges support from the Bundesministerium fuer Bildung und Forschung (BMBF) award 05A20WM4.

The Australian SKA Pathfinder is part of the Australia Telescope National Facility which is funded by the Commonwealth of Australia for operation as a National Facility managed by CSIRO. This scientific work uses data obtained from the Murchison Radio-astronomy Observatory (MRO), which is jointly funded by the Commonwealth Government of Australia and State Government of Western Australia. The MRO is managed by the CSIRO, who also provide operational support to ASKAP. We acknowledge the Wajarri Yamatji people as the traditional owners of the Observatory site. This work was supported by resources provided by the Pawsey Supercomputing Centre with funding from the Australian Government and the Government of Western Australia.

This research was supported by the Australian Research Council Centre of Excellence for All Sky Astrophysics in 3 Dimensions (ASTRO 3D) through project number CE170100013. Parts of this research were conducted by the Australian Research Council Centre of Excellence for All-sky Astrophysics (CAASTRO), through project number CE110001020. 

GAMA is a joint European-Australasian project based around a spectroscopic campaign using the Anglo-Australian Telescope. The GAMA input catalogue is based on data taken from the Sloan Digital Sky Survey and the UKIRT Infrared Deep Sky Survey. Complementary imaging of the GAMA regions is being obtained by a number of independent survey programmes including {\it GALEX} MIS, VST KiDS, VISTA VIKING, {\it WISE}, {\it Herschel}-ATLAS, GMRT and ASKAP providing UV to radio coverage. GAMA is funded by the STFC (UK), the ARC (Australia), the AAO, and the participating institutions. The GAMA website is http://www.gama-survey.org/.
Based on observations made with ESO Telescopes at the La Silla Paranal Observatory under programme ID 179.A-2004.
Based on observations made with ESO Telescopes at the La Silla Paranal Observatory under programme ID 177.A-3016. 
We gratefully acknowledge DUG Technology for their support and HPC services.

This research made use of Astropy, a community-developed core Python package for Astronomy \citep{Astropy-Collaboration:2013}, Matplotlib \citep{Hunter:2007}, and Numpy \citep{Harris:2020}.

\section*{Data Availability}
The data underlying this article will be shared upon reasonable request to the corresponding author, JR.




\bibliographystyle{mnras}
\bibliography{references_dingo}







\bsp	
\label{lastpage}
\end{document}